\documentclass[aps]{revtex4}
\usepackage{eurosym}
\usepackage{amsfonts}
\usepackage{amsmath}
\usepackage{amssymb,epsf}
\usepackage{color}
\usepackage{hyperref}
\usepackage{orcidlink}

\begin{document}
\title{Born-Infeld AdS Black Holes Surrounded by Perfect Fluid Dark Matter}
\author{Behzad Eslam Panah\,\orcidlink{0000-0002-1447-3760}}
\email{eslampanah@umz.ac.ir}
\affiliation{Department of Theoretical Physics, Faculty of Basic Sciences, University of
Mazandaran, P. O. Box 47416-95447, Babolsar, Iran}
\author{Bilel Hamil\,\orcidlink{0000-0002-7043-6104}}
\email{hamilbilel@gmail.com/bilel.hamil@umc.edu.dz}
\affiliation{Laboratoire de Physique Math\'{e}matique et Physique Subatomique,LPMPS,
Facult\'{e} des Sciences Exactes, Universit\'{e} Constantine 1, Constantine,
Algeria}
\author{Manuel E. Rodrigues\,\orcidlink{0000-0001-8586-0285}}
\email{esialg@gmail.com}
\affiliation{Faculdade de F\'{\i}sica, Programa de P\'{o}s-Gradua\c{c}\~{a}o em F\'{\i}%
sica, Universidade Federal do Par\'{a}, 66075-110, Bel\'{e}m, Par\'{a},
Brazill}
\affiliation{Faculdade de Ci\^{e}ncias Exatas e Tecnologia, Universidade Federal do Par\'{a}, Campus Universit\'{a}rio de Abaetetuba, 68440-000, Abaetetuba, Par\'{a}, Brazil}

\begin{abstract}
With an intent to study the combined effect of Born-Infeld field as a
nonlinear electrodynamics (BI-NED) and Perfect Fluid Dark Matter (PFDM) to
Einstein-$\Lambda $ theory of gravity, we obtain the exact charged AdS black
hole solutions. Next, we study the effects of the parameters of PFDM and BI
on the event horizon of these black holes. Then, we compute the conserved
and thermodynamical quantities and evaluate that these thermodynamical
quantities can satisfy the first law of thermodynamics. In addition, we
investigate thermal stability conditions for these black hole solutions in
context of canonical ensemble by using the heat capacity and the Helmholtz
free energy. We study the effects of the parameters of PFDM and BI-NED on
the local and global stability areas. Next, we extend our study to
investigate thermodynamical properties of BI-NED AdS black holes surrounded
by PFDM in extended phase space by considering the cosmological constant as
a thermodynamics pressure. We, obtain the conserved and thermodynamic
quantities which can satisfy the first law of thermodynamics in extended
phase space. We then perform a rigorous analytical verification of the
Ehrenfest equations to determine whether the critical behavior of the
charged BI-NED AdS black holes surrounded by PFDM corresponds to a
second-order phase transition. Our findings indicate that both Ehrenfest
relations are satisfied, confirming that the black hole system undergoes a
second-order phase transition at the critical point. we obtain heat engines
corresponding to these black holes. The goal is to see how black holes'
parameters such as parameters of PFDM and BI-NED would affect efficiency of
the heat engines. Finally, we study the geodesic structure of the spacetime
by analyzing both timelike and null geodesics. The effective potential is
examined to determine the conditions for stable and unstable circular
orbits, as well as the behavior of the innermost stable circular orbit and
the photon sphere. We show that the PFDM parameter significantly modifies
the orbital structure and potential barrier, while the Born-Infeld parameter
produces comparatively weaker corrections. The existence of bound, critical,
and unstable trajectories highlights the dynamical impact of dark matter and
nonlinear electrodynamics on particle and photon motion around the black
hole.
\end{abstract}

\maketitle

\section{Introduction}

A primary motivation for studying Nonlinear Electrodynamics (NED) stems from
its inherent richness compared to the standard Maxwell theory; notably, NED
encompasses the linear Maxwell field theory as a specific limiting case.
Furthermore, theoretical shortcomings of the Maxwell field$-$such as its
inability to adequately describe radiation within certain materials \cite%
{Lorenci2001,Lorenci2002,Novello2012} and the challenge of modeling the
self-interaction of virtual electron-positron pairs \cite%
{Heisenberg,Yajima,Schwinger}$-$provide strong impetus for adopting NED
formalisms \cite{DelphenichQED}.

Crucially, the incorporation of NED formalism offers a pathway to resolve
the cosmological singularity at the Big Bang and the spacetime singularity
associated with black holes \cite{AyonLETT,Lorenci20022,Dymnikova,Corda2011}%
. Specifically, solutions for regular black holes within Einstein gravity
can be derived when coupled to a suitable NED \cite%
{AyonLETT,Soleng,Oliveira,Palatnik,Ayon1998}. Moreover, the influence of NED
effects becomes profoundly significant in highly magnetized compact
astrophysical objects, including neutron stars and pulsars \cite%
{Cuesta2004,Cuesta20042,Bialynicka}. Investigation into horizonless magnetic
solutions resulting from various nonlinear electromagnetic fields has also
been documented in the literature \cite{Magnetic1,Magnetic2}. A universal
and noteworthy characteristic shared by all NED models is that solutions
describing black holes coupled to these theories inherently satisfy both the
zeroth and first laws of thermodynamics.

The inherent ultraviolet divergence associated with the electric field of a
classical point charge at the origin necessitates the adoption of nonlinear
extensions to Maxwell's electrodynamics. To regularize this singularity,
Born and Infeld proposed an influential formulation in 1934, known as
Born-Infeld NED (BI-NED) \cite{Born,BornInfeld}. Subsequently, this
framework was extended by Hoffmann through its coupling to General
Relativity (GR) \cite{Hoffmann}. The resulting gravitational solutions
coupled to BI-NED have been extensively explored across various theoretical
domains. These investigations encompass the properties of static black hole
metrics \cite{Dehghani2006,Cai2008,Mazharimousavi}, the analysis of rotating
compact objects \cite{Hendi2008,Rastegar}, the construction of wormhole
geometries \cite{Lu,Dehghani2009,Eiroa,Hendi2014}, and modeling phenomena
within superconductors \cite{Jing2011,Gangopadhyay2012,Yao}. Moreover, the
theoretical relevance of BI-NED has been significantly renewed by its
natural emergence as the effective low-energy limit of open string theory 
\cite{Fradkin,Matsaev,Bergshoeff,Callan,Andreev,Leigh}.

A pivotal and enduring challenge in contemporary physics lies in the
definitive elucidation of the nature of dark matter (DM). Despite its
inferred contribution constituting approximately $27\%$ of the universe's
total critical energy density, direct empirical detection remains
unsuccessful, thereby cementing its status as one of the foremost unresolved
enigmas within the domains of cosmology and high-energy particle physics. A
comprehensive understanding of DM's structural properties is indispensable,
not solely for accurately characterizing the formation and evolutionary
mechanisms of large-scale cosmic structures and galactic morphologies, but
also for rigorously investigating its dynamic interplay with dark energy,
the cosmological constant responsible for driving the observed accelerated
expansion of the cosmos.

The Perfect Fluid Dark Matter (PFDM) model provides a compelling framework
for investigating the role of DM in astrophysical and cosmological
scenarios, particularly in relation to black holes. In contrast to standard
particle-based models, the perfect fluid model conceptualizes DM as a
continuous, non-viscous fluid governed by specific equations of state \cite%
{Kiselev2003}. This fluid-like representation enables the rigorous modeling
of DM distributions surrounding black holes and allows for an examination of
how these distributions modify the observable characteristics of black hole
spacetimes \cite{Qiao2023}.

The introduction of PFDM model fundamentally modifies the spacetime
geometries derived from GR. Extensive theoretical investigations have
established that incorporating PFDM alters the characteristic metrics of
established black hole solutions. Specifically, metric perturbations and
their associated physical consequences have been rigorously predicted and
analyzed for the vacuum Schwarzschild metric \cite{SchPFDM1,SchPFDM2}, the
charged Reissner-Nordstr\"{o}m (RN) solution \cite{RNPFDM}, the rotating
Kerr metric \cite{KerrPFDM1,KerrPFDM2}, the singularity-free Bardeen
solution \cite{BardeenPFDM}, and solutions derived from Euler-Heisenberg
electrodynamics \cite{EHPFDM1,EHPFDM2}. These metric deformations carry
significant physical ramifications, influencing observables such as
gravitational wave signatures \cite{GWPFDM1,GWPFDM2}, modifying geodesic
motion \cite{GeodesicPFDM}, and altering the conditions for spacetime
stability and thermodynamic phase transitions \cite%
{PhTranPFDM1,PhTranPFDM2,PhTranPFDM3}. Furthermore, the presence of PFDM
critically impacts several astrophysical signatures associated with black
holes, including the analysis of gravitational lensing \cite{GLPFDM}, the
size and appearance of their shadows \cite{ShadowPFDM1,ShadowPFDM2}, the
characteristics of accretion disk structures \cite{accdiskPFDM}, and the
deflection angles of light \cite{deflectionPFDM}.

The thermodynamic framework of black holes immersed in a PFDM background has
been a focal point of study \cite{ThermodynamicPFDM1,ThermodynamicPFDM2},
alongside investigations into their quasinormal modes \cite%
{QNMPFDM1,QNMPFDM2} and the calculation of their Gibbs free energy functions 
\cite{GFEPFDM1,GFEPFDM2,GFEPFDM3}. Moreover, both the structure of event
horizons \cite{EventPFDM} and the viability of wormhole solutions \cite%
{ModMaxPFDM} have demonstrated sensitivity to the PFDM background. More
recently, specific studies have focused on complex configurations, for
example; the thermodynamics and phase transitions of asymptotically ModMax
anti-de Sitter (AdS) black holes under PFDM influence have been examined in
Ref. \cite{ModMaxPFDM}, the impact of PFDM on particle dynamics surrounding
a static black hole situated within an external magnetic field have been
explored in Ref. \cite{MagneticPFDM}. Further investigations have
characterized the phase structure and critical behavior of charged-AdS black
holes \cite{AdSPFDM}, the photon orbits and phase transitions in Letelier
AdS black holes \cite{LADSPFDM}, and the combined thermodynamics and
geodesic structure \cite{TGLAdSPFDM} for these Letelier AdS solutions, all
within the context of a PFDM background.

The investigation of black hole phase transitions utilizes methodologies
from classical thermodynamics as well as the more modern field of
thermodynamic geometry. In the classical framework, notable analogues
include the relationship between charged AdS black holes and liquid-gas
systems, along with the application of Ehrenfest relations. Banerjee et al. 
\cite{Banerjee2011A,Banerjee2011B} enhanced this approach by developing a
novel formalism based on the Ehrenfest analogy between thermodynamic
variables and black hole parameters $\left( V\leftrightarrow Q, \text{ and}%
~P\leftrightarrow -U\right) $. Their analysis of black hole phase
transitions treated these systems as grand-canonical ensembles and
rigorously examined the Ehrenfest relations using both analytical and
graphical techniques.

Among the recent thermodynamical advances in the field of black holes
thermodynamics, one can point out to extended phase space and geometrical
thermodynamics. In the extended phase space, one consider the cosmological
constant in AdS black holes to be a thermodynamical quantity known as
pressure \cite{PLambda1,PLambda2,PLambda3,PLambda4}. Such consideration
results into introduction of a van der Waals behavior, the reentrant of
phase transition \cite{reentrant1,reentrant2}, existence of the triple point 
\cite{triple1,triple2} and possibility of having classical heat engine \cite%
{Heat1,Heat2,Heat3,Heat4,Heat5,Heat6,Heat7,Heat8,Heat9,Heat10,Heat11,Heat12,Heat13,Heat14,Heat15,Heat16,Heat17,Heat18,Heat19}%
. Furthermore, a new method for classification and analysis of the thermodynamic system was constructed based on topology \cite{TTD}. In this method, topological invariant numbers are calculated, which, when summed, yield a global topological invariant that classifies the thermodynamics as exhibiting a predominance of stable states, unstable states, or an equilibrium between them. We shall employ this method to classify our new solution.

The investigation of geodesic motion around compact objects constitutes an
essential tool for understanding the physical and geometrical properties of
neutron stars and black holes. The trajectories of test particles encode
valuable information about the underlying spacetime structure. A well-known
observational example is provided by the motion of stars in the vicinity of
Sagittarius A$^{\ast }$ at the center of the Milky Way galaxy \cite%
{Eckart,Gillessen}, which offers strong evidence for the presence of a
supermassive black hole. In black hole spacetimes, radial and circular
geodesics represent two fundamental classes of motion with direct
astrophysical relevance. Consequently, geodesic dynamics in various
gravitational backgrounds have been widely explored in the literature \cite%
{Hamil,Hamil1,Hamil2,Ahmad,Hamil3,Hamil4,Hackmann,Jamil,Fernando}. More
generally, spacetime curvature effects can be systematically analyzed
through the geodesic equation \cite{Synge,Pirani}, which governs the
relative acceleration between neighboring geodesics under different physical
conditions \cite{Kar,Kole1,Uniy1}. The geodesic equation therefore provides
a powerful framework for characterizing spacetime geometry, and its
solutions for timelike, null, and spacelike congruences yield important
insights into gravitational dynamics.

These considerations motivate us to investigate topological AdS black holes in the presence of Born–Infeld nonlinear electrodynamics and perfect fluid dark matter. Our aim is to explore how the corresponding parameters, together with the topological constant, influence the properties of the black hole solutions. In particular, we examine their impact on the horizon structure, thermodynamic quantities, and phase behavior. Furthermore, by treating the black holes as heat engines, we analyze how these parameters affect the engine efficiency. Finally, we study the timelike and null geodesic structures to understand how nonlinear electrodynamics, and dark matter, modify particle and photon motion in the resulting spacetime.

\section{Action and Field Equations}

The action of Einstein-$\Lambda $ gravity in the presence of BI-NED field
and DM is given by 
\begin{equation}
\mathcal{I}=\int d^{4}x\sqrt{-g}\left[ \left( \frac{R-2\Lambda +\mathcal{L}%
\left( \mathcal{F}\right) }{2\kappa ^{2}}\right) +\mathcal{L}_{PFDM}\right] ,
\label{action}
\end{equation}%
where $R$ is the Ricci scalar. Also, $\kappa ^{2}=8\pi $ because we set $%
c=G=1$ in the action (\ref{action}), where $c$, and $G$, respectively, are
the speed of light and the gravitational constant. In addition, $\mathcal{L}%
_{PFDM}$ is the Lagrangian density of PFDM. In the above action, $\Lambda $
is the cosmological constant, and $g$ refers to the determinant of the
metric tensor $g_{\mu \nu }$ (i.e., $g=\det \left( g_{\mu \nu }\right) $).
Furthermore, the term $\mathcal{L}\left( \mathcal{F}\right) $ is related to
the Lagrangian of BI-NED theory as 
\begin{equation}
L(\mathcal{F})=4\beta ^{2}\left( 1-\sqrt{1+\frac{\mathcal{F}}{2\beta ^{2}}}%
\right) ,  \label{LBI}
\end{equation}%
in the above Lagrangian, $\beta $ is the BI parameter. Also, $\mathcal{F}$
is the Maxwell invariant in the form $\mathcal{F}=F_{\mu \nu }F^{\mu \nu }$,
in which $F_{\mu \nu }$ is the electromagnetic field tensor and is defined
as $F_{\mu \nu }=\partial _{\mu }A_{\nu }-\partial _{\nu }A_{\mu }$, where $%
A_{\mu }$ is the gauge potential.

Now, we are in a position to obtain field equations. Applying variational
principle to the action (\ref{action}), one can find 
\begin{eqnarray}
G_{\mu }^{~\nu }-\frac{1}{2}g_{\mu }^{~\nu }L(\mathcal{F})+2L_{\mathcal{F}%
}F_{\mu \sigma }F^{\sigma \nu } &=&8\pi T_{\mu }^{~\nu },  \label{FE} \\
&&  \notag \\
\nabla _{\mu }\left( \sqrt{-g}L_{\mathcal{F}}F^{\mu \nu }\right) &=&0,
\label{Ftr}
\end{eqnarray}%
where $L_{\mathcal{F}}=\frac{dL(\mathcal{F})}{d\mathcal{F}}$ and $G_{\mu
}^{~\nu }=R_{\mu }^{~\nu }-\frac{1}{2}g_{\mu }^{~\nu }R$.

In Eq. (\ref{FE}), $T_{\mu }^{~\nu }$ is the energy momentum tensor of PFDM,
and it can be expressed in the following form 
\begin{equation}
T_{\mu }^{~\nu }=\left( 
\begin{array}{cccc}
-\rho & 0 & 0 & 0 \\ 
0 & p_{r} & 0 & 0 \\ 
0 & 0 & p_{\theta } & 0 \\ 
0 & 0 & 0 & p_{\varphi }%
\end{array}%
\right) =\left( 
\begin{array}{cccc}
\frac{b}{8\pi r^{3}} & 0 & 0 & 0 \\ 
0 & \frac{b}{8\pi r^{3}} & 0 & 0 \\ 
0 & 0 & \frac{-b}{16\pi r^{3}} & 0 \\ 
0 & 0 & 0 & \frac{-b}{16\pi r^{3}}%
\end{array}%
\right) ,
\end{equation}

We consider a static spherical symmetric in $4$-dimensional spacetime in the
following form 
\begin{equation}
ds^{2}=-\psi \left( r\right) dt^{2}+\frac{dr^{2}}{\psi \left( r\right) }%
+r^{2}(d\theta ^{2}+\sin ^{2}\theta d\varphi ^{2}),  \label{metric}
\end{equation}%
where $\psi \left( r\right) $ is the metric function.

Now, we consider a radial electric field which its related gauge potential
is 
\begin{equation}
A_{\mu }=h\left( r\right) \delta _{\mu }^{t},  \label{A}
\end{equation}

Using Eqs. (\ref{Ftr}), (\ref{metric}) and (\ref{A}), we obtain following
differential equations 
\begin{equation}
r\beta ^{2}H^{\prime }\left( r\right) -2H^{3}\left( r\right) +2\beta
^{2}H\left( r\right) =0,  \label{h(r)}
\end{equation}%
where $H\left( r\right) =h^{\prime }\left( r\right) =\frac{dh\left( r\right) 
}{dr}$, and prime denotes derivation with respect to radial coordinate. It
is a matter of calculation to show that 
\begin{equation}
H\left( r\right) =\frac{q}{r^{2}\Gamma },  \label{H(r)}
\end{equation}%
where $\Gamma =\sqrt{1+\frac{q^{2}}{\beta ^{2}r^{4}}}$ and $q$ is an
integration constant related to the electric charge.

By employing Eqs. (\ref{metric}), and (\ref{H(r)}) within Eq. (\ref{FE}), we
can find the components of the equations of motion (Eq. (\ref{FE})), which
are 
\begin{eqnarray}
eq_{tt} &=&\psi \left( r\right) +r\psi ^{\prime }\left( r\right) -\frac{%
2\beta ^{2}r^{3}+r+b}{r}+\frac{2\beta ^{2}r^{2}}{\Delta },  \label{eqtt} \\
&&  \notag \\
eq_{\theta \theta } &=&r\psi ^{\prime }\left( r\right) +\frac{r^{2}\psi
^{\prime \prime }\left( r\right) }{2}-2\beta ^{2}r^{2}\left( 1-\Delta
\right) +\frac{b}{2r},  \label{eqthethe}
\end{eqnarray}%
where $\Delta =\sqrt{1-\left( \frac{H\left( r\right) }{\beta }\right) ^{2}}$%
. In addition, $eq_{tt}=eq_{rr}$, and $eq_{\theta \theta }=eq_{\varphi
\varphi }$, where $eq_{tt}$, $eq_{rr}$, $eq_{\theta \theta }$ and $%
eq_{\varphi \varphi }$ are related to components of $tt$, $rr$, $\theta
\theta $ and $\varphi \varphi $ of the equations of motion (Eq. (\ref{FE})).

\section{Black Hole Solutions}

By considering equations of motion (Eqs. (\ref{eqtt}) and (\ref{eqthethe})),
we can obtain the metric function $\psi \left( r\right) $ in the following
form 
\begin{equation}
\psi \left( r\right) =1-\frac{2m_{0}}{r}-\frac{\Lambda r^{2}}{3}+\frac{%
2\beta ^{2}r^{2}\left( 1-\mathfrak{F}_{1}\right) }{3}+\frac{b}{r}\ln \left( 
\frac{r}{\left\vert b\right\vert }\right) ,  \label{f(r)}
\end{equation}%
where $\mathfrak{F}_{1}={}_{2}F_{1}\left( \left[ \frac{-1}{2},\frac{-3}{4}%
\right] ,\left[ \frac{1}{4}\right] ,1-\Gamma ^{2}\right) $ is the
hypergeometric function. Also $m_{0}$ is an integration constant related to
geometrical mass of the solution.

It is notable that the obtained solutions in Eq. (\ref{f(r)}) reduce to the
charged AdS black hole solutions in the presence of PFDM when $\beta
\rightarrow \infty $, i.e., 
\begin{equation}
\psi \left( r\right) =1-\frac{2m_{0}}{r}-\frac{\Lambda r^{2}}{3}+\frac{q}{%
r^{2}}+\frac{b}{r}\ln \left( \frac{r}{\left\vert b\right\vert }\right) .
\end{equation}

In addition, by considering $\beta \rightarrow \infty $, and $b=0$, the
metric function $\psi \left( r\right) $ (Eq. (\ref{f(r)})) turns to the
charged AdS black hole solutions in the following form 
\begin{equation}
\psi \left( r\right) =1-\frac{2m_{0}}{r}-\frac{\Lambda r^{2}}{3}+\frac{q}{%
r^{2}}.
\end{equation}

Our next step is to examine the geometrical structure of the solutions.
First, we will investigate the existence of essential singularities. For
this purpose, we calculate the Ricci and Kretschmann scalars of these
spacetimes.

We can get the Ricci scalar ($R$) of the solutions in the following form 
\begin{equation}
R=4\Lambda -8\beta ^{2}\left( 1-\mathfrak{F}_{1}\right)-\frac{b}{r^{3}}+%
\frac{12q^{2}\mathfrak{F}_{2}}{r^{4}}+\frac{8q^{4}\mathfrak{F}_{3}}{5\beta
^{2}r^{8}},  \label{R}
\end{equation}%
where $\mathfrak{F}_{2}$, and $\mathfrak{F}_{3}$ defined as 
\begin{eqnarray}
\mathfrak{F}_{2} &=&{}_{2}F_{1}\left( \left[ \frac{1}{2},\frac{1}{4}\right] ,%
\left[ \frac{5}{4}\right] ,1-\Gamma ^{2}\right) ,  \notag \\
&&  \notag \\
\mathfrak{F}_{3} &=&{}_{2}F_{1}\left( \left[ \frac{3}{2},\frac{5}{4}\right] ,%
\left[ \frac{9}{4}\right] ,1-\Gamma ^{2}\right) .
\end{eqnarray}

The Kretschmann scalar ($K=R_{\mu \nu \rho \sigma }R^{\mu \nu \rho \sigma }$%
) of this spacetime is%
\begin{eqnarray}
K &=&\frac{32\left( \beta ^{2}\left( \mathfrak{F}_{1}-1\right) +\frac{%
\Lambda }{2}\right) \left( \beta ^{2}\left( \mathfrak{F}_{1}-1\right) +\frac{%
2\Lambda r^{3}-b}{4r^{3}}\right) }{3}+\frac{32q^{2}\beta ^{2}\mathfrak{F}%
_{2}\left( \mathfrak{F}_{1}-1\right) +4\left( 1+4\Lambda q^{2}\mathfrak{F}%
_{2}\right) -4}{r^{4}}  \notag \\
&&  \notag \\
&&+\frac{8m_{0}\left( 5b-6\mathcal{B}_{1,0,1}\right) }{r^{6}}+\frac{\left(
12\ln ^{2}\left( \frac{r}{\left\vert b\right\vert }\right) -20\ln \left( 
\frac{r}{\left\vert b\right\vert }\right) +13\right) b^{2}}{r^{6}}+\frac{%
8q^{2}\mathfrak{F}_{2}\mathcal{B}_{6,7,12}}{r^{7}}  \notag \\
&&  \notag \\
&&+\frac{16q^{4}\left( 75\mathfrak{F}_{2}^{2}+4\mathfrak{F}_{3}\left( 
\mathfrak{F}_{1}-1+\frac{\Lambda }{2\beta ^{2}}\right) \right) }{15r^{8}}-%
\frac{16q^{4}\mathfrak{F}_{3}\mathcal{B}_{2,3,4}}{5\beta ^{2}r^{11}}-\frac{%
64q^{6}\mathfrak{F}_{2}\mathfrak{F}_{3}}{5\beta ^{2}r^{12}}+\frac{64q^{8}%
\mathfrak{F}_{3}^{2}}{25\beta ^{4}r^{16}},  \label{K}
\end{eqnarray}%
where $\mathcal{B}_{i,j,k}=ib\ln \left( \frac{r}{\left\vert b\right\vert }%
\right) -jb-km_{0}$. Whereas these scalar diverge at $r=0$, i.e., 
\begin{eqnarray}
\underset{r\rightarrow 0}{\lim }R &\rightarrow &\infty ,  \notag \\
&&  \notag \\
\underset{r\rightarrow 0}{\lim }R_{\mu \nu \rho \sigma }R^{\mu \nu \rho
\sigma } &\rightarrow &\infty ,
\end{eqnarray}%
\bigskip so there is an essential curvature singularity at $r=0$.

On the other hand, the asymptotical behavior of these solutions is (A)dS,
because 
\begin{eqnarray}
\underset{r\rightarrow \infty }{\lim }R &\rightarrow &4\Lambda ,  \notag \\
&&  \notag \\
\underset{r\rightarrow \infty }{\lim }R_{\mu \nu \rho \sigma }R^{\mu \nu
\rho \sigma } &\rightarrow &\frac{8\Lambda ^{2}}{3},
\end{eqnarray}

\begin{figure}[h]
\centering
\includegraphics[width=70mm]{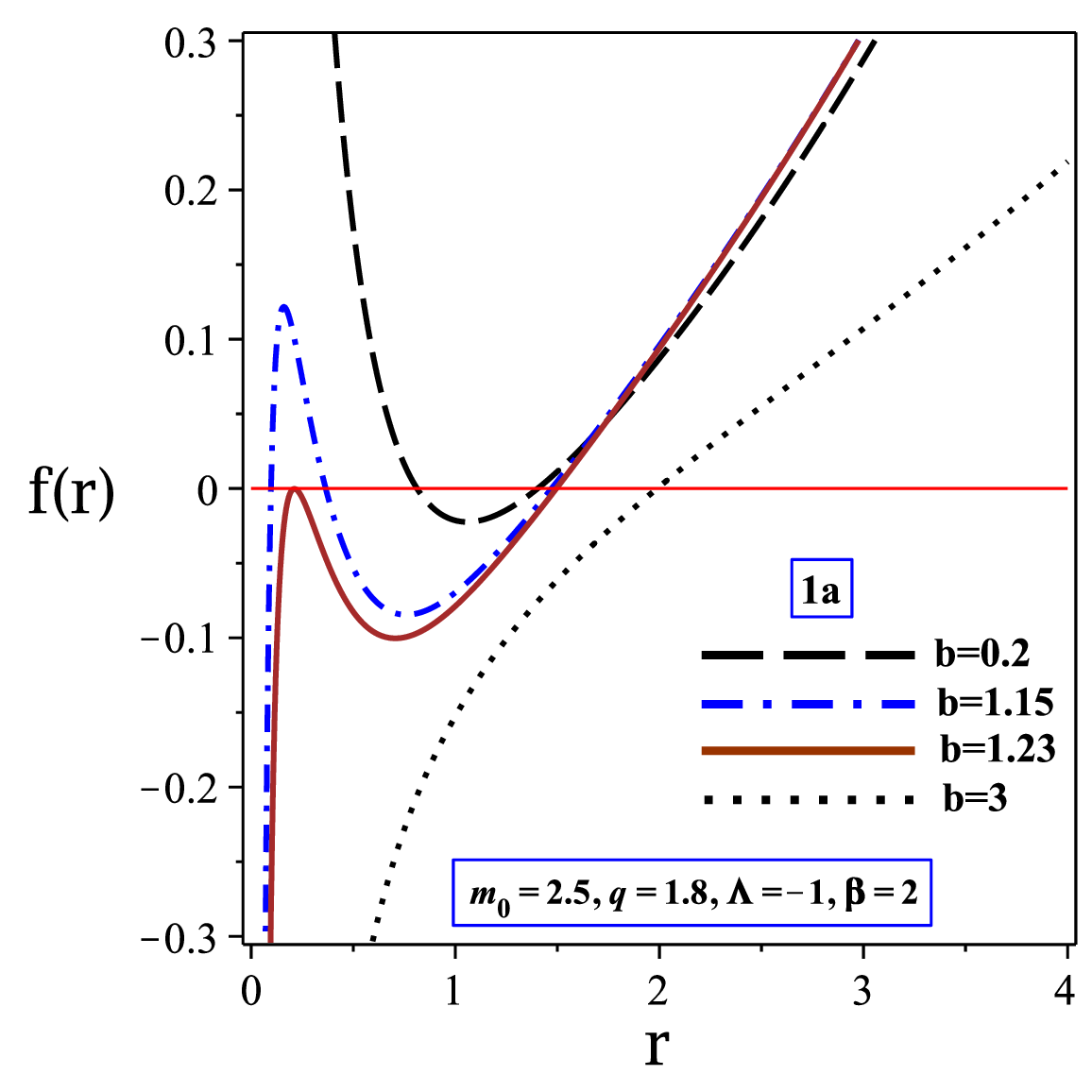} \includegraphics[width=70mm]{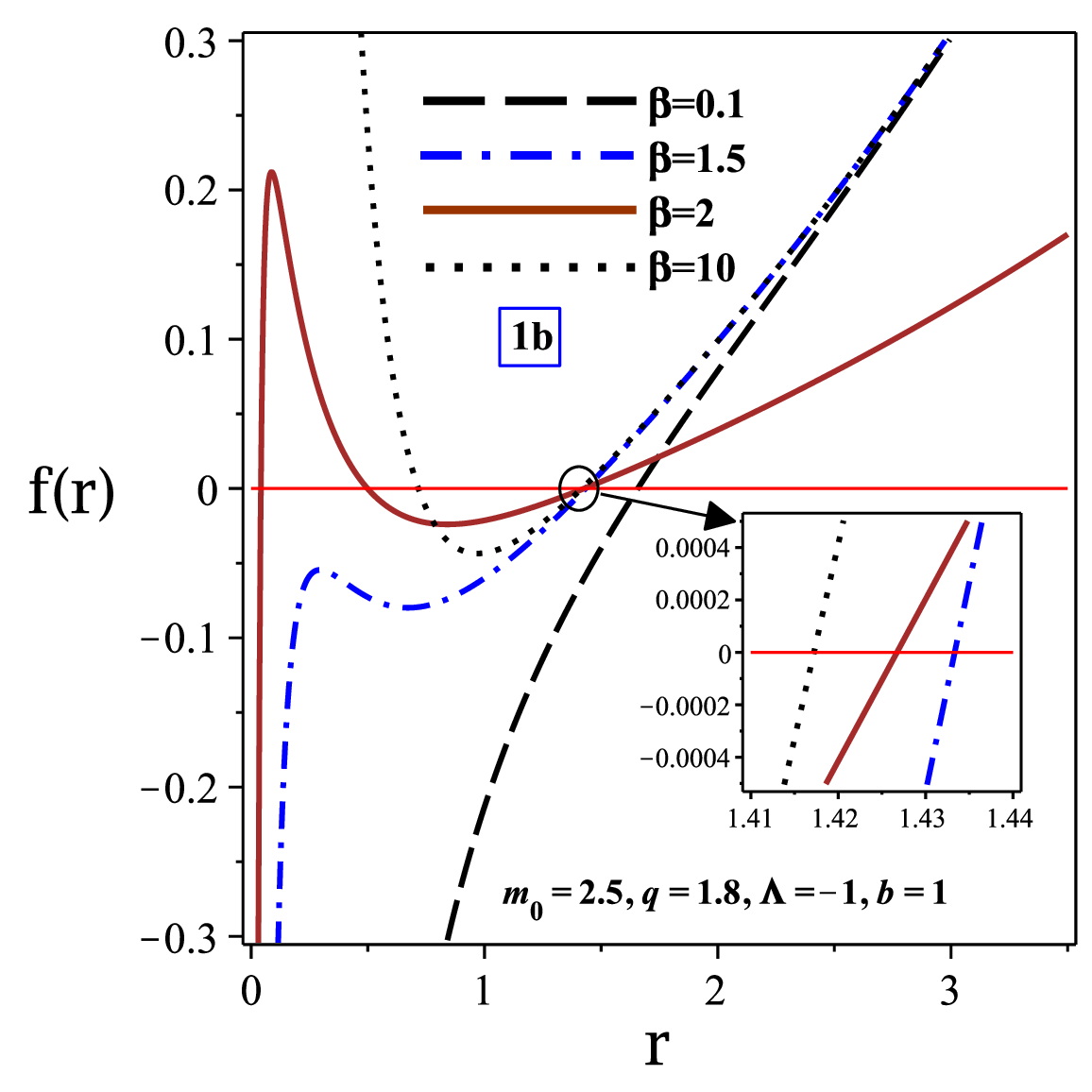}
\caption{The mertic function $\protect\psi(r)$ versus $r$ for different
values of parameters $b$ in the left panel, and $\protect\beta$ in the right
panel.}
\label{fig1}
\end{figure}

Our analysis indicate that there is an essential curvature singularity at $%
r=0$. To understand that the singularity covered by an event horizon, we
plot the metric fucntion $\psi (r)$ versus $r$ in two panels in Figure. \ref%
{fig1}. As one can see, the singularity is covered by an event horizon. So,
the obtained solutions are black hole solution. In addition, we study the
effects of the parameters of PFDM ($b$) and BI-NED ($\beta $) on the root of
metric function ($\psi (r)$). Our findings reveal that there are three
different behaviors for root of BI-NED AdS black holes where only depend on
the parameters $b$ and $\beta$. Indeed, by varying $b$ and $\beta$ the
number of roots of the metric function change when other parameters are
fixed. These behaviors are:

i) For small values of $b$, there are two roots. The small root is related
to the inner horizon and the large root is the event horizon. Such behavior
of the metric fucntion was reported for Rissner-Nordestron (RN) black holes
(see dashed line in the left panel of Fig. \ref{fig1} (Fig. \ref{fig1}a)).
However, for small values of $\beta$, there is only one root and it is
related to the event horizon ((see dashed and dsahsed-dotted lines in the
right panel of Fig. \ref{fig1} (Fig. \ref{fig1}b)))

ii) For medium values of $b$ and $\beta$, there are three roots. In other
words, we encounter with multi-horizons black holes (two inner horizons and
one event horizon), see dashed-dotted and continuous lines in the left panel
of Fig. \ref{fig1} (Fig. \ref{fig1}a) and continuous line in the right panel
of Fig. \ref{fig1} (Fig. \ref{fig1}b).

iii) For large values of $b$, there is only one root and it is related to
the event horizon. This black holes behaves similar to the Schwarzschild
black holes (see the dotted line in the left panel of Fig. \ref{fig1} (Fig. %
\ref{fig1}a)). However, for large value of $\beta$, there are two roots. The
small root is related to the inner horizon and the large root is the event
horizon. Such behavior of the metric fucntion was reported for
Rissner-Nordestron (RN) black holes (see the dotted line in the right panel
of Fig. \ref{fig1} (Fig. \ref{fig1}b)).

There are two distinct behaviors observed when varying the parameters of
PFDM ($b$) and BI-NED ($\beta$):

\textbf{i) Number of Roots:} The number of roots of the metric function
first increases from two to three as the PFDM parameter increases, then
decreases from three to one (see the left panel in Fig. \ref{fig1}). In
contrast, when varying the BI parameter, the number of roots of $f(r)$
increases from one to three and then decreases to two as the value of $\beta$
increases (see the right panel in Fig. \ref{fig1}).

\textbf{ii) Event Horizon:} Increasing the PFDM parameter results in an
increase in the event horizon (or black hole size), see the left panel in
Fig. \ref{fig1}, for more details. Conversely, increasing the BI-NED
parameter leads to a decrease in the event horizon (see the right panel in
Fig. \ref{fig1}).

\section{Thermodynamic Properties in Non-Extended Phase Space}

Considering BI-NED AdS black holes surrounded by PFDM, we calculate the
conserved and thermodynamical quantities such as Hawking temperature,
entropy, charge, electric potential, and total mass. Then we evaluate the
effect of PFDM parameters on these quantities. Using these thermodynamical
quantities, we check that these quantities satisfy the first law of
thermodynamics. Then, we are going to expand our study to local and global
stability conditions of the solutions in canonical ensemble by studying the
heat capacity and the Helmholtz free energy, respectively.

\subsection{Conserved and Thermodynamic Quantities}

Due to the fact that employed metric only contains one temporal killing
vector ($\chi _{\mu }=\left( 1,0,0,0\right) $), one can use the concept of
surface gravity ($\kappa =\sqrt{\nabla _{\mu }\chi _{\nu }\nabla ^{\mu }\chi
^{\nu }}$) for calculating the temperature on the event horizon ($r_{+}$)
which leads to 
\begin{equation}
T=\frac{\kappa }{2\pi }=\frac{1}{2\pi }\sqrt{\nabla _{\mu }\chi _{\nu
}\nabla ^{\mu }\chi ^{\nu }}=\frac{1}{4\pi }\frac{d\psi \left( r\right) }{dr}%
|_{r=r_{+}},  \label{Temp1}
\end{equation}%
by considering Eqs. (\ref{f(r)}) and (\ref{Temp1}), we can get the Hawking
temperature for these black holes as 
\begin{equation}
T=\frac{\left( 1+\frac{b}{r_{+}}\right) }{4\pi r_{+}}+\frac{\left( 2\beta
^{2}\left( 1-\mathfrak{F}_{1+}\right) -\Lambda \right) r_{+}}{4\pi }-\frac{%
q^{2}\mathfrak{F}_{2+}}{\pi r_{+}^{3}},  \label{Temp2}
\end{equation}%
where $\mathfrak{F}_{1+}=\mathfrak{F}_{1}\left\vert _{r=r_{+}}\right. $, $%
\mathfrak{F}_{2+}={}_{2}F_{1}\left( \left[ \frac{1}{2},\frac{1}{4}\right] ,%
\left[ \frac{5}{4}\right] ,1-\Gamma _{+}^{2}\right) $, and $\Gamma
_{+}=\Gamma \left\vert _{r=r_{+}}\right. $. It is clear that the parameters
of PFDM and BI-NED influence the Hawking temperature of these black holes.

The entropy of these black holes can be obtained by using the area law, in
the following form 
\begin{equation}
S=\frac{A}{4},  \label{S}
\end{equation}%
where $A$\ is the horizon area and is obatined 
\begin{equation}
A=\left. \int_{0}^{2\pi }\int_{0}^{\pi }\sqrt{g_{\theta \theta }g_{\varphi
\varphi }}\right\vert _{r=r_{+}}=4\pi \left. r^{2}\right\vert
_{r=r_{+}}=4\pi r_{+}^{2},  \label{Area}
\end{equation}%
so the entropy of BI-NED AdS black holes surrounded by PFDM is given by 
\begin{equation}
S=\pi r_{+}^{2}.  \label{entropy}
\end{equation}%
We can get the electric charge of the solutions by employ the Gauss law in
the following form 
\begin{equation}
Q=\frac{F_{tr}}{4\pi }\int_{0}^{2\pi }\int_{0}^{\pi }\sqrt{g}d\theta
d\varphi =q,  \label{Q}
\end{equation}%
where for case $t=$ constant and $r=$constant, the determinant of metric
tensor $g$ is $r^{4}\sin ^{2}\theta $.

In order to obtain electric potential, we can calculate it on the horizon
with respect to a reference which leads to 
\begin{equation}
\Phi =A_{\mu }\chi ^{\mu }\left\vert _{r\rightarrow \infty }\right. -A_{\mu
}\chi ^{\mu }\left\vert _{r\rightarrow r_{+}}\right. =\frac{q}{r_{+}}%
\mathfrak{F}_{2+},  \label{U}
\end{equation}

Applying Ashtekar-Magnon-Das (AMD) approach, we find the total mass of these
black holes in the following form%
\begin{equation}
M=m_{0},  \label{M}
\end{equation}%
where $m_{0}$ is the geometrical mass and is given by solving ($\psi
(r=r_{+})=0$). So the total mass is given by 
\begin{equation}
M=\frac{r_{+}+b\ln \left( \frac{r_{+}}{\left\vert b\right\vert }\right) }{2}%
+\left( \frac{\beta ^{2}\left( 1-\mathfrak{F}_{1+}\right) }{3}-\frac{\Lambda 
}{6}\right) r_{+}^{3}.  \label{MM}
\end{equation}

\begin{figure}[h]
\centering
\includegraphics[width=70mm]{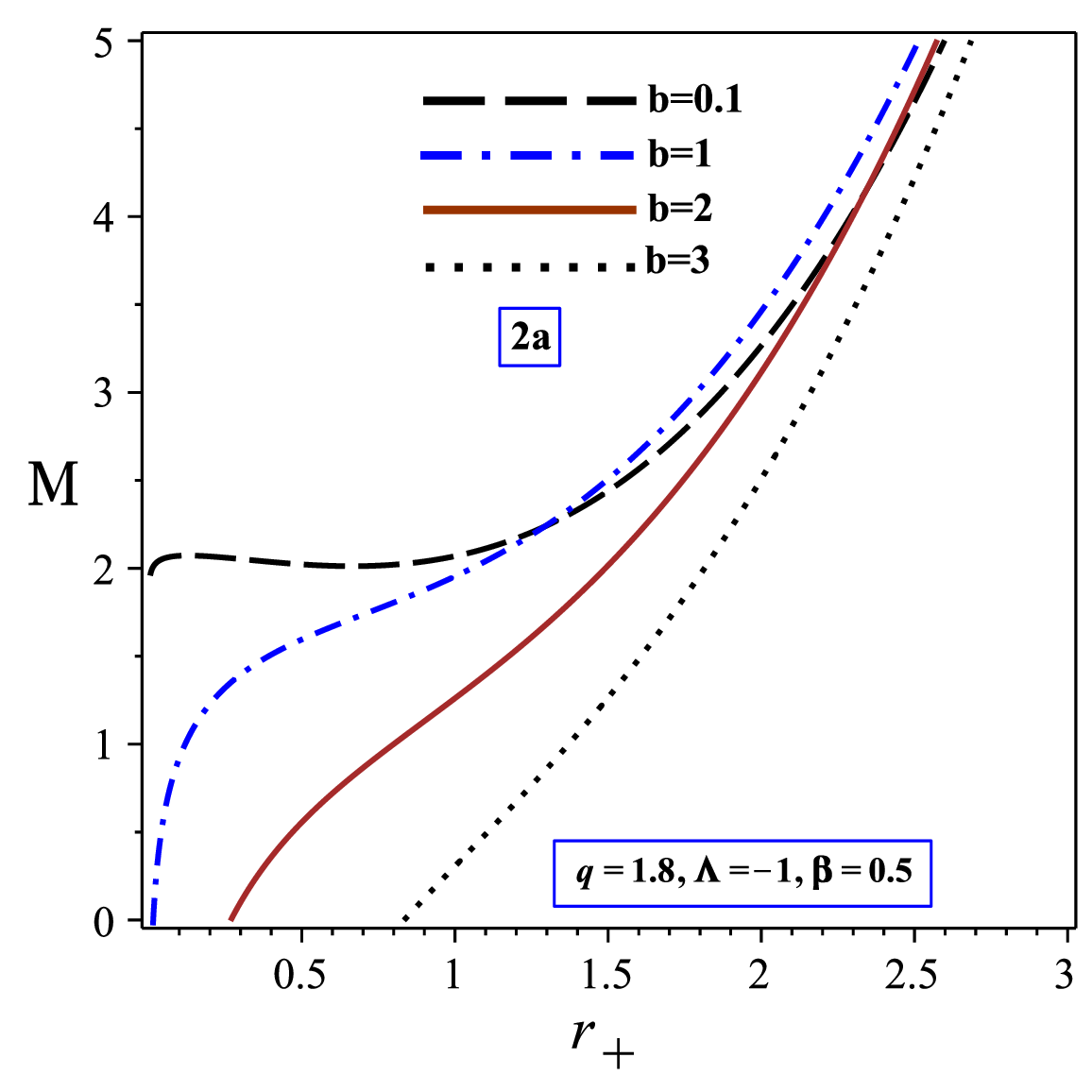} \includegraphics[width=70mm]{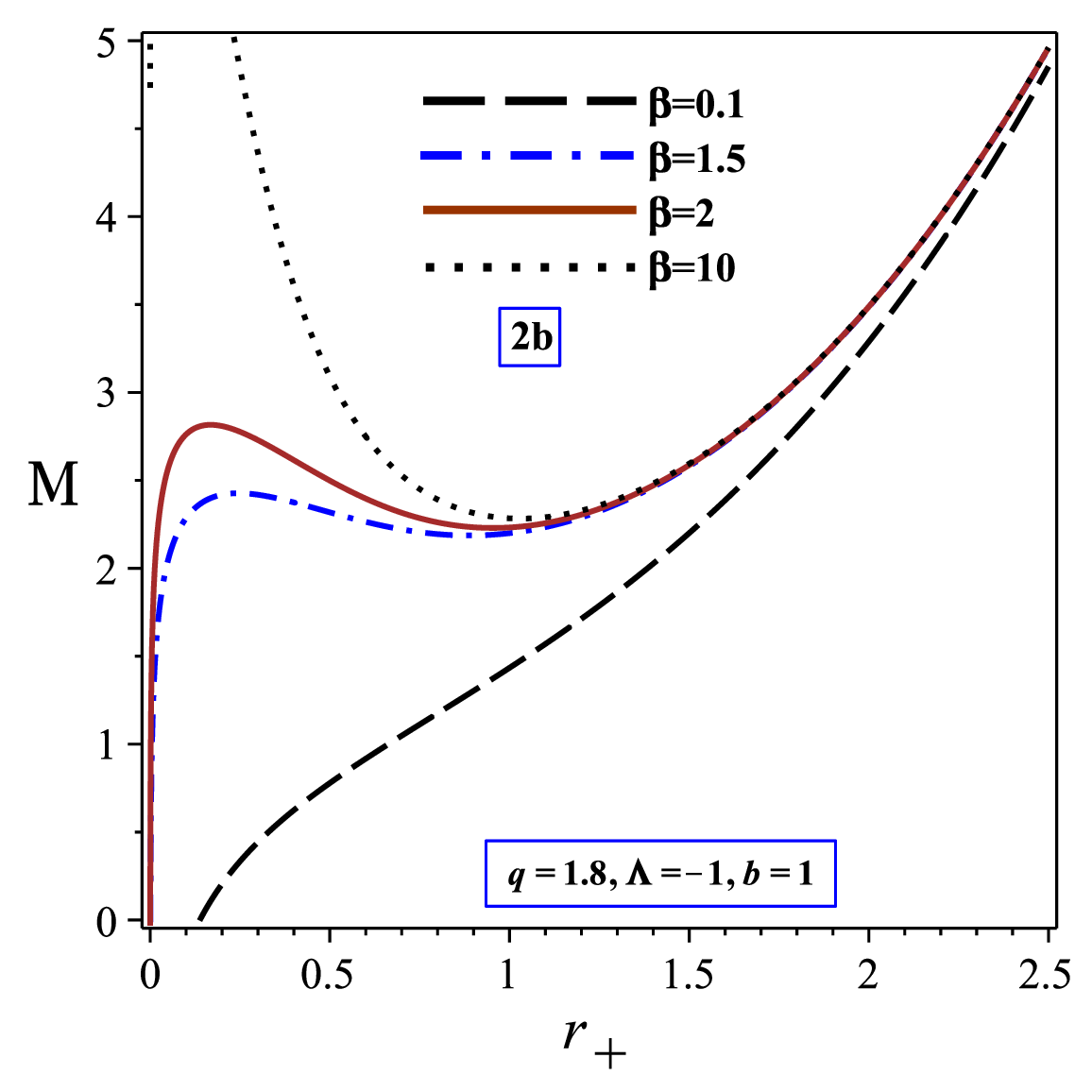}
\caption{The total mass $M$ versus $r_{+}$ for different values of $b$ (the
left panel) and $\protect\beta$ (the right panel).}
\label{Fig2}
\end{figure}

\newpage

Now, we can examine the effect of the parameters of PFDM and BI-NED on total
mass. For this purpose, we sereies $M$ versus $q$ up to order ($q^{6}$)
which leads to 
\begin{equation}
M\simeq \frac{r_{+}}{2}+\frac{b\ln \left( \frac{r_{+}}{\left\vert
b\right\vert }\right) }{2}-\frac{\Lambda r_{+}^{3}}{6}+\frac{q^{2}}{2r_{+}}-%
\frac{q^{4}}{40\beta ^{2}r_{+}^{5}}+\frac{q^{6}}{144\beta ^{4}r_{+}^{9}}.
\end{equation}

Our findings indicate that the asymptotically behavior of $M$ is the same
for various values of $b$, $q$ and $\beta $ because the asymptotical
behavior of $M$ depends on the cosmological cosnatnt ($\underset{%
r\rightarrow \infty }{\lim }M\propto \frac{-\Lambda r_{+}^{3}}{6}$). So, for
the negative of the cosmological constant, the total mass increases by
increasing the size of black hole (or $r_{+}$).

High-energy limit of $M$ in the absence of the electric charge only depends
on the PFDM's parameter as $\underset{r\rightarrow 0}{\lim }M\propto \frac{%
b\ln \left( \frac{r_{+}}{\left\vert b\right\vert }\right) }{2}$. In the
presence of $q$, and for small value of $b$, the total mass is the finite
positive at $r\rightarrow 0$ (except $r=0$), see the dashed line in the left
panel of Fig. \ref{Fig2} (Fig. \ref{Fig2}a). For medium value of $b$, $%
\underset{r\rightarrow 0}{\lim }M\rightarrow 0$, see the dashed-dotted line
in the left panel of Fig. \ref{Fig2} (Fig. \ref{Fig2}a). For large value of $%
b$, there are black holes with $M=0$ at finite radius (see continuous and
dotted lines in the left panel of Fig. \ref{Fig2} (Fig. \ref{Fig2}a)). It is
notable that the finite radius shfits to large radius when $b$ increases.
Our analysis indicate that the behavior of mass of the small black holes is
very sensitive to the parameter of PFDM ($b$). Such behaviors are different
of ususal black holes such as Schwarzschild and Reissner-Nordstr\"{o}m black
holes.

On the other hand, the effect of BI-NED parameter on the total mass is
different of PFDM's parameter. In other words, by decreasing of values of $%
\beta $, the divergence of total mass removes and it reach to a finite value
(see the right panel in Fig. \ref{Fig2} (Fig. \ref{Fig2}b)). In addition,
for very large values of $\beta $, at $r_{+}=0$, the total mass behaves
similar to Reissner-Nordstr\"{o}m black holes beacuse $\underset{%
r\rightarrow 0}{\lim }M\rightarrow \infty $.

As a result, the divergence of total mass (at $r_{+}=0$) of BI-NED AdS black
holes surrounded by PFDM removes when the parameter of BI-NED decreases or
the parameter of PFDM includes large value. In other words, the effects of
these parameters on the total mass act opposite each other.

Replacing Eqs. (\ref{entropy}) and (\ref{Q}) within Eq. (\ref{MM}), we can
obtain $M=M(S,Q)$ in the following form 
\begin{equation}
M(S,Q)=\left( \frac{1}{2}-\frac{\Lambda S}{6\pi }+\frac{S\left( 1-\mathfrak{F%
}_{1_{SQ}}\right) \beta ^{2}}{3\pi }+\frac{b\sqrt{\pi }\ln \left( \frac{%
\sqrt{S}}{\left\vert b\right\vert \sqrt{\pi }}\right) }{2\sqrt{S}}\right) 
\sqrt{\frac{S}{\pi }},  \label{MS}
\end{equation}%
where $\mathfrak{F}_{1_{SQ}}=_{2}F_{1}\left( \left[ \frac{-1}{2},\frac{-3}{4}%
\right] ,\left[ \frac{1}{4}\right] ,-\frac{\pi ^{2}Q^{2}}{\beta ^{2}S^{2}}%
\right) $.

Now, we are in a position to study the first law of thermodynamics. Here, we
should calculate $\left( \frac{\partial M}{\partial S}\right) _{Q}$ and $%
\left( \frac{\partial M}{\partial Q}\right) _{S}$, and then use Eqs. (\ref%
{entropy}) and (\ref{Q}) to convert them as functions of $r_{+}$ and $q$.
After some simplifications, one finds that these quantities are,
respectively, the same as temperature ( $T=\left( \frac{\partial M}{\partial
S}\right) _{Q}$) and electric potential ($\Phi =\left( \frac{\partial M}{%
\partial Q}\right) _{S}$) which were obtained in Eqs. (\ref{Temp2}) and (\ref%
{U}). Hence, although the parameters of PFDM and BI-NED affected
thermodynamic and conserved quantities, the first law remains valid as 
\begin{equation}
dM=TdS+\Phi dQ.
\end{equation}

\subsection{Heat Capacity: Local Stability}

In the context of canonical ensemble, thermal stability (or local stability)
conditions of topological BI-NED AdS black holes surrounded by PFDM
determine by the sign of heat capacity. The positivity of heat capacity
guarantees thermally stable solutions, whereas the negative heat capacity is
denoted as an unstable state. On the other hand, positive and negative of
the temperature determine the physical and non-physics thermal system. So,
we study the sign of heat capacity and temperature to obtain physical and
stable areas.

The heat capacity is defined as 
\begin{equation}
C=T\left( \frac{\partial S}{\partial T}\right) _{Q}=T\frac{\left( \frac{%
\partial S}{\partial r_{+}}\right) _{Q}}{\left( \frac{\partial T}{\partial
r_{+}}\right) _{Q}}.  \label{CQ}
\end{equation}

Using obtained temperature (Eq. (\ref{Temp2})) and entropy (\ref{entropy}),
we can obtain the heat capacity in the following form 
\begin{equation}
C=\frac{\pi \left( \mathfrak{F}_{1+}+2\left( \Gamma _{+}^{2}-1\right) 
\mathfrak{F}_{2+}-\frac{\left( 2\beta ^{2}-\Lambda \right) r_{+}^{3}+b+r_{+}%
}{2\beta ^{2}r_{+}^{3}}\right) r_{+}^{2}}{\mathfrak{F}_{1+}+\frac{4\left(
\Gamma _{+}^{2}-1\right) ^{2}\mathfrak{F}_{3+}}{5}-\frac{\left( 2\beta
^{2}-\Lambda \right) r_{+}^{3}-2b-r_{+}}{2\beta ^{2}r_{+}^{3}}},  \label{CQ1}
\end{equation}%
in which where $\mathfrak{F}_{3+}=\mathfrak{F}_{3}\left\vert
_{r=r_{+}}\right. $.

\begin{figure}[h]
\centering 
\includegraphics[width=70mm]{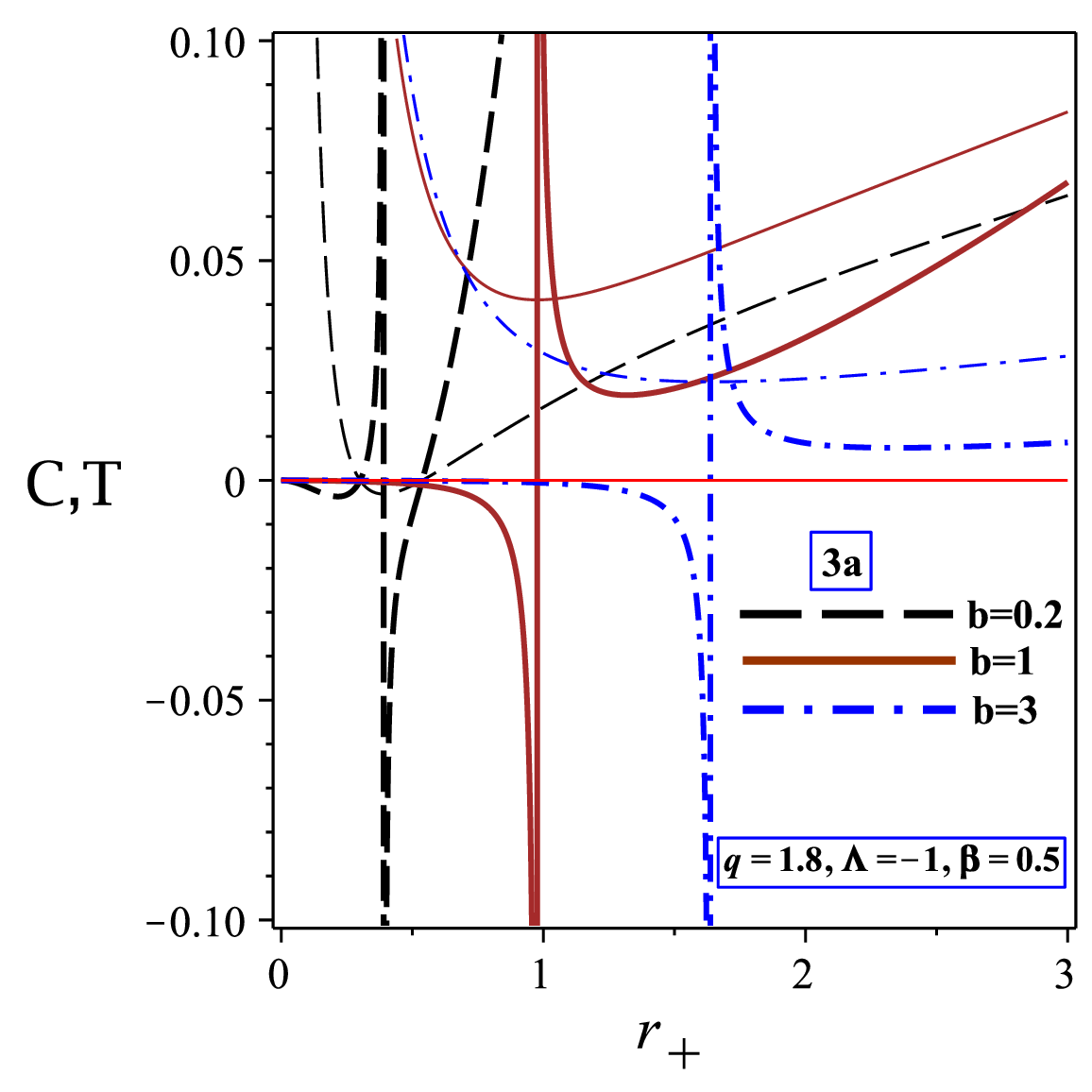} \includegraphics[width=70mm]{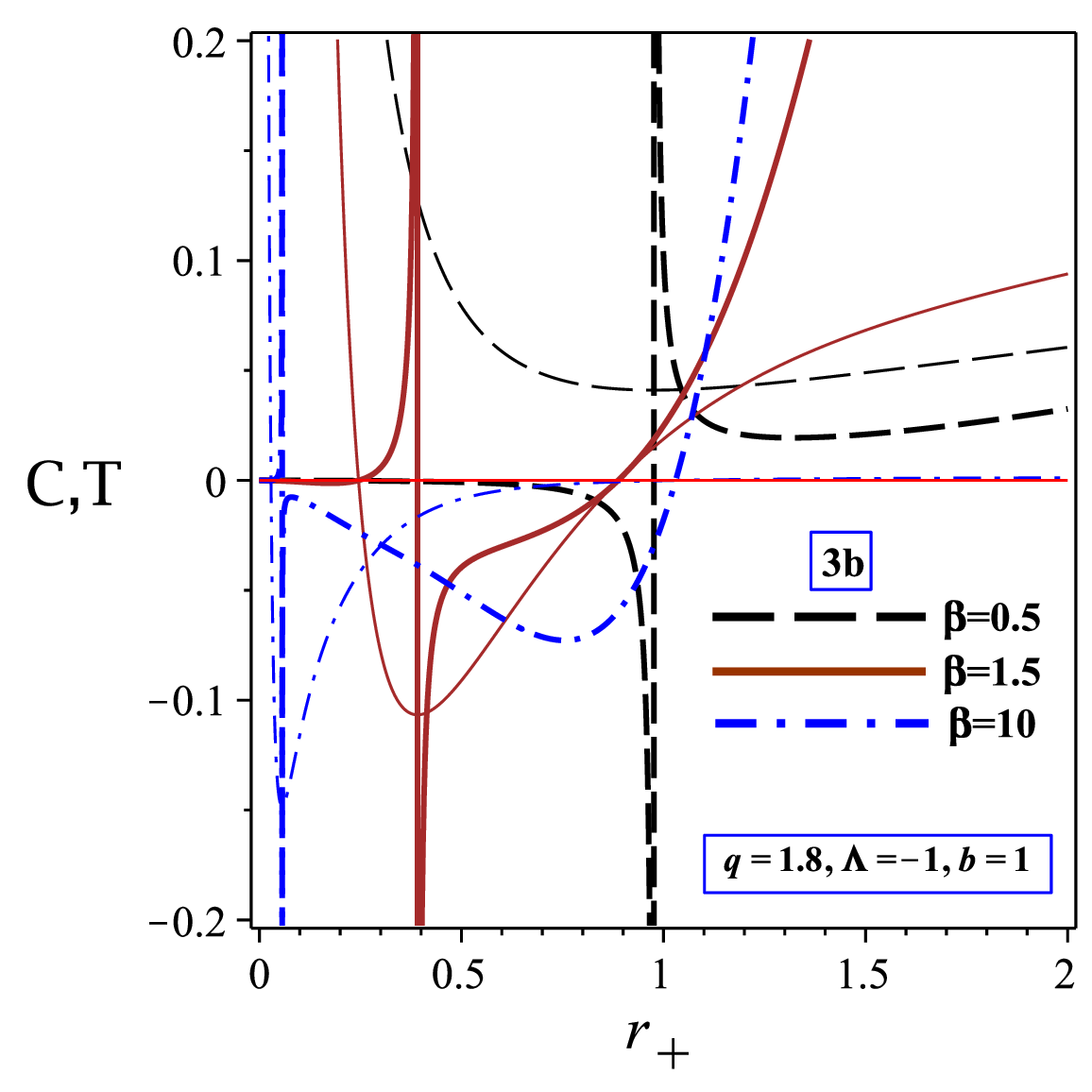}
\caption{The heat capacity $C$ (thick lines) and temperature $T$ (thin
lines) are shown as functions of $r_{+}$ for different values of $b$ in the
left panel and $\protect\beta$ in the right panel.}
\label{Fig3}
\end{figure}

Now, we examine the local thermodynamic stability by plotting the heat
capacity (Eq. (\ref{CQ1})) and temperature (Eq. (\ref{Temp2})) as functions
of the event horizon radius $r_{+}$. To this end, two panels are presented
in Fig. \ref{Fig3} to analyze the influence of the parameters of PFDM and
BI-NED on the stability regions. Our results reveal that:

\begin{itemize}
\item The effect of the PFDM parameter is illustrated in the left panel of
Fig. \ref{Fig3} (Fig. \ref{Fig3}a). Our findings indicate that: i) There is
a critical value for $b$, at which we can find two roots (the first root is
denoted as $r_{{0}_{1}}$ and the second root as $r_{{0}_{2}}$) and one
divergence point (denoted as $r_{+_{{div}}}$) for the heat capacity when $%
b<b_{critical}$ (see the dashed line in the left panel of Fig. \ref{Fig3}).
Notably, the divergence point is located between the two roots, i.e., $r_{{0}%
_{1}}< r_{+_{{div}}}<r_{{0}_{2}}$. Both the heat capacity and temperature
are positive when $r_{+}> r_{{0}_{2}}$, indicating that large black holes
can satisfy the local stability condition. ii) For $b> b_{critical}$, as $b$
increases, there is only one divergence point in the heat capacity. As the
PFDM parameter increases, this divergence point shifts toward larger horizon
radii. In this scenario, the heat capacity is negative for $r_{+}<r_{+_{{div}%
}}$ and becomes positive for $r_{+}>r_{+_{{div}}}$. Thus, the BI-NED AdS
black holes surrounded by PFDM satisfy the local thermodynamic stability
condition, as both $C$ and $T$ are positive when $r_{+}>r_{+_{{div}}}$.
Additionally, the local stability area decreases as $b$ increases.

\item Considering the different values for the BI-NED parameter, the heat
capacity and temperature of these black holes are plotted in the right panel
of Fig. \ref{Fig3} (Fig. \ref{Fig3}b). i) There is a critical value for the
BI-NED parameter at which the heat capacity exhibits two roots and one
divergence point when $\beta> \beta_{critical}$ (see the continuous and
dashed-dotted lines in the right panel of Fig. \ref{Fig3}). The heat
capacity and temperature are simultaneously positive when $r_{+}>r_{0_{2}}$.
Indeed, the local stability area is located in the range $r_{+}>r_{0_{2}}$.
ii) For $\beta< \beta_{critical}$, there is only one divergence point where $%
C$ and $T$ are positive when $r_{+}>r_{+_{div}}$, and they become negative
when $r_{+}<r_{+_{div}}$. In other words, large BI-NED AdS black holes can
satisfy the local stability conditions. Conversely, as $\beta$ increases,
the local stability area initially increases and then decreases.
\end{itemize}

\newpage

\subsection{Helmholtz free energy: Global Stability}

In the canonical ensemble context, the negative of Helmholtz free energy ($F$%
) determines the global stability. So, we employ the Helmholtz free energies
to evaluate the global stability of the BI-NED AdS black holes surrounded by
PFDM.

While the conventional thermodynamic definition for the Helmholtz free
energy is given by $F=U-TS$, it is crucial to note that within the context
of black hole thermodynamics, this potential is redefined as follows 
\begin{equation}
F=M-TS,  \label{F}
\end{equation}%
by applying Eqs. (\ref{Temp2}), (\ref{entropy}) and (\ref{MM}) within Eq. (%
\ref{F}), we find the Helmholtz free energy of these black holes in the
following form 
\begin{equation}
F=\frac{\left( \mathfrak{F}_{1+}+6\left( \Gamma _{+}^{2}-1\right) \mathfrak{F%
}_{2+}+\frac{6b\ln \left( \frac{r_{+}}{\left\vert b\right\vert }\right)
+3\left( r_{+}-b\right) +\Lambda r_{+}^{3}}{2\beta ^{2}r_{+}^{3}}-1\right)
\beta ^{2}r_{+}^{3}}{6},  \label{FF}
\end{equation}%
where it depends on all the system parameters, especially those of PFDM and
BI-NED.

\begin{figure}[h]
\centering
\includegraphics[width=70mm]{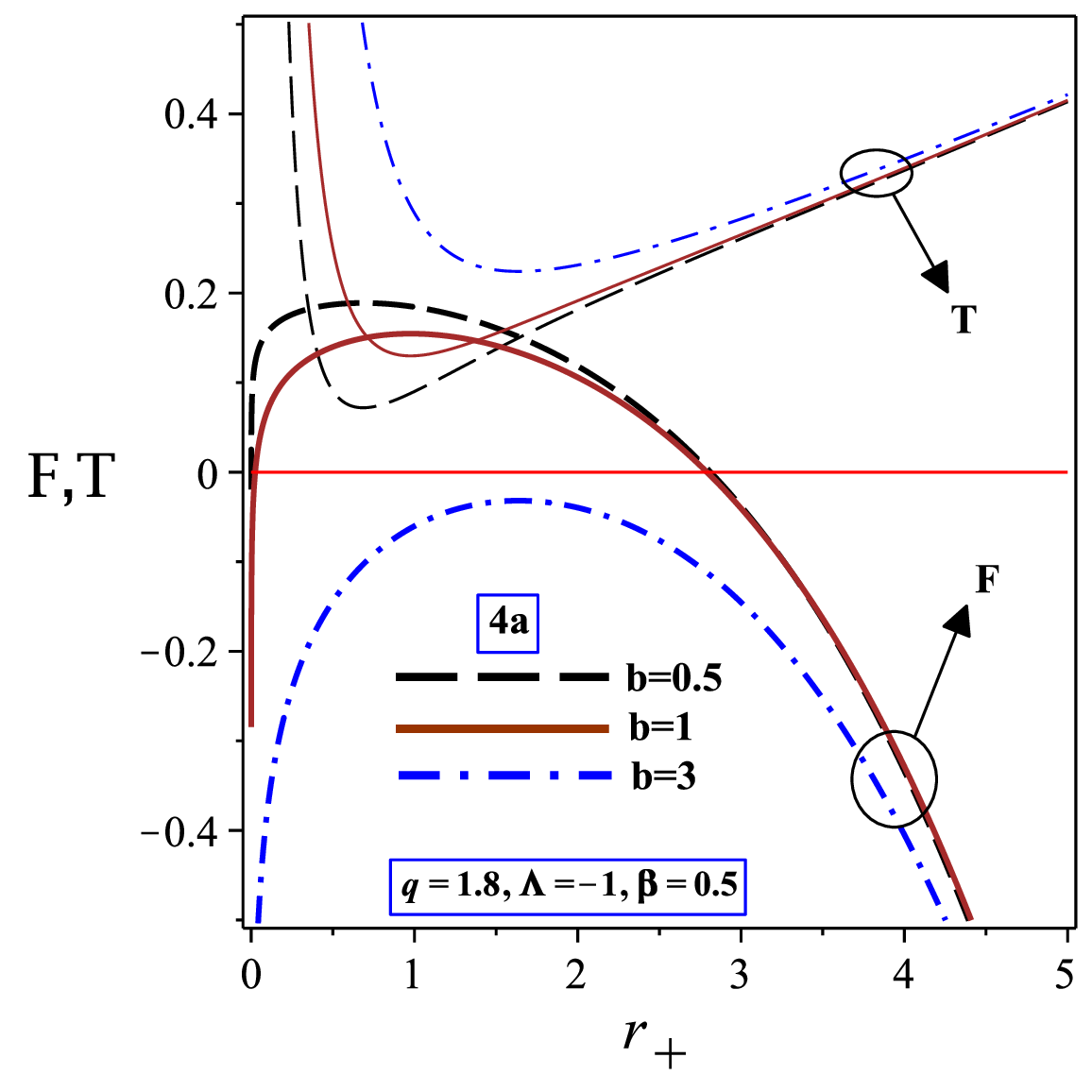} \includegraphics[width=70mm]{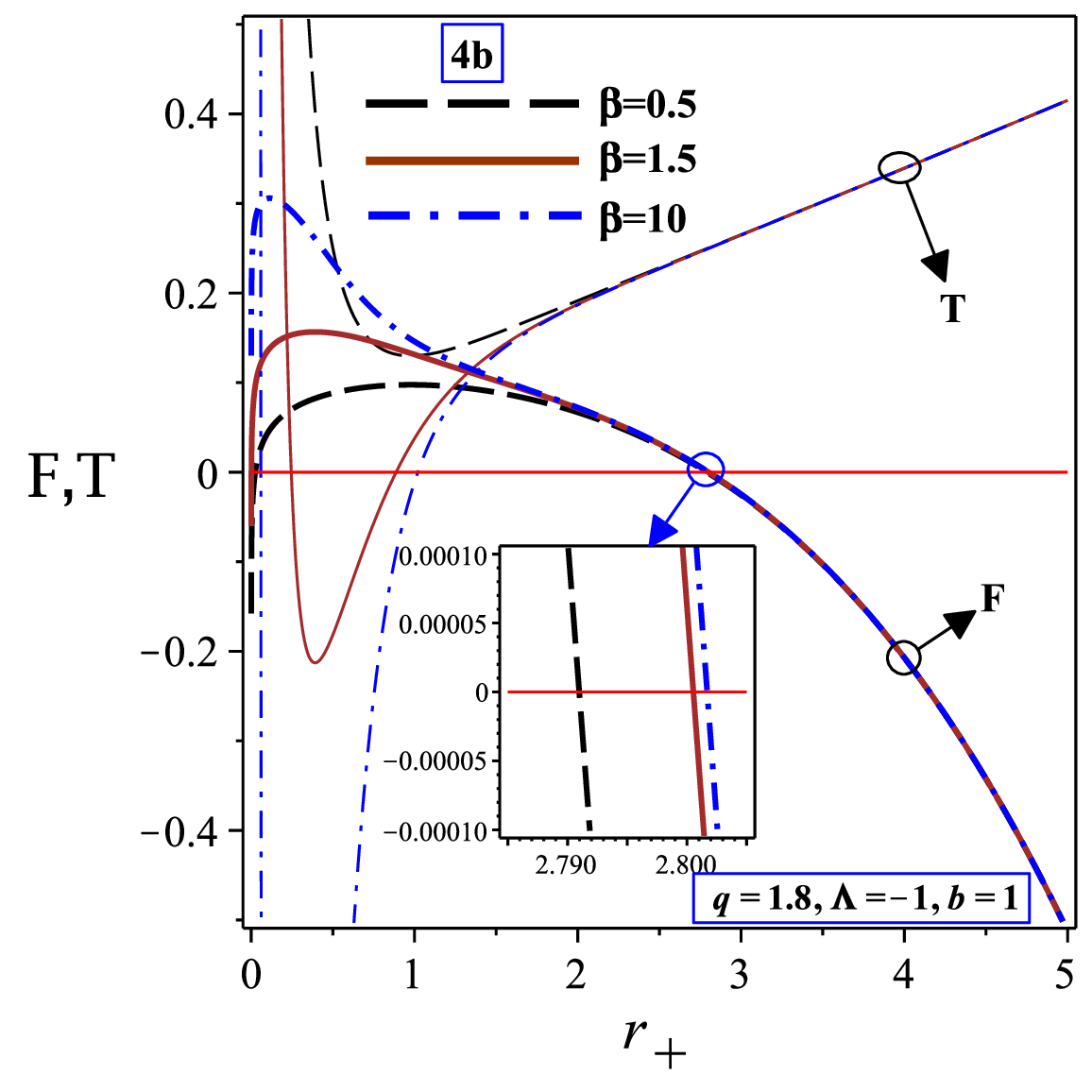}
\caption{The mertic function $\protect\psi (r)$ versus $r$ for different
parameters $b$ in the left panel and $\protect\beta$ in the right panel.}
\label{fig4}
\end{figure}

Now, we can evaluate the global stability condition by study the Helmholtz
free energy. As we mentioned, in the canonical ensemble context, black holes
satisfy the global stability condition when $F$ is negative (i.e., $F<0$).
Due to complexity of the Helmholtz free energy, we plot $F$ versus $r_{+}$
by varying the parameters of PFDM and BI-NED when the other parameters are
fixed in Fig. \ref{fig4}. Our analysis indicate that

\begin{itemize}
\item The effect of the PFDM parameter on $F$ and $T$ is plotted in the left
panel of Fig. \ref{fig4} (Fig. \ref{fig4}a). We find that: i) There is a
critical value of the PFDM parameter. For $b <b_{critical}$, two roots for $%
F $ appear; before the first root and after the second root, the Helmholtz
free energy is negative, while it is positive between these roots (see the
continuous line in the left panel of Fig. \ref{fig4}). This indicates that
both small and large BI-NED AdS black holes can satisfy the global stability
condition. ii) For $b>b_{critical}$, the black holes satisfy the global
stability condition at every radius (see the dashed-dotted line in the panel
of Fig. \ref{fig4}). iii) For very large values of $b \ll b_{critical}$,
only large black holes can satisfy the condition $F < 0$ (see the dashed
line in the left panel of Fig. \ref{fig4}).

\item The variation of $\beta$ with respect to heat capacity and temperature
versus the event horizon ($r_{+}$) is illustrated in the right panel of Fig. %
\ref{fig4} (Fig. \ref{fig4}b). Our findings reveal the following: i)
Increasing $\beta$ results in a decrease in the global stability area. ii)
There are two roots where the Helmholtz free energy is positive between
these roots. Therefore, medium black holes (i.e., those in the range $%
r_{0_{1}}<r_{+}<r_{0_{2}}$) cannot satisfy the global stability conditions
because $F$ is positive in this region. iii) As $\beta \to \infty$, there is
one root where $F$ is positive before this root and negative afterward. In
other words, large BI-NED AdS black holes can satisfy the global stability
conditions when $\beta \to \infty$.
\end{itemize}

\section{Thermodynamic Properties in Extended Phase Space}

Here, we extend our study by considering the cosmological constant as a
thermodynamics pressure. Indeed, in the extended phase space, the
cosmological constant acts as a thermodynamic variable corresponding to
pressure \cite{PLambda1,PLambda2,PLambda3,PLambda4} in the following form 
\begin{equation}
P=\frac{-\Lambda }{8\pi },  \label{P}
\end{equation}%
where this postulate leads to an interpretation of the black hole mass as
enthalpy \cite{PLambda1}.

By substituting $\Lambda =-8\pi P$ within the equation (\ref{MS}), we find 
\begin{equation}
M\left( S,Q,P,b\right) =\left( \frac{1}{2}+\frac{4PS}{3}+\frac{S\left( 1-%
\mathfrak{F}_{1_{SQ}}\right) \beta ^{2}}{3\pi }+\frac{b\sqrt{\pi }\ln \left( 
\frac{\sqrt{S}}{\left\vert b\right\vert \sqrt{\pi }}\right) }{2\sqrt{S}}%
\right) \sqrt{\frac{S}{\pi }}.  \label{MSQPb}
\end{equation}

Using Eq. (\ref{MSQPb}), we can the conjugate quantities associated with the
intensive parameters $S$, $Q$, $P$, and $b$, which are 
\begin{eqnarray}
T &=&\left( \frac{\partial M\left( S,Q,P,b\right) }{\partial S}\right)
_{Q,P,b}=\frac{\left( \beta ^{2}\left( 1-\mathfrak{F}_{1_{SQ}}\right) +4\pi
P\right) \sqrt{\frac{S}{\pi }}}{2\pi }+\frac{b+2\sqrt{\frac{S}{\pi }}}{4S}-%
\frac{Q^{2}\mathfrak{F}_{2_{SQ}}\sqrt{\frac{\pi }{S}}}{S},  \label{TSQPb} \\
&&  \notag \\
V &=&\left( \frac{\partial M\left( S,Q,P,b\right) }{\partial P}\right)
_{S,Q,b}=\frac{4S^{3/2}}{3\pi ^{1/2}},  \label{VSQPb} \\
&&  \notag \\
\Phi &=&\left( \frac{\partial M\left( S,Q,P,b\right) }{\partial Q}\right)
_{S,P,b}=\frac{\sqrt{\pi }Q\mathfrak{F}_{2_{SQ}}}{\sqrt{S}},  \label{PhiSQPb}
\\
&&  \notag \\
B &=&\left( \frac{\partial M\left( S,Q,P,b\right) }{\partial b}\right)
_{S,Q,P}=\frac{\ln \left( \frac{\sqrt{S}}{\left\vert b\right\vert \sqrt{\pi }%
}\right) -1}{2},  \label{BSQPb}
\end{eqnarray}%
where $\mathfrak{F}_{2_{SQ}}=_{2}F_{1}\left( \left[ \frac{1}{2},\frac{1}{4}%
\right] ,\left[ \frac{5}{4}\right] ,-\frac{\pi ^{2}Q^{2}}{\beta ^{2}S^{2}}%
\right) $. Also, the conserved and thermodynamic quantities (Eqs. (\ref%
{MSQPb})-(\ref{BSQPb})) satisfy the first law of thermodynamics in extended
phase space as 
\begin{equation}
dM=TdS+\Phi dQ+VdP+Bdb.
\end{equation}

In Refs. \cite{Cong2019,Johnson2020}, it was suggested that there is a link
between super-entropy black holes and thermodynamic instability. Cofirming
this connection would be of great importance. It should be mentioned that
super-entropy black holes are a result of the violation of the inverse
isoperimetric inequality \cite{Ratio}. Indeed, a black hole can be a
super-entropy black hole if it meets the condition $\mathcal{R}<1$,
otherwise, it cannot be a super-entropy black hole. It is notable that $%
\mathcal{R}$ defined as the isoperimetric ratio in the following form \cite%
{Ratio} 
\begin{equation}
\mathcal{R}=\left( \frac{3V}{\mathcal{V}}\right) ^{1/3}\left( \frac{\mathcal{%
V}}{A}\right) ^{1/2},  \label{Ratio}
\end{equation}%
where $\mathcal{V}=4\pi $ for the static spherical symmetric in $4-$%
dimensional spacetime. By replacing Eqs. (\ref{Area}), and (\ref{VSQPb})
within the above relation, we find that%
\begin{equation}
\mathcal{R}=1,  \label{RR}
\end{equation}%
so these black holes cannot be super-entropy black holes, however the
reverse isoperimetric inequality ($\mathcal{R}\geq 1$) can be satisfied \cite%
{Ratio}.

\subsection{Classical Ehrenfest Relations}

In classical thermodynamics, phase transitions are classified as either
first-order or higher-order transitions using the Clausius-Clapeyron and
Ehrenfest equations. A first-order transition satisfies the
Clausius-Clapeyron equation, while a second-order transition follows the
Ehrenfest relations.

Banerjee et al. \cite{Banerjee2012} presented a compelling analogy of the
Ehrenfest equations by comparing thermodynamic state variables$-$%
specifically, $V\leftrightarrow Q$ and $P\leftrightarrow -\Phi-$with black
hole parameters. This comparison leads to 
\begin{eqnarray}
-\left( \frac{\partial \Phi}{\partial T}\right) _{S} &=&\frac{%
C_{\Phi_{2}}-C_{\Phi_{1}}}{TQ\left( \alpha _{2}-\alpha _{1}\right) }=\frac{%
\Delta C_{\Phi}}{TQ\Delta \alpha }, \\
&&  \notag \\
-\left( \frac{\partial \Phi}{\partial T}\right) _{Q} &=&\frac{\alpha
_{2}-\alpha _{1}}{\kappa _{T_{2}}-\kappa _{T_{1}}}=\frac{\Delta \alpha }{%
\Delta \kappa _{T}},
\end{eqnarray}%
where $\Phi$ is the electric potential. Also, $V$ is the thermodynamic
volume of the black hole. In addition, $\alpha $ and $\kappa _{T}$ are the
analogs of the volume expansion coefficient and the isothermal
compressibility, respectively, and are defined as 
\begin{eqnarray}
\alpha &=&\frac{1}{Q}\left( \frac{\partial Q}{\partial T}\right) _{\Phi}, \\
&&  \notag \\
\kappa _{T} &=&\frac{1}{Q}\left( \frac{\partial Q}{\partial \Phi}\right)
_{T}.
\end{eqnarray}

The established link between $P-V$ criticality and the specific heat of
black holes at constant pressure ($C_{P}$) allows for the direct application
of the classical Ehrenfest equations in black hole research. Therefore, in
this section, we will analytically verify the following Ehrenfest equations 
\begin{eqnarray}
\left( \frac{\partial P}{\partial T}\right) _{S} &=&\frac{C_{P_{2}}-C_{P_{1}}%
}{VT\left( \alpha _{2}-\alpha _{1}\right) }=\frac{\Delta C_{P}}{VT\Delta
\alpha },  \label{Ehrenfest1} \\
&&  \notag \\
\left( \frac{\partial P}{\partial T}\right) _{V} &=&\frac{\alpha _{2}-\alpha
_{1}}{\kappa_{T_{2}}-\kappa_{T_{1}}}=\frac{\Delta \alpha }{\Delta \kappa_{T}}%
,  \label{Ehrenfest2}
\end{eqnarray}%
where $\alpha$ and $\kappa_{T}$ are given by \cite{Mo2013} 
\begin{eqnarray}
\alpha &=&\frac{1}{V}\left( \frac{\partial V}{\partial T}\right) _{P},
\label{alpha} \\
&&  \notag \\
\kappa_{T} &=&\frac{-1}{V}\left( \frac{\partial V}{\partial P}\right) _{T}.
\label{kT}
\end{eqnarray}

We will calculate the relevant quantities in equations (\ref{Ehrenfest1})-(%
\ref{Ehrenfest2}) to analyze the $P-V$ criticality in the extended phase
space of BI-NED AdS black holes surrounded by PFDM.

Using Eq. (\ref{MSQPb}), the specific heat of black holes at constant
pressure ($C_{P}$) is obatined as 
\begin{equation}
C_{P}=T\left( \frac{\partial S}{\partial T}\right) _{P}=\frac{2\left( 
\mathfrak{F}_{1_{SQ}}+\frac{2\pi ^{2}Q^{2}}{\beta ^{2}S^{2}}\mathfrak{F}%
_{2_{SQ}}-\frac{\left( b+\sqrt{\frac{s}{\pi }}\right) \pi ^{3/2}}{2\beta
^{2}S^{3/2}}-\frac{4\pi P}{\beta ^{2}}-1\right) S}{\mathfrak{F}_{1_{SQ}}+%
\frac{4\pi ^{4}Q^{4}}{5\beta ^{4}S^{4}}\mathfrak{F}_{3_{SQ}}+\frac{\left( 2b+%
\sqrt{\frac{s}{\pi }}\right) \pi ^{3/2}}{2\beta ^{2}S^{3/2}}-\frac{4\pi P}{%
\beta ^{2}}-1}.  \label{CP}
\end{equation}%
where $\mathfrak{F}_{3_{SQ}}$ is defined as 
\begin{equation*}
\mathfrak{F}_{3_{SQ}}=_{2}F_{1}\left( \left[ \frac{3}{2},\frac{5}{4}\right] ,%
\left[ \frac{9}{4}\right] ,-\frac{\pi ^{2}Q^{2}}{\beta ^{2}S^{2}}\right) .
\end{equation*}

We can obtain the volume expansion coefficient and the isothermal
compressibility coefficient by applying Eqs. (\ref{TSQPb}) and (\ref{VSQPb})
in conjunction with Eqs. (\ref{alpha}) and (\ref{kT}). This results in 
\begin{eqnarray}
\alpha &=&\frac{1}{V}\left( \frac{\partial V}{\partial T}\right) _{P}=\frac{%
-6\pi ^{3/2}}{\beta ^{2}S^{1/2}\left( \mathfrak{F}_{1_{SQ}}+\frac{4\pi
^{4}Q^{4}}{5\beta ^{4}S^{4}}\mathfrak{F}_{3_{SQ}}+\frac{\left( 2b+\sqrt{%
\frac{S}{\pi }}\right) \pi ^{3/2}}{2\beta ^{2}S^{3/2}}-\frac{4\pi P}{\beta
^{2}}-1\right) },  \label{alpha2} \\
&&  \notag \\
\kappa _{T} &=&\frac{-1}{V}\left( \frac{\partial V}{\partial P}\right) _{T}=%
\frac{-12\pi }{\beta ^{2}\left( \mathfrak{F}_{1_{SQ}}+\frac{4\pi ^{4}Q^{4}}{%
5\beta ^{4}S^{4}}\mathfrak{F}_{3_{SQ}}+\frac{\left( 2b+\sqrt{\frac{S}{\pi }}%
\right) \pi ^{3/2}}{2\beta ^{2}S^{3/2}}-\frac{4\pi P}{\beta ^{2}}-1\right) }.
\label{kT2}
\end{eqnarray}

It is important to note that $C_{P}$, $\alpha $, and $\kappa _{T}$ have a
shared denominator: $\mathfrak{F}_{1_{SQ}}+\frac{4\pi ^{4}Q^{4}}{5\beta
^{4}S^{4}}\mathfrak{F}_{3_{SQ}}+\frac{\left( 2b+\sqrt{\frac{S}{\pi }}\right)
\pi ^{3/2}}{2\beta ^{2}S^{3/2}}-\frac{4\pi P}{\beta ^{2}}-1$. This implies
that both $\alpha $ and $\kappa _{T}$ may diverge at the critical point,
similar to how the specific heat at constant pressure behaves when $%
\mathfrak{F}_{1_{SQ}}+\frac{4\pi ^{4}Q^{4}}{5\beta ^{4}S^{4}}\mathfrak{F}%
_{3_{SQ}}+\frac{\left( 2b+\sqrt{\frac{S}{\pi }}\right) \pi ^{3/2}}{2\beta
^{2}S^{3/2}}-\frac{4\pi P}{\beta ^{2}}-1=0$.

We will now examine the validity of the Ehrenfest equations (Eqs. (\ref%
{Ehrenfest1}) and (\ref{Ehrenfest2})) at the critical point. To do this, we
will utilize the definition of the volume expansion coefficient $\alpha $
(Eq. (\ref{alpha})), leading to the following relationship 
\begin{equation}
V\alpha =\left( \frac{\partial V}{\partial T}\right) _{P}=\left( \frac{%
\partial V}{\partial S}\right) _{P}\left( \frac{\partial S}{\partial T}%
\right) _{P}=\left( \frac{\partial V}{\partial T}\right) _{P}\frac{C_{P}}{T},
\end{equation}%
where $\frac{C_{P}}{T}=\left( \frac{\partial S}{\partial T}\right) _{P}$.
Applying the above equation, we rewrite the R.H.S of Eq. (\ref{Ehrenfest1})
as the following form 
\begin{equation}
\frac{\Delta C_{P}}{VT\Delta \alpha }=\left. \left( \frac{\partial S}{%
\partial V}\right) _{P}\right\vert _{c},  \label{RHS}
\end{equation}%
where the footnote "$c$" denotes the values of physical quantities at the
critical point. Applying Eqs. (\ref{entropy}), and (\ref{VSQPb}), within (%
\ref{RHS}), we find that 
\begin{equation}
\frac{\Delta C_{P}}{VT\Delta \alpha }=\frac{\sqrt{\pi }}{2\sqrt{S}}.
\label{RHS2}
\end{equation}

Using Eq. (\ref{TSQPb}), the L.H.S of Eq. (\ref{Ehrenfest1}) is given by 
\begin{equation}
\left( \frac{\partial P}{\partial T}\right) _{S}=\frac{\sqrt{\pi }}{2\sqrt{S}%
}.  \label{LHS2}
\end{equation}

Equations (\ref{RHS2}) and (\ref{LHS2}) demonstrate that the first Ehrenfest
equation remains valid at the critical point.

We will verify the validity of the second Ehrenfest equation. To do this, we
utilize the thermodynamic identity $\left( \frac{\partial V}{\partial P}%
\right) {T}\left( \frac{\partial P}{\partial T}\right) {V}\left( \frac{%
\partial T}{\partial V}\right) _{P}=-1$, as reported in Ref. \cite{Mo2013}.
By applying this identity alongside Eqs. (\ref{alpha})-(\ref{kT2}), we can
express the R.H.S of Eq. (\ref{Ehrenfest2}) in the following form 
\begin{equation}
\frac{\Delta \alpha }{\Delta \kappa _{T}}=\left. \left( \frac{\partial P}{%
\partial T}\right) _{P}\right\vert _{c}=\frac{\sqrt{\pi }}{2\sqrt{S}}.
\label{RHS3}
\end{equation}

The L.H.S of Eq. (\ref{Ehrenfest2}), is given by 
\begin{equation}
\left. \left( \frac{\partial P}{\partial T}\right) _{V}\right\vert _{c}=%
\frac{\sqrt{\pi }}{2\sqrt{S}},  \label{LHS3}
\end{equation}%
where we consider the equation (\ref{TSQPb}) to obtain the above relation.
By comparing Eqs. (\ref{RHS3}) and (\ref{LHS3}), we find that the second
Ehrenfest equation valids at the critical point.

The Prigogine--Defay (PD) ratio is defined as \cite{PD1,PD2}%
\begin{equation}
\Pi =\frac{\Delta C_{P}\Delta \kappa _{T}}{VT\left( \Delta \alpha \right)
^{2}},  \label{PD}
\end{equation}%
by replacing Eqs. (\ref{RHS2}) and (\ref{RHS3}) within PD ratio (Eq. (\ref%
{PD})) , we find that 
\begin{equation}
\Pi =1,  \label{PDfinal}
\end{equation}

The PD ratio quantitatively assesses deviations from the second Ehrenfest
equation. For a typical second-order phase transition, this ratio is equal
to unity \cite{Banerjee2010}. Based on this analysis, the conclusions drawn
from Eq. (\ref{PDfinal}) and the inherent consistency of the Ehrenfest
relations lead us to conclude that the transition at the $P-V$ critical
point in the extended phase space of the BI-NED AdS black hole surrounded by
PFDM is fundamentally a second-order transition.

\section{Heat Engine}

Considering BI-NED AdS black hole surrounded by PFDM as a thermodynamic
system in the extended phase space, it is natural to assume it as a heat
engine \cite%
{Heat1,Heat2,Heat3,Heat4,Heat5,Heat6,Heat7,Heat8,Heat9,Heat10,Heat11,Heat12,Heat13,Heat14,Heat15,Heat16,Heat17,Heat18,Heat19}%
. A heat engine is a physical system that takes some heat ($Q_{H}$) from a
warm reservoir converts some of this thermal energy to useful work ($W$),
and transfers the remaining heat energy ($Q_{C}$) to the cold reservoir.

\begin{figure}[tbph]
\centering
\includegraphics[width=0.7\linewidth]{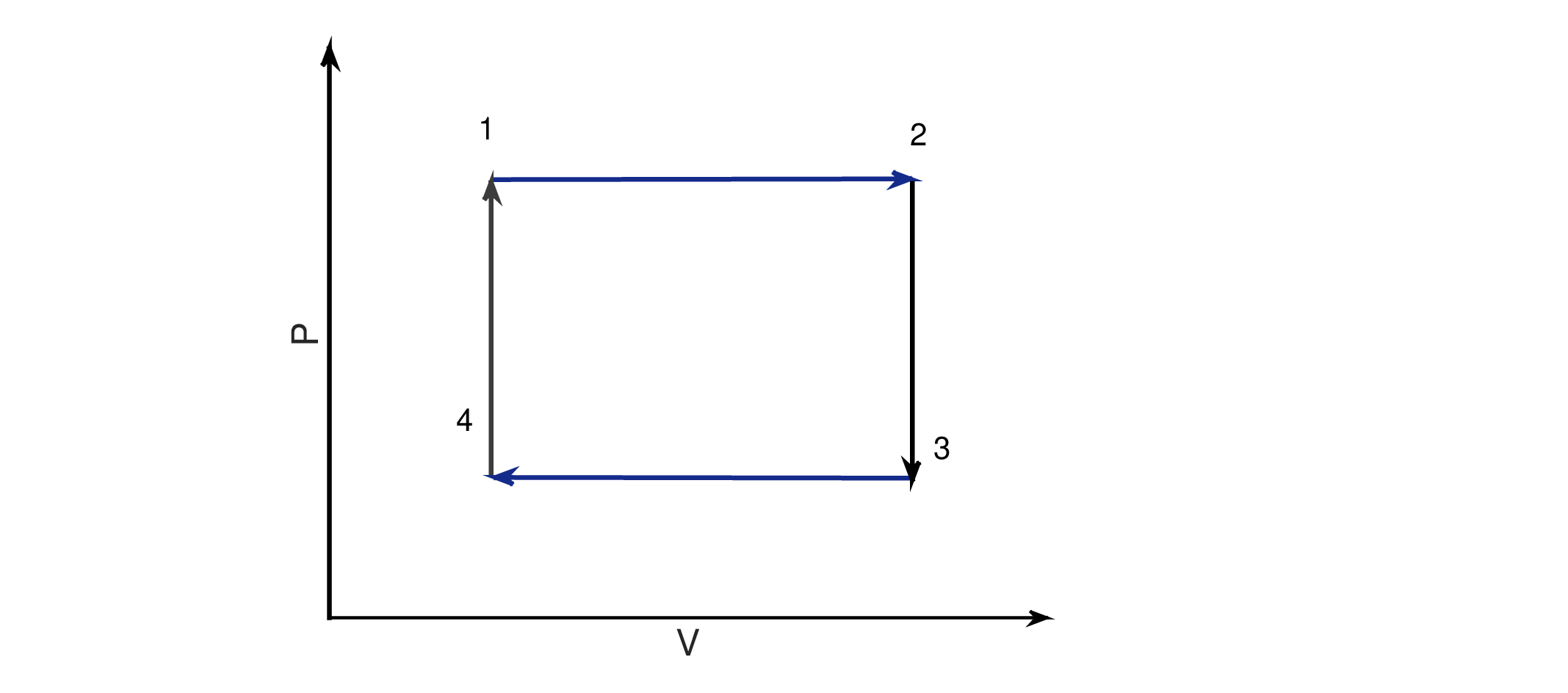}
\caption{$P-V$ diagram for Carnot cycle which consists two isobars (paths of 
$1\rightarrow 2$ and $3\rightarrow 4$) and two isochores (paths of $%
2\rightarrow 3$ and $4\rightarrow 1$). }
\label{Fig5}
\end{figure}

The efficiency of heat engine is defined as 
\begin{equation}
\eta =\frac{W}{Q_{H}}=1-\frac{Q_{C}}{Q_{H}},  \label{eta1}
\end{equation}%
where $Q_{H}=Q_{C}+W$. Also, the heat engine efficiency depends on the
choice of a closed path in the $P-V$ diagram (which is known as a cycle) and
the equation of the state of the black hole in question. So, there are
different classical cycles, where some of these cycles involve a pair of
isotherms at temperatures $T_{H}$ and $T_{C}$ (where $T_{H}>T_{C}$). For
example, in the Carnot cycle (the simplest cycle), there is a pair of
isotherms with different temperatures. Using different methods, one can
connect these two systems: i) The first method is an isochoric path (like
the classical Stirling cycle). ii) The second one is an adiabatic path (like
the classical Carnot cycle). Indeed, the form of the path for the definition
of the cycle is important. According to this fact, the thermodynamic volume
is proportional to entropy (Eq. \ref{VSQPb}), which leads to the same
behavior between adiabatic and isochoric processes. In other words, Carnot
and Stirling methods coincide with each other, and calculating efficiency is
straightforward in this case \cite{Heat1,Heat2}. So, we consider the
mentioned cycle in Fig. \ref{Fig5}.

We calculate the work done along the heat cycle as 
\begin{eqnarray}
W &=&PdV=W_{1\rightarrow 2}+W_{2\rightarrow 3}+W_{3\rightarrow
4}+W_{4\rightarrow 1}=W_{1\rightarrow 2}+W_{3\rightarrow 4}  \notag \\
&&  \notag \\
&=&P_{1}\left( V_{2}-V_{1}\right) +P_{4}\left( V_{4}-V_{3}\right) ,
\label{WW}
\end{eqnarray}%
where $W_{2\rightarrow 3}=W_{4\rightarrow 1}=0$. It means that the work done
along paths of $2\rightarrow 3$ and $4\rightarrow 1$ are zero because $%
V_{3}-V_{2}=0$, and $V_{4}-V_{1}=0$. Using Eq. (\ref{VSQPb}), we can get the
work done along the heat cycle as 
\begin{equation}
W=\frac{4}{3\sqrt{\pi }}\left( P_{1}-P_{4}\right) \left(
S_{2}^{3/2}-S_{1}^{3/2}\right) .  \label{Wfinal}
\end{equation}

The upper isobar will give the heat input in the following form 
\begin{equation}
Q_{H}=\int_{T_{1}}^{T_{2}}P\left( P_{1},T\right)
dT=\int_{S_{1}}^{S_{2}}C_{P}\left( P_{1},T\right) \left( \frac{\partial T}{%
\partial S}\right) dS.  \label{QHH}
\end{equation}%
where $C_{P}$ is the heat capacity and is obtained as $C_{P}\left(
P_{1},T\right) =\frac{2\left( \mathfrak{F}_{1_{SQ}}+\frac{2\pi ^{2}Q^{2}}{%
\beta ^{2}S^{2}}\mathfrak{F}_{2_{SQ}}-\frac{\left( b+\sqrt{\frac{s}{\pi }}%
\right) \pi ^{3/2}}{2\beta ^{2}S^{3/2}}-\frac{4\pi P_{1}}{\beta ^{2}}%
-1\right) S}{\mathfrak{F}_{1_{SQ}}+\frac{4\pi ^{4}Q^{4}}{5\beta ^{4}S^{4}}%
\mathfrak{F}_{3_{SQ}}+\frac{\left( 2b+\sqrt{\frac{s}{\pi }}\right) \pi ^{3/2}%
}{2\beta ^{2}S^{3/2}}-\frac{4\pi P_{1}}{\beta ^{2}}-1}$. So, the heat input
is given by 
\begin{eqnarray}
Q_{H} &=&\left. \left( \frac{\beta ^{2}}{3}\left( 1-\mathfrak{F}%
_{1_{SQ}}\right) +\frac{4\pi P_{1}}{3}\right) \left( \frac{S}{\pi }\right)
^{3/2}+\frac{b\ln \left( \sqrt{\frac{S}{\pi }}\right) +\sqrt{\frac{S}{\pi }}%
}{2}\right\vert _{S_{1}}^{S_{2}}  \notag \\
&&  \notag \\
&=&\frac{\left( \frac{\beta ^{2}}{3}+\frac{4\pi P_{1}}{3}\right) \left(
S_{2}^{3/2}-S_{1}^{3/2}\right) }{\pi ^{3/2}}+\frac{S_{2}^{1/2}-S_{1}^{1/2}}{%
2\pi ^{1/2}}+\frac{b}{4}\ln \left( \frac{S_{2}}{S_{1}}\right) +\frac{\beta
^{2}}{3\pi ^{3/2}}\left( S_{1}^{3/2}\mathfrak{F}_{1_{S_{1}Q}}-S_{2}^{3/2}%
\mathfrak{F}_{1_{S_{2}Q}}\right) ,  \label{QHfinal}
\end{eqnarray}%
where $\mathfrak{F}_{1_{S_{1}Q}}=\left. \mathfrak{F}_{1_{SQ}}\right\vert
_{S=S_{1}}$, and $\mathfrak{F}_{1_{S_{2}Q}}=\left. \mathfrak{F}%
_{1_{SQ}}\right\vert _{S=S_{2}}$.

\begin{figure}[h]
\centering
\includegraphics[width=70mm]{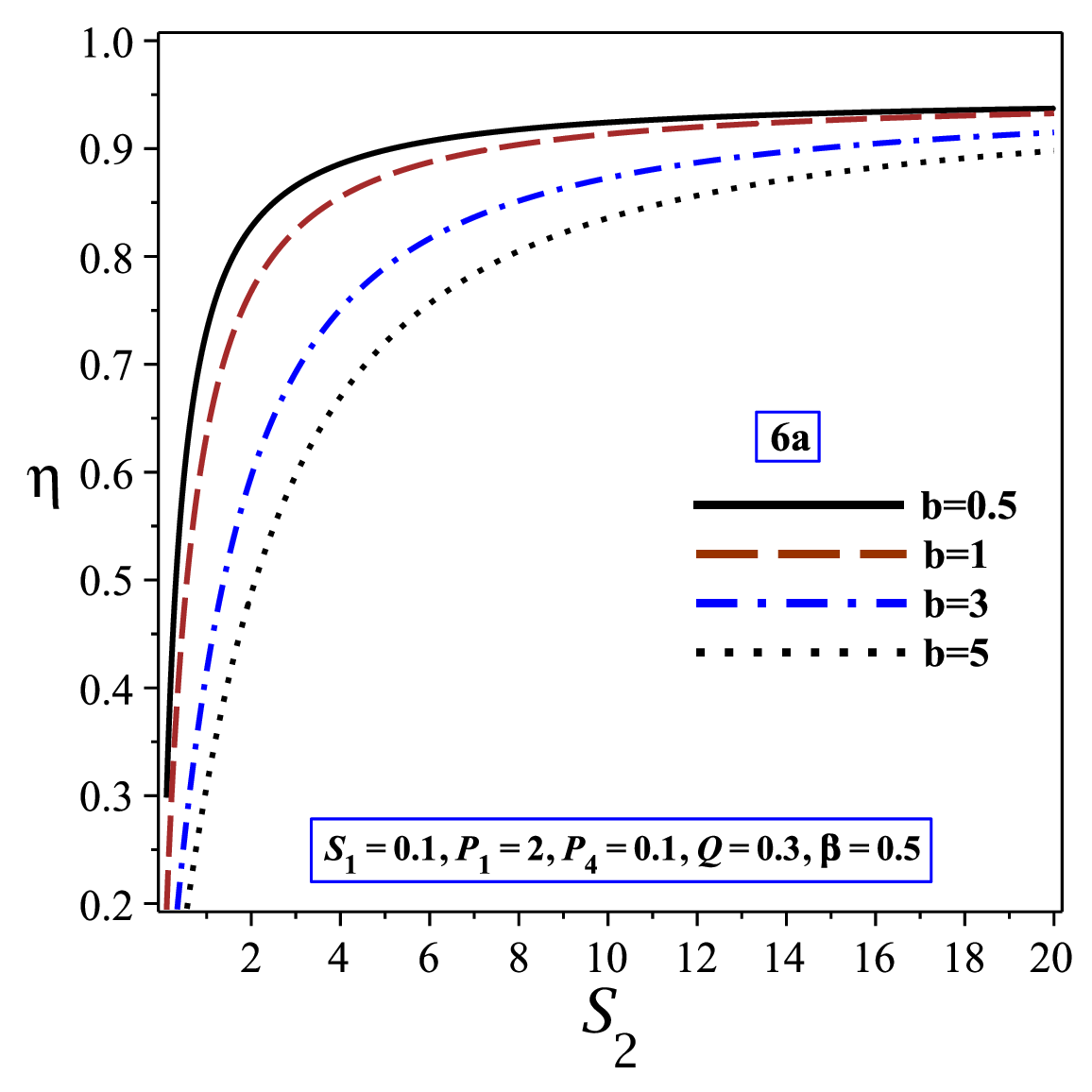} %
\includegraphics[width=70mm]{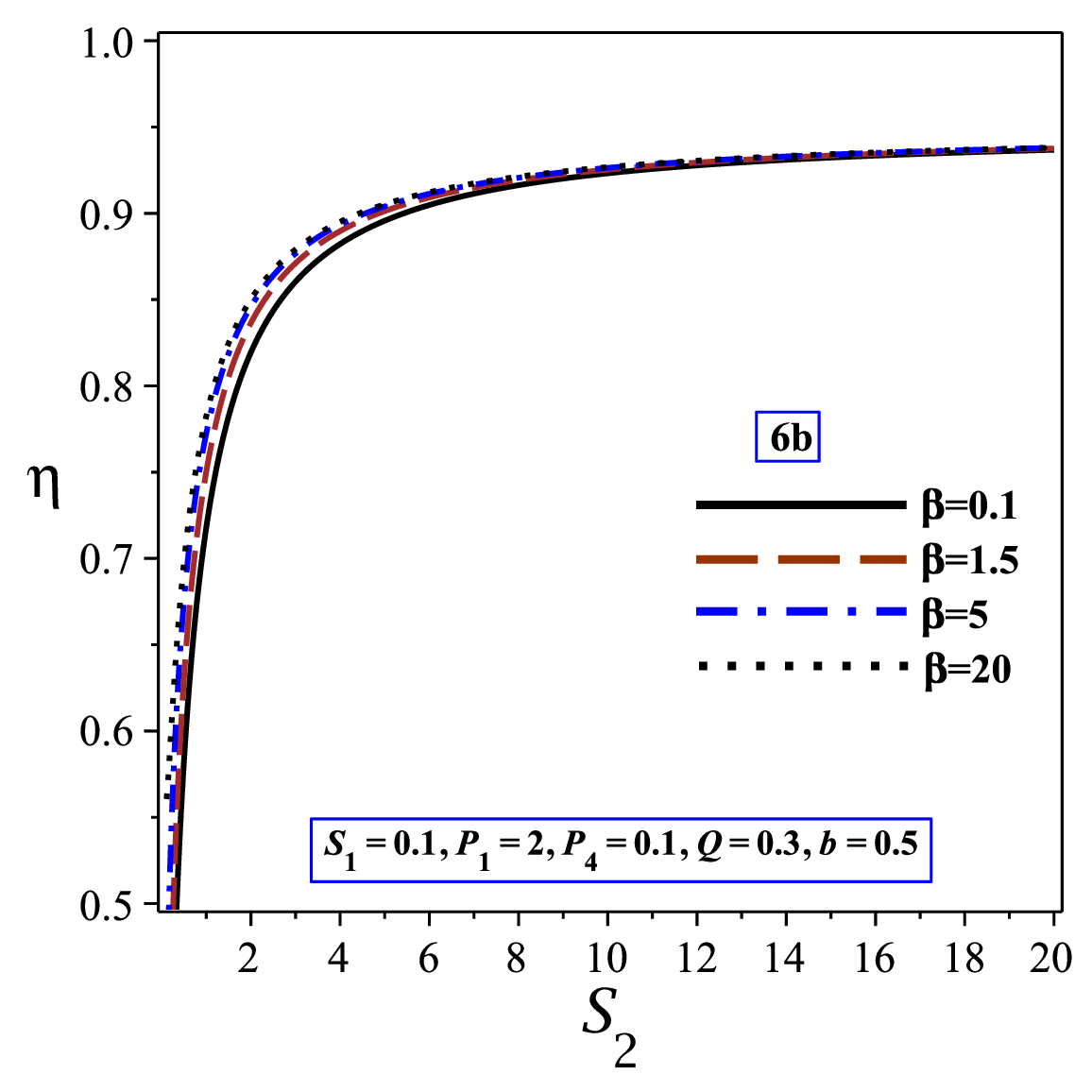}
\caption{The heat engine efficiency $\protect\eta $ versus $S_{2}$ for
different parameters of $b$ in the left panel and $\protect\beta$ in the
right panel.}
\label{Fig6}
\end{figure}

After some calculations, one can get the heat engine efficiency of BI-NED
AdS black holes surrounded by PFDM by inserting Eqs. (\ref{Wfinal}) and (\ref%
{QHfinal}) into Eq. (\ref{eta1}), as 
\begin{equation}
\eta =\frac{\left( P_{1}-P_{4}\right) \left( S_{2}^{3/2}-S_{1}^{3/2}\right) 
}{\frac{\left( \beta ^{2}+4\pi P_{1}\right) \left(
S_{2}^{3/2}-S_{1}^{3/2}\right) }{4\pi }+\frac{3\left(
S_{2}^{1/2}-S_{1}^{1/2}\right) }{8}+\frac{3b\pi ^{1/2}}{16}\ln \left( \frac{%
S_{2}}{S_{1}}\right) +\frac{\beta ^{2}}{4\pi }\left( S_{1}^{3/2}\mathfrak{F}%
_{1_{S_{1}Q}}-S_{2}^{3/2}\mathfrak{F}_{1_{S_{2}Q}}\right) },
\label{etatotal}
\end{equation}%
where it is dependent on both the parameters of PFDM and BI-NED. According
to the equation (\ref{etatotal}), the PFDM' parameter is in the denominator
of $\eta$. Therefore, when $b$ is increased, the heat engine efficiency
decreases if $S_{2}$ remains constant (see the left panel of Fig. \ref{Fig6}%
, (Fig. \ref{Fig6}a)). Additionally, the heat engine efficiency is sensitive
to changes in the BI-NED parameter due to the complexity of the
hypergeometric function located in the denominator of $\eta$. To illustrate
this, we plot $\eta$ versus $S_{2}$ in the right panel of Fig. \ref{Fig6}
(Fig. \ref{Fig6}b), considering different values of $\beta$ while keeping
the other parameters fixed. Our analysis reveals that $\eta$ increases with
an increase in $\beta$ at the same value of $S_{2}$. Given that $S_{2}>S_{1}$%
, an increase in $S_{2}$ results in higher heat engine efficiency for all
values of $b$ and $\beta$.

\newpage

The Carnot efficiency is determined as 
\begin{equation}
\eta _{c}=1-\frac{T_{C}}{T_{H}},
\end{equation}%
this the maximum efficiency any heat engine can have and any higher
efficiency would violate the second law of thermodynamics. To calculate
Carnot efficiency, we consider the $T_{H}$ and $T_{C}$ in our cycle
correspond to $T_{2}$ and $T_{4}$, respectively. So, Carnot efficiency is
given by 
\begin{equation}
\eta _{c}=1-\frac{S_{1}^{1/2}\left[ \mathfrak{F}_{1_{S_{1}Q}}+\frac{2\pi
^{2}Q^{2}\mathfrak{F}_{2_{S_{1}Q}}}{\beta ^{2}S_{1}^{2}}-\frac{\left( b+%
\sqrt{\frac{S_{1}}{\pi }}\right) \pi ^{3/2}}{2\beta ^{2}S_{1}^{3/2}}-\frac{%
4\pi P_{4}}{\beta ^{2}}-1\right] }{S_{2}^{1/2}\left[ \mathfrak{F}%
_{1_{S_{2}Q}}+\frac{2\pi ^{2}Q^{2}\mathfrak{F}_{2_{S_{2}Q}}}{\beta
^{2}S_{2}^{2}}-\frac{\left( b+\sqrt{\frac{S_{2}}{\pi }}\right) \pi ^{3/2}}{%
2\beta ^{2}S_{2}^{3/2}}-\frac{4\pi P_{1}}{\beta ^{2}}-1\right] },
\label{etac}
\end{equation}

\begin{figure}[h]
\centering
\includegraphics[width=70mm]{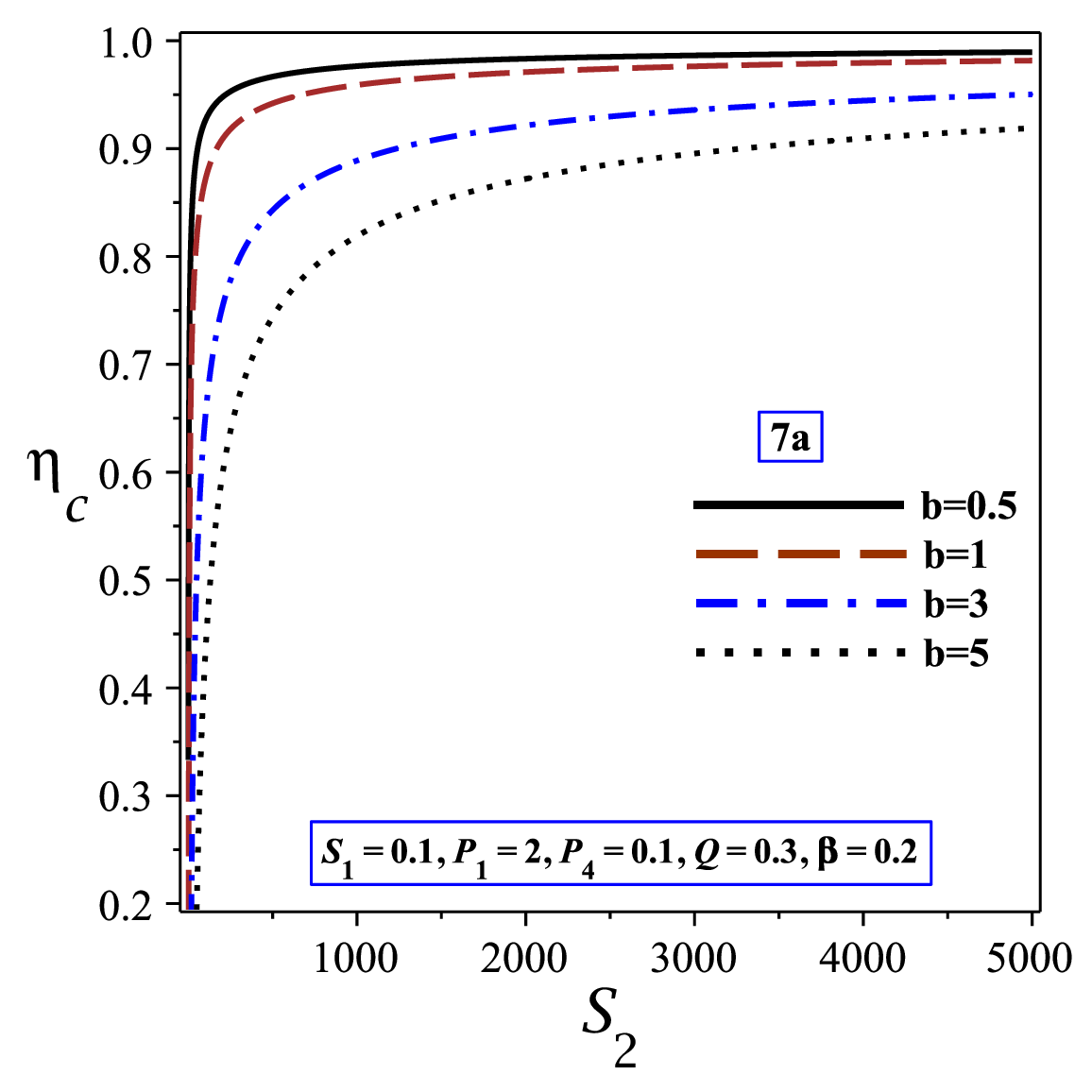} %
\includegraphics[width=70mm]{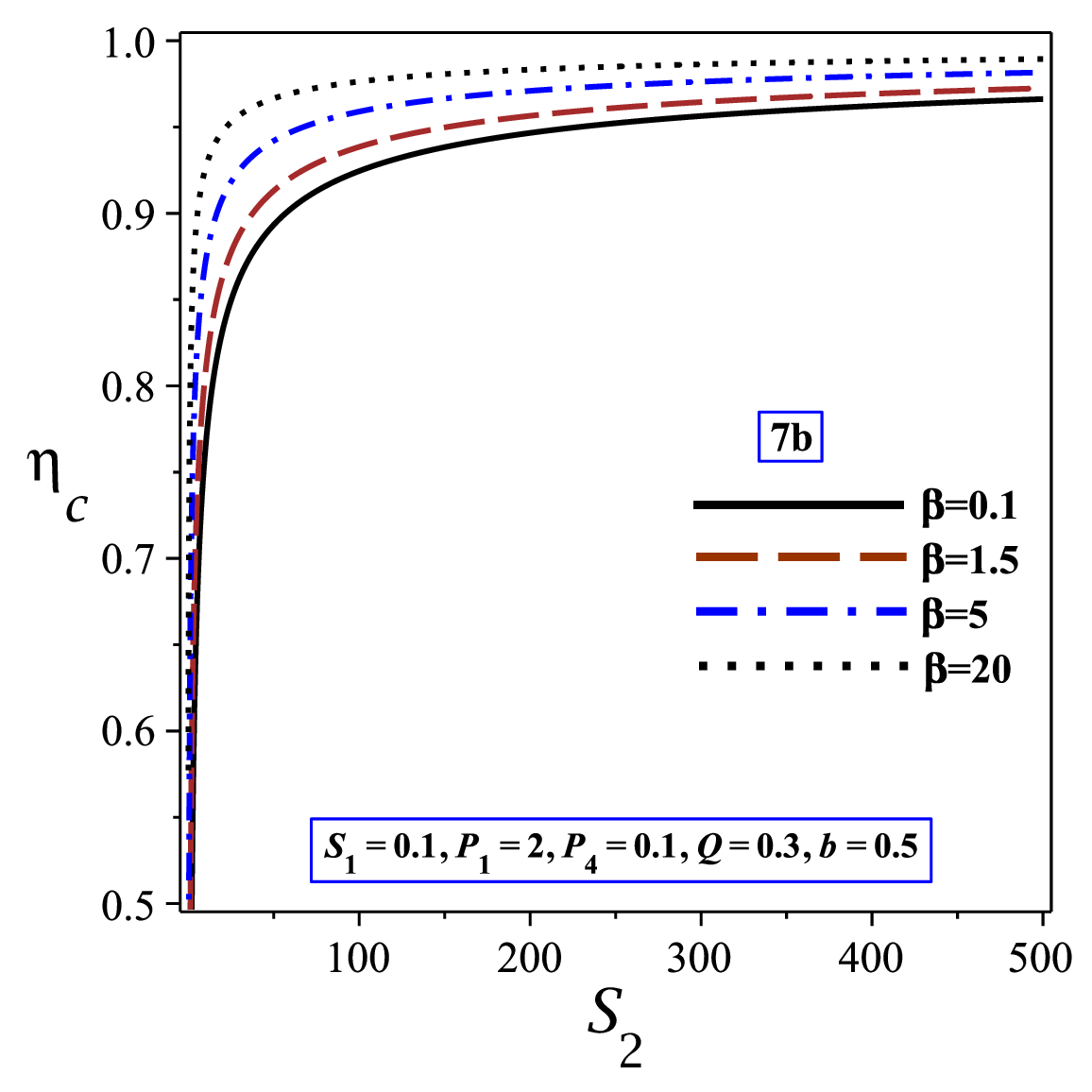}
\caption{The Canot efficiency $\protect\eta _{c}$ versus $S_{2}$ for
different parameters of $b$ in the left panel and $\protect\beta$ in the
right panel.}
\label{Fig7}
\end{figure}

From the equation (\ref{etac}), it is evident that the behavior of Carnot
efficiency is critically dependent on both the parameters of PFDM and
BI-NED. To study the effect of PFDM's parameter, we plot Carnot efficiency ($%
\eta_{c}$) versus $S_{2}$ for different values of $b$ in Fig. \ref{Fig7}a.
Our findings indicate that as $b$ increases, the heat engine efficiency
decreases when $S_{2}$ remains constant (see Fig. \ref{Fig7}a). However, for
different values of $b$, the Carnot efficiency increases and approaches $1$
as $S_{2}$ becomes very large. In Fig. \ref{Fig7}b, we evaluate the effect
of the BI-NED parameter on $\eta_{c}$ for various values of $\beta$. The
results show that $\eta_{c}$ increases with increasing $\beta$ when $S_{2}$
remains constant. Additionally, $\eta_{c}$ approaches $1$ as $S_{2}$ tends
to very large values.

\newpage

\begin{figure}[h]
\centering
\includegraphics[width=70mm]{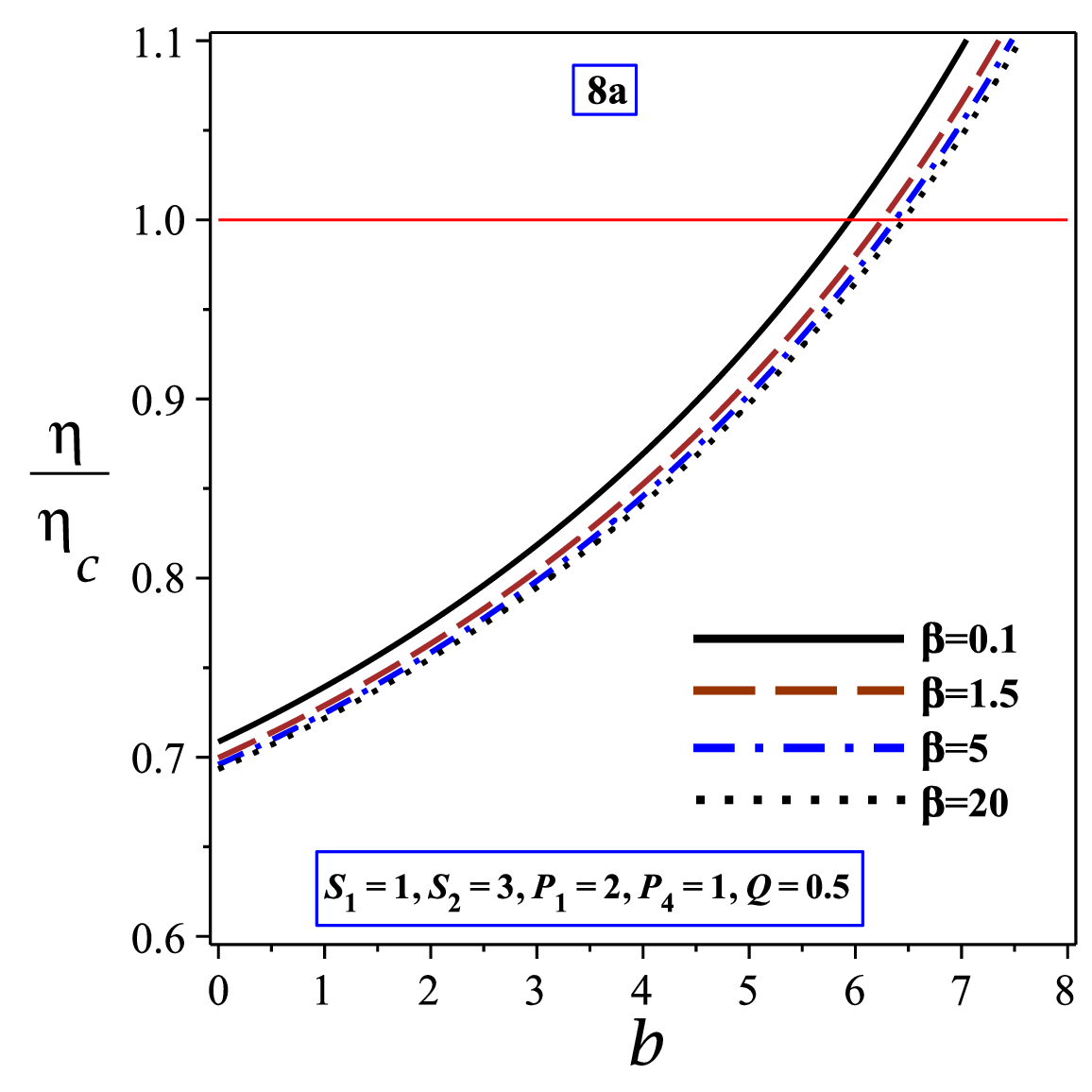} %
\includegraphics[width=70mm]{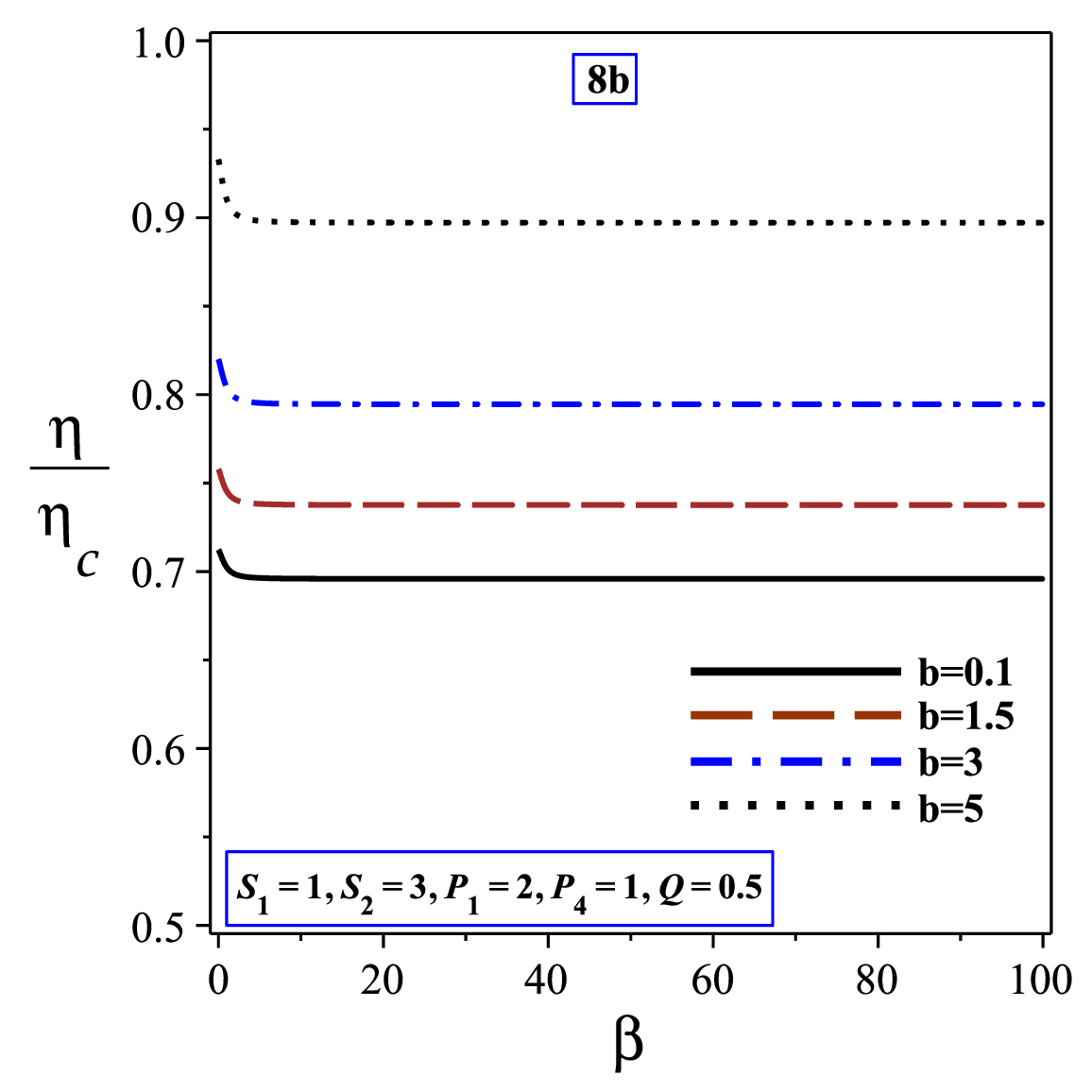}
\caption{$\frac{\protect\eta }{\protect\eta _{c}}$ versus $b$ (in the left
panel) and $\protect\beta$ (in the right panel) are plotted.}
\label{Fig8}
\end{figure}

To compare the heat engine efficiency of BI-NED AdS black holes surrounded
by PFDM with the Carnot efficiency, we plot the ratio of the efficiencies ($%
\eta$) to the Carnot efficiency ($\eta_c$) (i.e., $\frac{\eta}{\eta_c}$) for
different values of $b$ and $\beta$ in Fig. \ref{Fig8}a and Fig. \ref{Fig8}%
b, respectively. This ratio cannot exceed $1$ ($\frac{\eta}{\eta_c} <1$),
imposing a constraint on the parameters of PFDM. In other words, our
findings indicate that we cannot consider very large values for the PFDM
parameters because $\frac{\eta}{\eta_c}$ cannot exceed $1$. However, the
effect of the BI-NED parameter on this ratio shows that there is no
constraint for $\beta$, as $\frac{\eta}{\eta_c}$ remains less than $1$ for
all values of the BI-NED parameter.

\section{Topological Thermodynamic Defects}

In this section, we study the topological classification of the thermodynamic system of our new BI-PFDM-AdS solution. The method employed was developed in the well-known work of Ref.~\cite{TTD}. The approach proposed in \cite{TTD} introduces a topological classification of black hole solutions based on their thermodynamic structure, reinterpreting equilibrium states as topological defects in parameter space. The construction begins with the definition of a generalised \emph{off-shell} free energy, which depends explicitly on the horizon radius and on external thermodynamic variables, such as the reservoir temperature. Unlike the traditional analysis, the equilibrium condition is not imposed initially; instead, the free energy is regarded as a function defined on a two-dimensional parameter space. The physical black hole solutions correspond to the critical points of this free energy, namely the points where its gradient vanishes.

From this structure, a thermodynamic vector field is defined whose zeros coincide with each equilibrium solution. Each zero is treated as a topological defect, allowing the association of a winding number obtained from the rotation of the vector along a closed contour surrounding the critical point. This number is a topological invariant taking integer values, typically $+1$ or $-1$, which are directly related to local thermodynamic stability: positive values correspond to locally stable solutions (positive specific heat), whereas negative values indicate instability.

The central step of the method consists in defining a global topological number $W$, given by the sum of the winding numbers of all critical points present for a given thermodynamic configuration. This number does not depend on microscopic details or on specific parameter values, but only on the asymptotic behaviour of the temperature in extreme regimes of the horizon radius. In this way, different families of black holes can be classified into universal topological classes characterised by $W=-1$, $W=0$, or $W=+1$, reflecting respectively the predominance of unstable states, a balance between stable and unstable branches, or the predominance of stable states.

The method therefore establishes a conceptual bridge between black hole thermodynamics and the theory of topological defects, providing a global and robust characterisation of the phase structure that complements the traditional analysis based solely on specific heat or thermodynamic potentials.

Setting $\psi(r_h)=0$, taking into account $P=-\Lambda/(8\pi)$, and isolating the mass, we obtain
\begin{eqnarray}
M(r_h,q,P)=\frac{1}{6} \Bigg[3 r_h + 2 r_h^3 (4 P \pi + \beta^2) -
2 r_h^3\beta^2 ;;_2F_1[-(3/4), -(1/2), 1/
4, -(q^2/(r_h^4\beta^2))] + 3 b \ln\left(\frac{r_h}{|b|}\right)\Bigg].
\end{eqnarray}
The temperature is then
\begin{eqnarray}
T=\frac{1}{4\pi r_h^2}\Bigg[b + r_h +
2 r_h^3 \Bigg(4\pi  P +\beta^2\left(1 -
\sqrt{1 + \frac{q^2}{r_h^4\beta^2}}\right)\Bigg)\Bigg].
\end{eqnarray}
We then define the generalised free energy
\begin{eqnarray}
F(r_h,q,P)=M(r_h,q,P)-\frac{\pi r_h^2}{\tau}.
\end{eqnarray}
When the equations of motion are satisfied, imposing $\frac{\partial F}{\partial r_h}\equiv 0$, we obtain the identity $\tau=T^{-1}$.

\begin{figure}[h]
\centering
\includegraphics[width=70mm]{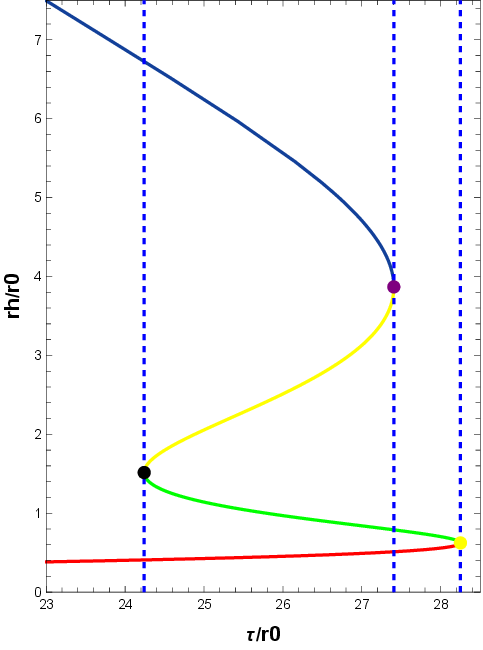} %
\caption{Representation of $\frac{\tau}{r_0}\times\frac{r_h}{r_0}$.}
\label{tau}
\end{figure}

Taking a particular example $b=0.05,\beta=0.5, q=r_0$, we find the values $r_1=0.6234565874733241r_0$, $r_2=1.515620115241819r_0$, $r_3=3.8684528125703905r_0$, after first solving $\frac{d\tau}{dr_h}=0$. The corresponding values of $\tau$ are respectively ${\tau_1=28.2502, \tau_2=24.2415, \tau_3=27.4052}$. We represent this example in Fig.~\ref{tau}. One can clearly see the vertical blue dashed lines, where the derivatives of $\tau$ vanish. We also observe four phases represented by the red, green, yellow, and dark-blue curves.
We now define a two-dimensional vector field
\begin{eqnarray}
&&\vec{\Phi}=\{\phi_r,\phi_{\theta}\},\\
&&\phi_r=\frac{\partial F}{\partial r_h}\;,\;\phi_{\theta}=-\cot\theta \csc\theta\;.
\end{eqnarray}
Normalising, we obtain
\begin{eqnarray}
&&\vec{N}=\{N_r,N_{\theta}\},\\
&&N_r=\frac{\phi_r}{\sqrt{\phi_r^2+\phi_{\theta}^2}}\;,\;N_{\theta}=\frac{\phi_{\theta}}{\sqrt{\phi_r^2+\phi_{\theta}^2}}\;.
\end{eqnarray}
The zeros of this vector field are the points separating the thermodynamic phases.

\begin{figure}[h]
\centering
\includegraphics[width=70mm]{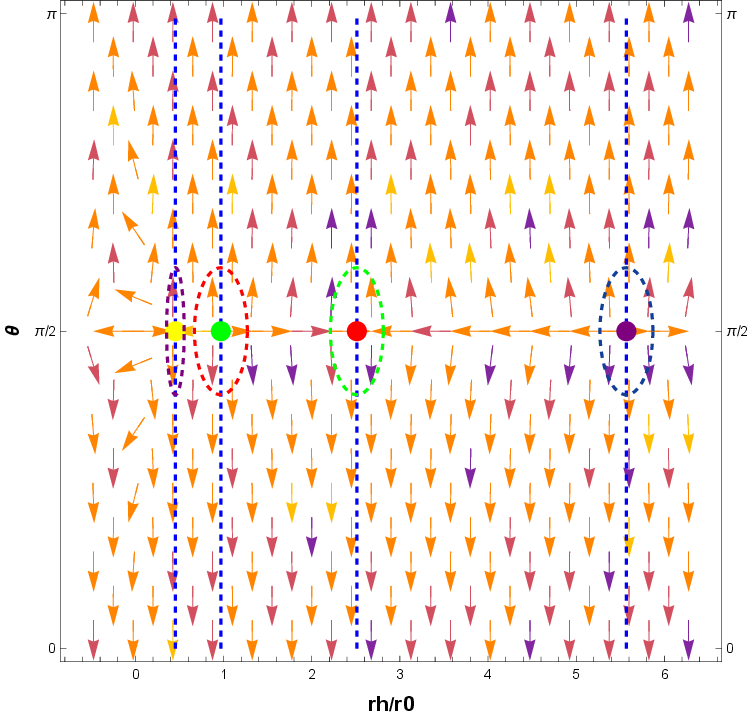} %
\caption{Representation of the vector field $\vec{N}$.}
\label{N}
\end{figure}

We now take the values $r_h=r_0;r,b=0.05,\beta=0.5,q=r_0,P=0.0022/r_0^2,\tau=26r_0$ (following the original reference \cite{TTD}). We can then represent the vector field as a function of $\theta\times\frac{r_h}{r_0}$ using the \texttt{VectorPlot} command in the Mathematica software, resulting in Fig.~\ref{N}. One can clearly see in the figure that at all phase transition points the field $\vec{N}$ vanishes and changes orientation. The vertical blue dashed curves represent the values of topological phase transition, namely $r_1=0.45348689451390667$, $r_2=0.9695037375490816$, $r_3=2.512179349141056$, $r_4=5.569267583344496$. The yellow, green, red, and purple points represent the zeros of the vector field.

We now integrate a curve around each point in order to calculate the winding numbers for each specific contour. These numbers are topological invariants and are independent of the contour. We use the following parametrisation for each contour
\begin{eqnarray}
&&r=r_i+k_i\cos(z),  \\
&&\theta=\frac{\pi}{2}+p_i\cos(z),
\end{eqnarray}
with $z\in[0,2\pi]$ and $k_i,p_i$ positive real numbers, and $r_i$ the values at which the vector field vanishes. Defining
\begin{eqnarray}
F(z)=\eta^{ij}N_{i}\frac{dN_{j}}{dz} \;,\; i,j=r,\theta\;,
\end{eqnarray}
where $\eta^{ij}$ is the Levi-Civita tensor in this space, we define the winding numbers as
\begin{eqnarray}
w_i=\frac{1}{2\pi}\int^{2\pi}_{0}F(z)dz\;.
\end{eqnarray}
Finally, summing all winding numbers, we obtain the topological invariant number that characterises the thermodynamic system of the black hole
\begin{eqnarray}
W=\Sigma_i w_i\,.
\end{eqnarray}
We now choose the contours such that $k_1=p_i=0.1$ and $k_2,k_3,k_4=0.3$. Performing the numerical integration, we obtain the winding numbers
\begin{eqnarray}
w_1=w_3=-1\;,\;w_2=w_4=1\,.
\end{eqnarray}
Therefore, the topological number is
\begin{eqnarray}
W=0.
\end{eqnarray}

We conclude that there exist four distinct phases, with two locally stable ones (second and fourth) and two locally unstable ones (first and third). This already occurred in the case of the BI-AdS black hole \cite{BIAdS}, which characterises the reentrant phase transition. The most important result is that this behaviour classifies the topology of the thermodynamic system of our BIPFDM-AdS solution as being in equilibrium between unstable and stable states, with topological number $W=0$.


\section{Timelike geodesics}

A free test particle moves around a black hole along time-like geodesics.
The corresponding geodesic equations can be obtained either via the
Hamilton-Jacobi formalism or by employing the Euler-Lagrange equations;
here, we follow the latter approach as described in detail by Chandrasekhar 
\cite{Chandra}. For a general static, spherically symmetric spacetime, the
Lagrangian for a point particle reads%
\begin{equation}
2\mathcal{L}=g_{\mu \nu }\dot{x}^{\mu }\dot{x}^{\nu }=-\psi \left( r\right) 
\dot{t}^{2}+\frac{\dot{r}^{2}}{\psi \left( r\right) }+r^{2}\left( \dot{\theta%
}^{2}+\sin ^{2}\theta \dot{\varphi}^{2}\right) ,
\end{equation}%
where the dot denotes differentiation with respect to the affine parameter $%
\tau $. The generalized momentum is then defined as 
\begin{equation}
p_{\mu }=\frac{\partial \mathcal{L}}{\partial \dot{x}^{\mu }}=g_{\mu \nu }%
\dot{x}^{\nu }.
\end{equation}%
The spacetime admits two Killing vectors, $\partial _{t}$\ and $\partial
_{\varphi }$, giving rise to two constants of motion: the energy $%
\varepsilon $ and the orbital angular momentum $L$, which satisfy

\begin{equation}
p_{t}=g_{00}\dot{t}=-\varepsilon ,
\end{equation}%
\begin{equation}
p_{\varphi }=g_{\varphi \varphi }\dot{\varphi}=L.
\end{equation}%
These equations yield%
\begin{equation}
\dot{t}=\frac{\varepsilon }{\left( 1-\frac{2m_{0}}{r}-\frac{\Lambda r^{2}}{3}%
+\frac{2\beta ^{2}r^{2}\left( 1-\mathfrak{F}_{1}\right) }{3}+\frac{b}{r}\ln
\left( \frac{r}{\left\vert b\right\vert }\right) \right) },
\end{equation}%
\begin{equation}
\dot{\varphi}=\frac{L}{r^{2}\sin ^{2}\theta }.
\end{equation}

\begin{figure}[ht]
\centering\includegraphics[width=80mm]{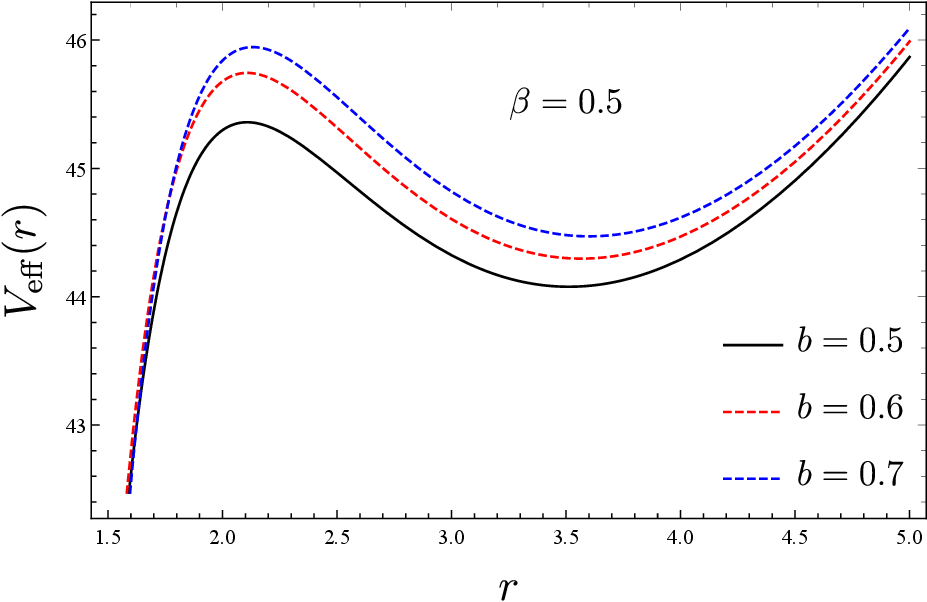} %
\includegraphics[width=80mm]{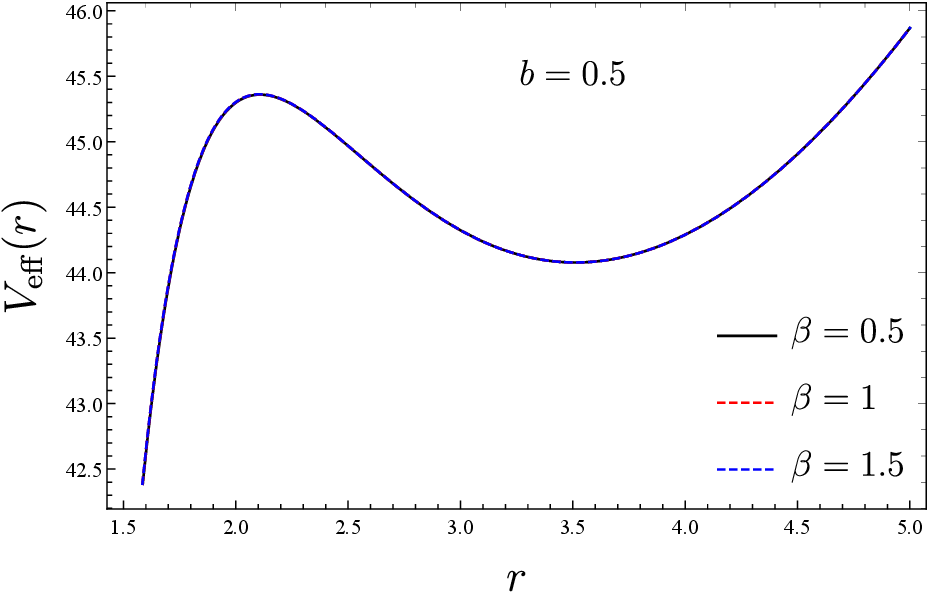}
\caption{Effective $V_{\mathrm{eff}}\left( r\right) $ potential for timelike
geodesics of a massive particle around the BI AdS black hole surrounded by
PFDM. Here we take: $m_{0}=1$, $q=0.5$, $L=10$ and $\Lambda =-1$}
\label{Figv1}
\end{figure}

For timelike geodesics, imposing $g_{\mu \nu }\dot{x}^{\mu }\dot{x}^{\nu
}=-1 $, and restricting motion to the equatorial plane $\theta =\frac{\pi }{2%
}$ and $\dot{\theta}=0$, the radial equation takes the form%
\begin{equation}
\dot{r}^{2}=\varepsilon ^{2}-\left( 1-\frac{2m_{0}}{r}-\frac{\Lambda r^{2}}{3%
}+\frac{2\beta ^{2}r^{2}\left( 1-\mathfrak{F}_{1}\right) }{3}+\frac{b}{r}\ln
\left( \frac{r}{\left\vert b\right\vert }\right) \right) \left( 1+\frac{L^{2}%
}{r^{2}}\right) .  \label{18}
\end{equation}%
This can be expressed as a one-dimensional motion equation:%
\begin{equation}
\dot{r}^{2}=\varepsilon ^{2}-V_{\mathrm{eff}}\left( r\right) ,  \label{e18}
\end{equation}

where the effective potential is%
\begin{equation}
V_{\mathrm{eff}}\left( r\right) =\left( 1-\frac{2m_{0}}{r}-\frac{\Lambda
r^{2}}{3}+\frac{2\beta ^{2}r^{2}\left( 1-\mathfrak{F}_{1}\right) }{3}+\frac{b%
}{r}\ln \left( \frac{r}{\left\vert b\right\vert }\right) \right) \left( 1+%
\frac{L^{2}}{r^{2}}\right) ,  \label{ef18}
\end{equation}%
The effective potential governs the geodesic structure, with circular orbits
corresponding to its local extrema. Figure \ref{Figv1} depicts $V_{\mathrm{%
eff}}\left( r\right) $ for Born-Infeld AdS black holes surrounded by perfect
fluid dark matter. For each combination of $b$ and $\beta $, the potential
exhibits a minimum and a maximum corresponding to stable and unstable
circular orbits, respectively. Increasing $b$ enhances the potential
barrier, reflecting the stronger gravitational effect of the surrounding
dark matter, whereas the BI parameter $\beta $ shows only a very weak
dependence, indicating a negligible contribution from the Born--Infeld
electromagnetic field.

\begin{figure}[ht]
\centering
\includegraphics[width=80mm]{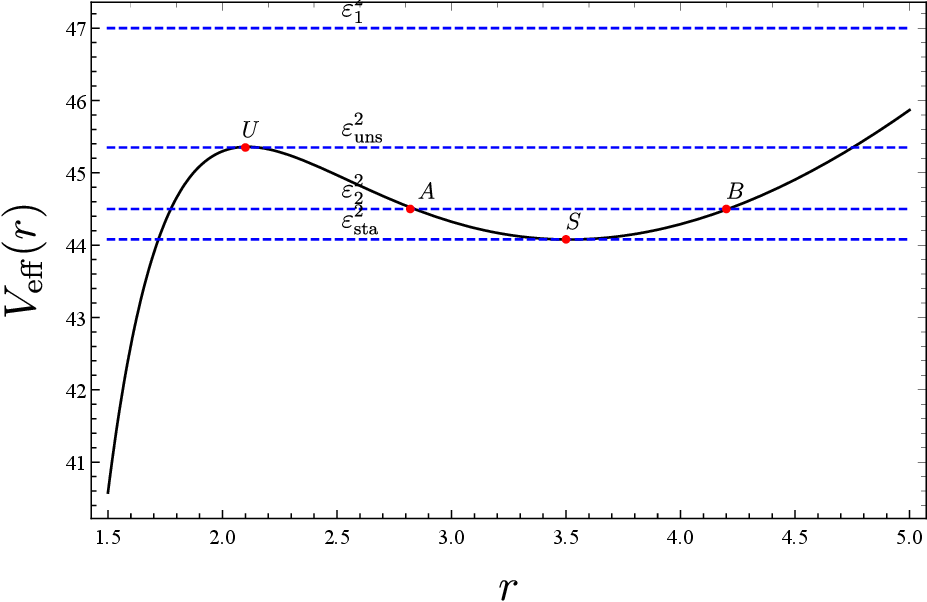}
\caption{The relation between the energy levels $\protect\varepsilon $ and
the effective potential $V_{\mathrm{eff}}\left( r\right) $.}
\label{Fig:7}
\end{figure}

We analyze the possible orbits of a massive test particle around Born-Infeld
AdS black holes surrounded by perfect fluid dark matter by solving Eq. (\ref%
{18}) numerically. In Fig.\ref{Fig:7}, we present the corresponding
effective potential. Depending on the initial energy and position of the
particle, different types of orbital motion are obtained:

\begin{itemize}
\item For $\varepsilon >\varepsilon _{uns}$, the particle falls directly
into the singularity from rest at a finite distance. In this case, a plunge
orbit occurs in which the particle approaches from infinity, partially
orbits the black hole, and ultimately spirals into the center.

\item For $\varepsilon =\varepsilon _{uns}$, the particle occupies an
unstable circular orbit; small perturbations in the initial conditions
determine whether it falls into the singularity or moves away.

\item For $\varepsilon =\varepsilon _{1}$, the particle exhibits bounded
motion between $A$ (perihelion) and $B$ (aphelion), corresponding to a
stable oscillatory orbit.

\item For $\varepsilon =\varepsilon _{sta}$, the particle remains in a
stable circular orbit.
\end{itemize}

\begin{figure}[ht]
\centering
\includegraphics[width=80mm]{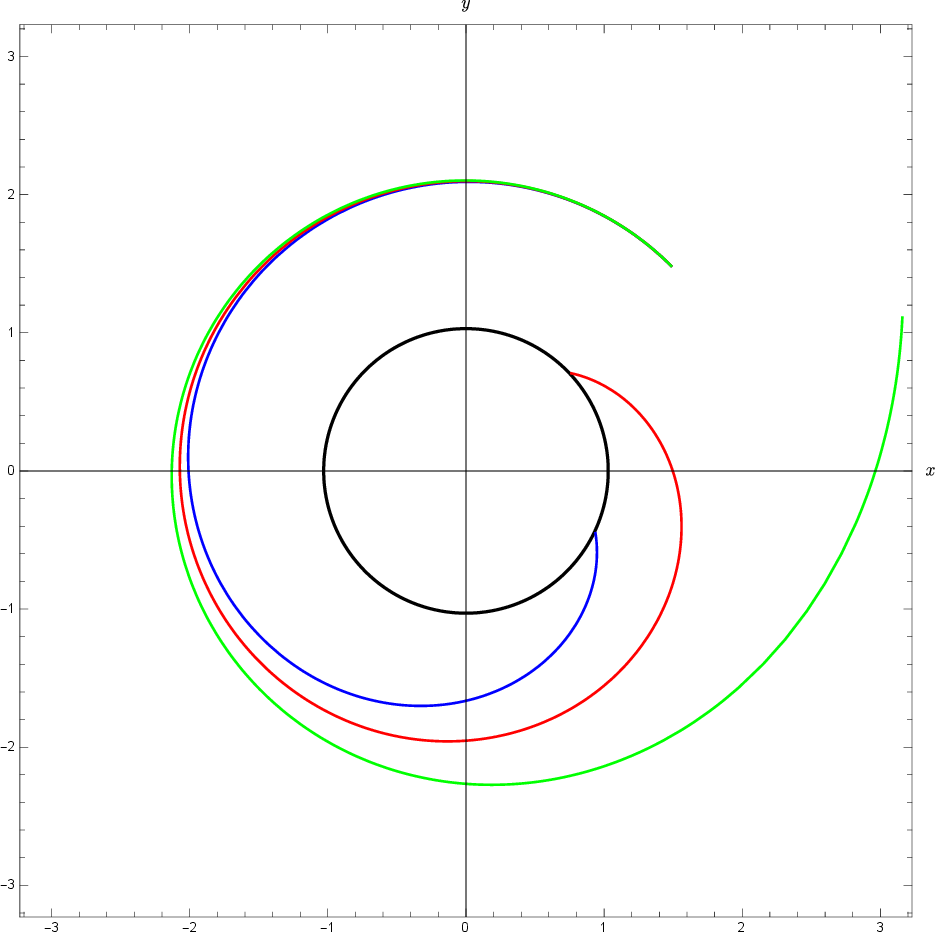} \includegraphics[width=80mm]{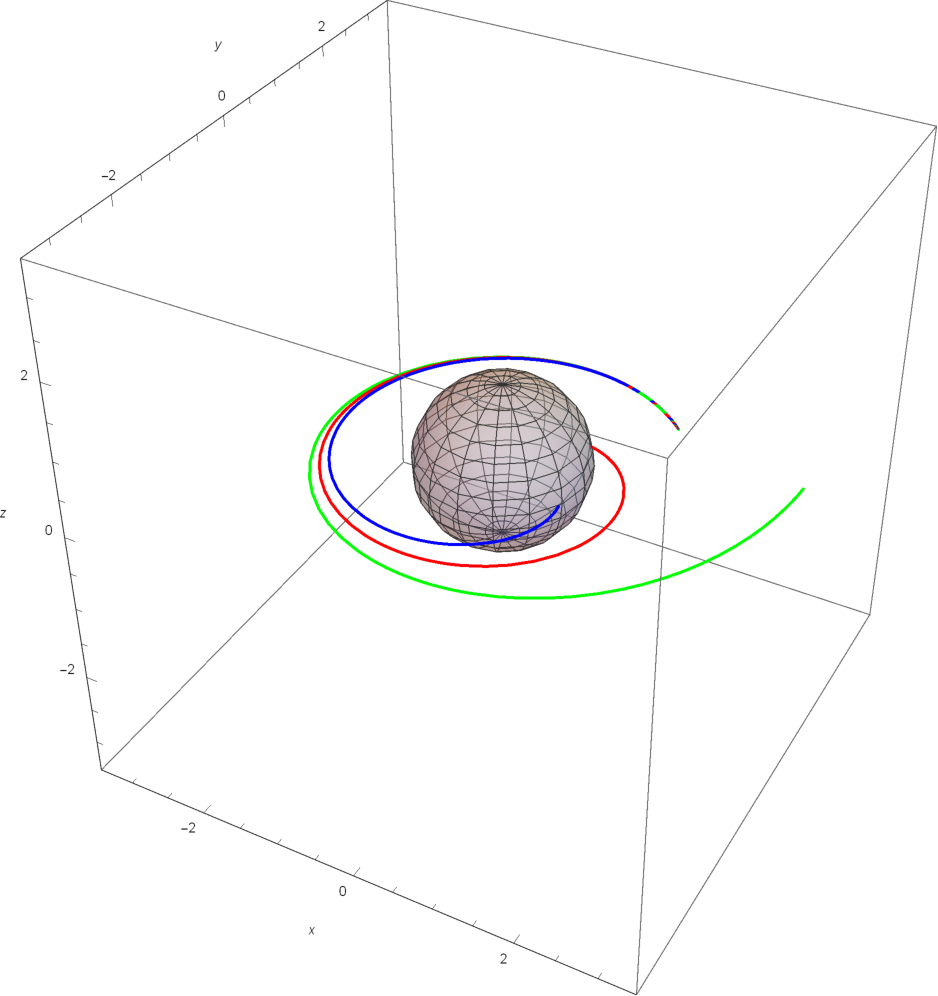}
\caption{The unstable circular orbits for different initial velocities. Blue
line $\dot{\protect\varphi}(0)=2.1678$, Red line $\dot{\protect\varphi}%
(0)=2.2678$, Green line $\dot{\protect\varphi}(0)=2.3678$}
\label{Figu1}
\end{figure}

In fig.\ref{Figu1} we illustrate the unstable circular orbit for different
initial angular velocities, with initial conditions $\left\{ {0,U,}\frac{{%
\pi }}{2}{,}\frac{{\pi }}{4}\right\} $. A critical value $\dot{\varphi}%
\left( 0\right) \simeq 2.2678$ is identified: for $\dot{\varphi}\left(
0\right) $ above this threshold, the particle does not cross the horizon,
whereas for smaller values it plunges inward.

\begin{figure}[ht]
\centering
\includegraphics[width=80mm]{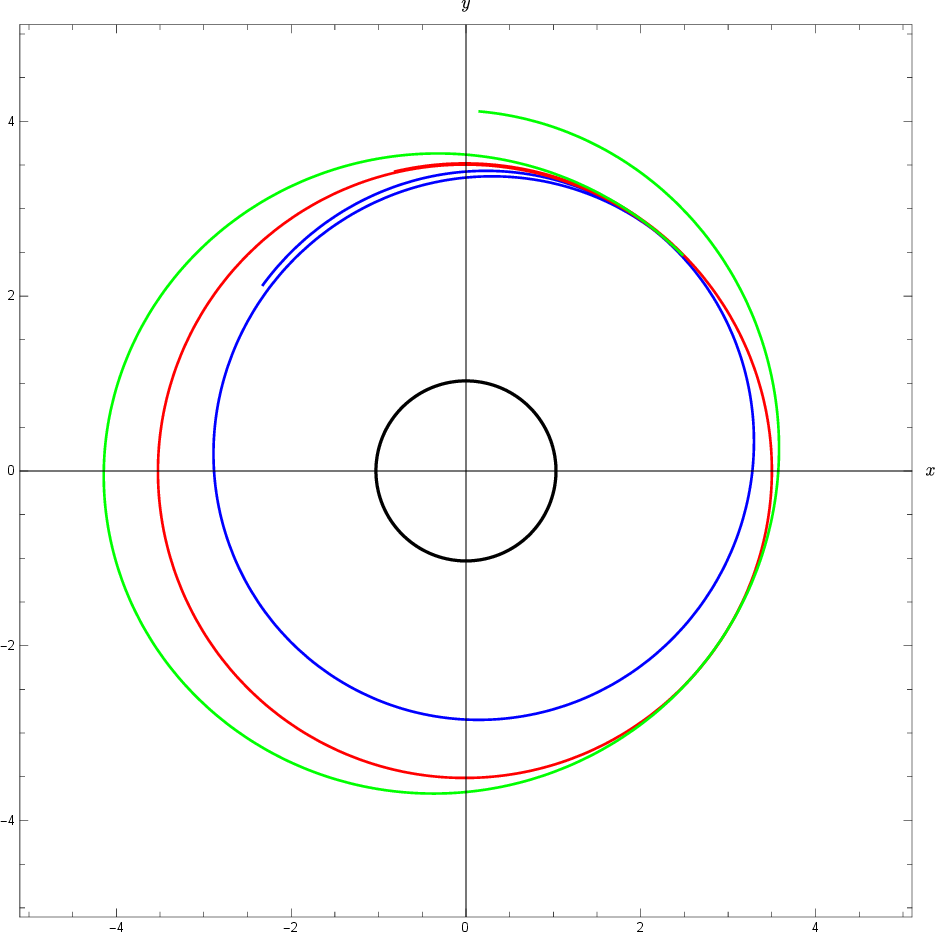} \includegraphics[width=80mm]{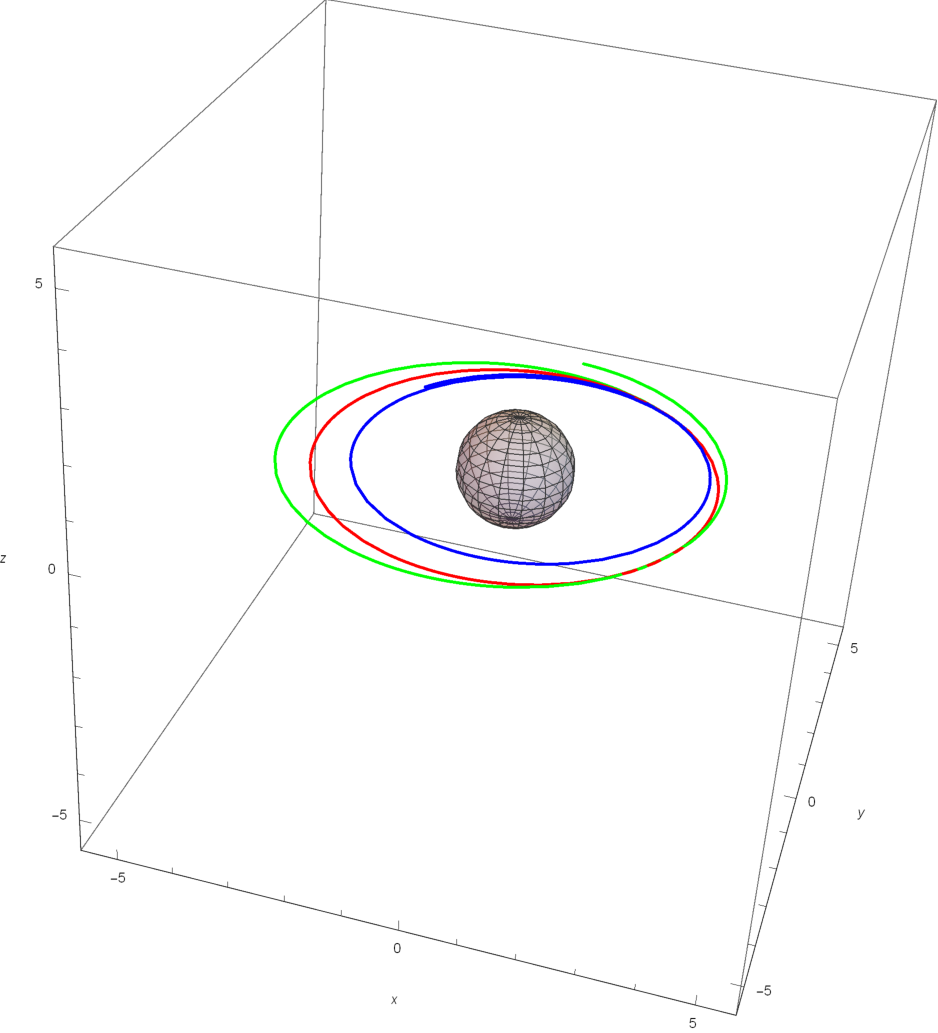}
\caption{The stable circular orbits for different initial velocities. Blue
line $\protect\varphi (0)=0.7165$, Red line $\protect\varphi (0)=0.7165$,
Green line $\protect\varphi (0)=0.9165$}
\label{Figs1}
\end{figure}

Fig.\ref{Figs1} shows the unstable circular orbit for the initial conditions 
$\left\{ {0,S,}\frac{{\pi }}{2}{,}\frac{{\pi }}{4}\right\} $ for different
initial angular velocities. The particle does not cross the horizon.
However, small perturbations destabilize the trajectory, leading either to
capture by the black hole or to escape.

\begin{figure}[ht]
\centering
\includegraphics[width=80mm]{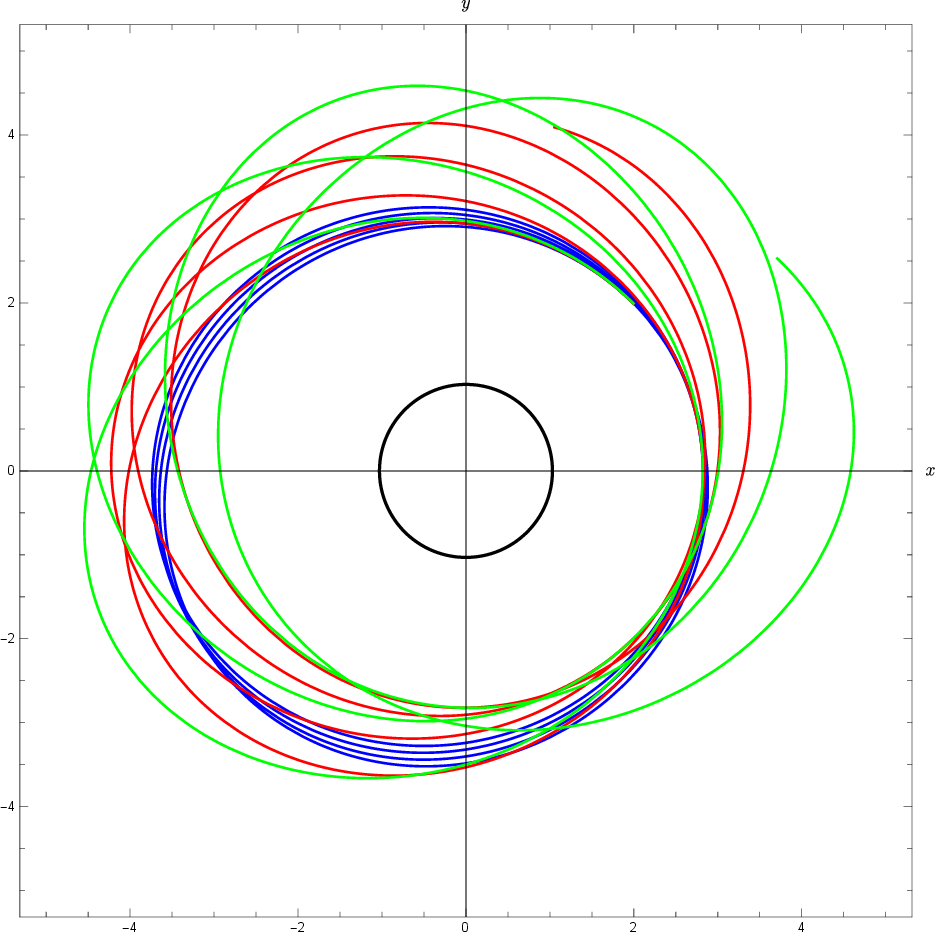} %
\includegraphics[width=80mm]{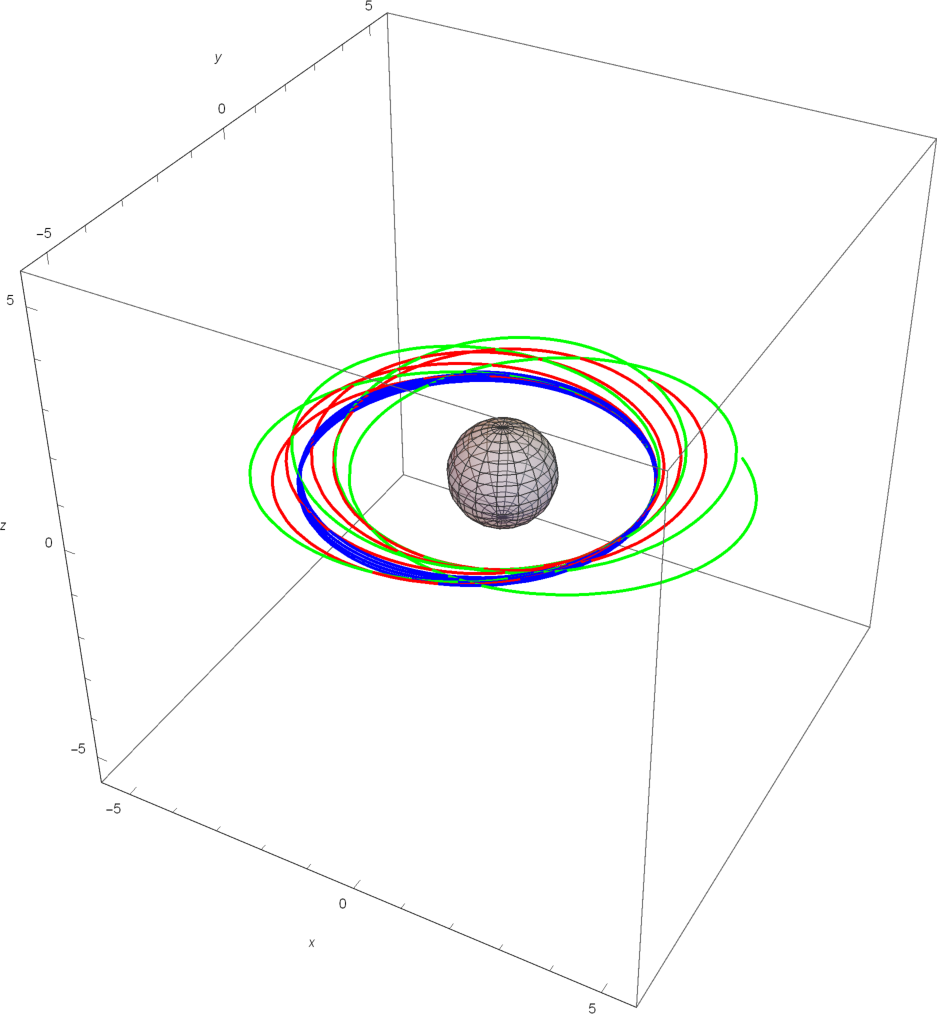}
\caption{The bound orbits for different initial velocities. Blue line $%
\protect\varphi (0)=1.158$, Red line $\protect\varphi (0)=1.258$, Green line 
$\protect\varphi (0)=1.358$}
\label{Figp1}
\end{figure}

In Fig.\ref{Figp1} we plot the bound orbits for the initial conditions $%
\left\{ {0,A,}\frac{{\pi }}{2}{,}\frac{{\pi }}{4}\right\} $ for different
initial angular velocities. In this case, the particle does not cross the
horizon but instead moves in a bounded trajectory within the radial range $%
A\leq r\leq B$, where $A$ represents the perihelion distance and $B$ the
aphelion distance. These correspond to planetary-type orbits.

\newpage \bigskip \bigskip \bigskip

We now consider the circular motion of a massive particle confined to the
equatorial plane. For a circular orbit, the radial coordinate remains
constant, implying $\dot{r}=0$ and $\ddot{r}=0$. Under these conditions, Eq.
(\ref{e18}) yields $V_{\mathrm{eff}}\left( r\right) =\varepsilon ^{2}$.
Furthermore, imposing the extremum condition $\frac{\partial }{\partial r}V_{%
\mathrm{eff}}\left( r\right) =0$. the specific energy and specific angular
momentum of the particle can be expressed as%
\begin{equation}
\varepsilon ^{2}=\frac{2\psi \left( r\right) ^{2}}{2\psi \left( r\right)
-r\psi ^{\prime }\left( r\right) },
\end{equation}%
\begin{equation}
L^{2}=\frac{r^{3}\psi ^{\prime }\left( r\right) }{2\psi \left( r\right)
-r\psi ^{\prime }\left( r\right) },
\end{equation}%
while the angular velocity of the particle is given by%
\begin{equation}
\Omega _{\varphi }=\frac{d\varphi }{dt}=\sqrt{\frac{\psi ^{\prime }\left(
r\right) }{2r}}.
\end{equation}

A circular orbit is stable if the effective potential exhibits a minimum and
unstable if it corresponds to a maximum. Tables \ref{tab:stab} summarize the
dependence of the stable and unstable circular orbit radii on the PFDM
parameter $b$ and BI-NED parameter $\beta $, respectively. For fixed $\beta $%
, the stable radius $r_{\mathrm{stable}}$ shows a non-monotonic behaviour
with increasing $b$ : it first increases, reaches a maximum around $b\sim 1$%
, and then decreases for larger values of $b$, indicating the existence of a
critical dark matter strength at which the orbital configuration undergoes a
qualitative modification. Meanwhile, the unstable circular orbit radius
first decreases, reaches a minimum and then increases. In contrast, for
fixed $b$, the variation of the unstable radius and and stable radius with
respect to the BI parameter $\beta $ shows very weak dependence. As $\beta $
increases, the unstable radius decreases slightly, whereas the stable radius
remains nearly constant with negligible change. Overall, the PFDM parameter
induces global and potentially bifurcation-like modifications in the
geodesic structure, whereas the BI-NED parameter has a negligible influence
on the stable and unstable orbit radii.

\begin{table}[ht]
\centering%
\begin{tabular}{|c|c|c||c|c|c|}
\hline
\multicolumn{3}{|c||}{$\beta =0.5$} & \multicolumn{3}{c|}{$b=0.5$} \\ \hline
$b$ & $r_{\mathrm{uns}}$ & $r_{\mathrm{sta}}$ & $\beta $ & $r_{\mathrm{uns}}$
& $r_{\mathrm{sta}}$ \\ \hline
0.2 & 2.54583 & 3.0049 & 0.2 & 2.11421 & 3.51165 \\ 
0.4 & 2.14641 & 3.43033 & 0.4 & 2.10863 & 3.51198 \\ 
0.6 & 2.10759 & 3.56575 & 0.6 & 2.10747 & 3.51204 \\ 
0.8 & 2.17506 & 3.62133 & 0.8 & 2.10706 & 3.51207 \\ 
1.0 & 2.30015 & 3.62803 & 1.0 & 2.10686 & 3.51208 \\ 
1.2 & 2.47019 & 3.5911 & 1.2 & 2.10676 & 3.51208 \\ 
1.4 & 2.69731 & 3.49522 & 1.4 & 2.10669 & 3.51208 \\ \hline
\end{tabular}%
\bigskip
\caption{Radii of stable and unstable circular orbits. Parameters $m_{0}=1$, 
$q=0.5$, $\Lambda =-1$ and $L=10$.}
\label{tab:stab}
\end{table}

\begin{figure}[ht]
\centering
\includegraphics[width=80mm]{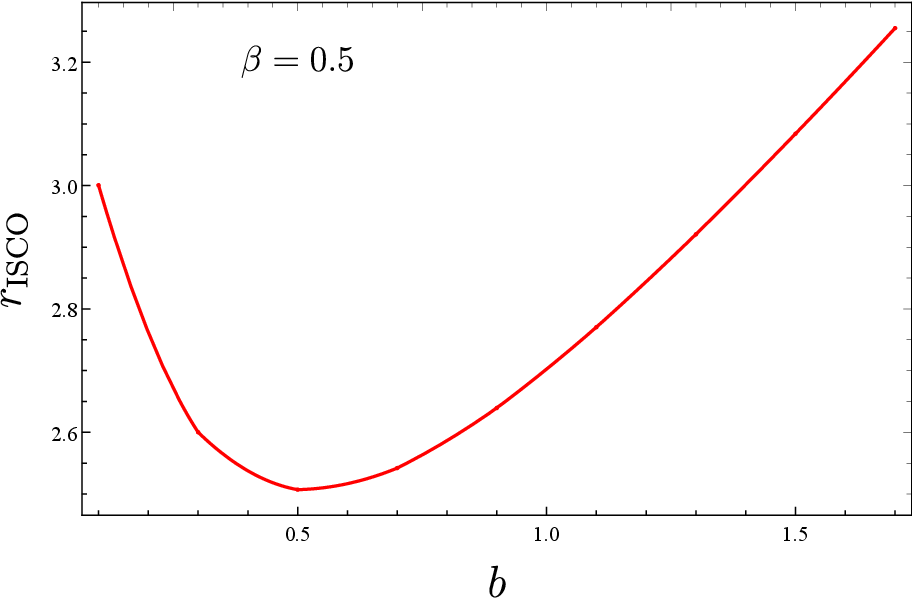} %
\includegraphics[width=80mm]{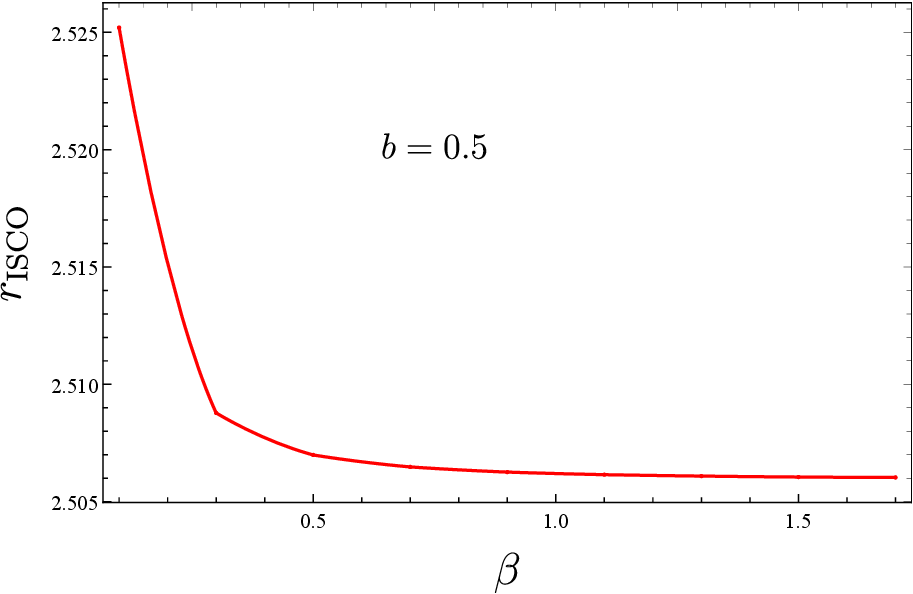}
\caption{Variation of the innermost stable circular orbit radius $r_{\mathrm{%
ISCO}}$ as a function of the PFDM parameter $b$ (left) and the BI parameter $%
\protect\beta $ (right). Here we take: $m_{0}=1$, $q=1.5$, $\protect\beta %
=0.55$.}
\label{Figv3}
\end{figure}

The most astrophysically relevant class of trajectories are the innermost
stable circular orbits (ISCOs), which determine the inner edge of accretion
disks. The ISCO corresponds to the inflection point of the effective
potential $V_{\mathrm{eff}}\left( r\right) $, and is therefore obtained from
the conditions%
\begin{equation}
\frac{\partial ^{2}}{\partial r^{2}}V_{\mathrm{eff}}\left( r\right) =\frac{%
\partial }{\partial r}V_{\mathrm{eff}}\left( r\right) =0,
\end{equation}%
Solving these equations simultaneously leads to the constraint

\begin{equation}
\left. r\psi \left( r\right) \psi ^{\prime \prime }\left( r\right) +3\psi
\left( r\right) \psi ^{\prime }\left( r\right) -2r\psi \left( r\right)
^{2}\right\vert _{r=r_{\mathrm{ISCO}}}=0.
\end{equation}%
Since this equation cannot be solved analytically in closed form, we
investigate it numerically. In Fig. \ref{Figv3} (left panel), we present the
variation of the ISCO radius as a function of the PFDM parameter $b$. The
curve exhibits a clear non-monotonic behaviour. For small values of $b$, $r_{%
\mathrm{ISCO}}$ decreases rapidly as $b$ increases, reaching a minimum at a
critical value $b_{crt}$. Beyond this point, $r_{\mathrm{ISCO}}$ increases
monotonically with further increase in $b$. This behaviour signals the
existence of a critical dark matter strength at which the ISCO radius
attains its smallest value. Physically, this indicates that for moderate
PFDM intensity, the effective gravitational attraction is enhanced in such a
way that stable circular motion can occur closer to the black hole. However,
for larger values of $b$, the dark matter contribution modifies the global
structure of the spacetime and shifts the ISCO outward. The resulting
U-shaped profile reflects the competing influence of the logarithmic PFDM
term in the metric function, which affects both the near-horizon and
intermediate radial regions of the effective potential. This non-monotonic
behaviour demonstrates the significant impact of PFDM on the strong-field
orbital structure. The dependence of the ISCO radius on the BI-NED parameter 
$\beta $ is shown in Fig. \ref{Figv3} (right panel). For very small values
of $\beta $, corresponding to the strongly nonlinear electrodynamic regime, $%
r_{\mathrm{ISCO}}$ decreases smoothly and reaches a minimum at an
intermediate value of the parameter.

\subsection{The effective force}

\begin{figure}[ht]
\centering
\includegraphics[width=80mm]{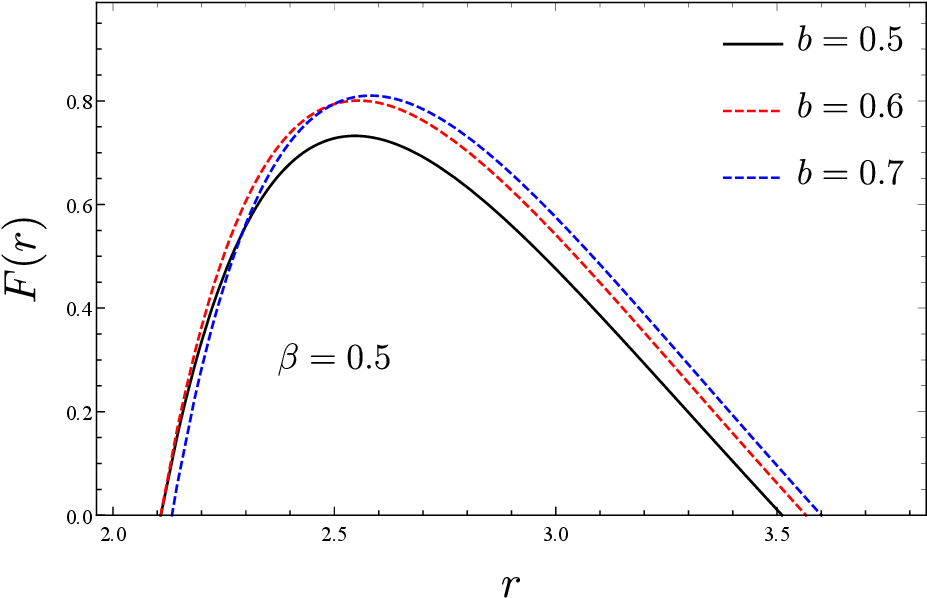} %
\includegraphics[width=80mm]{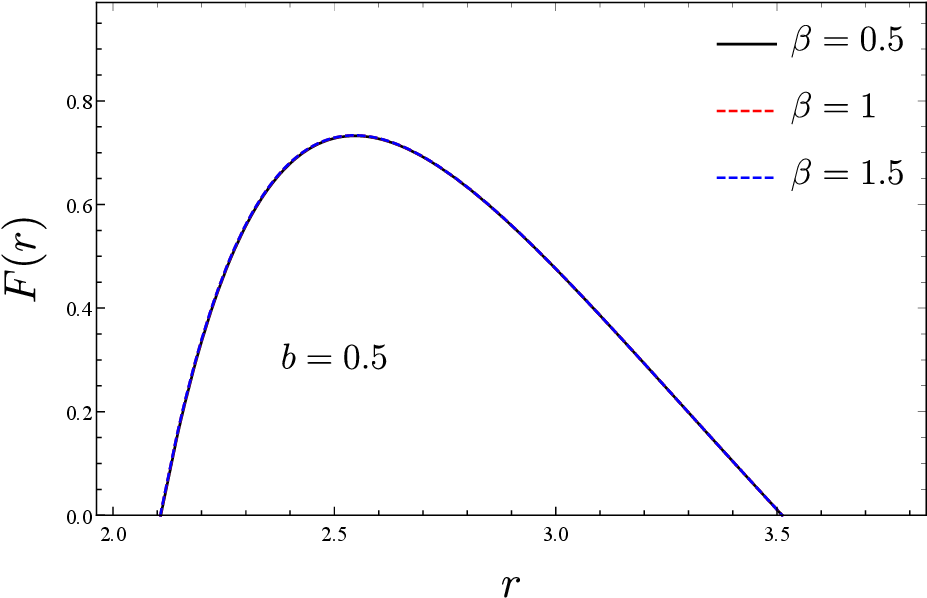}
\caption{Radial effective force $F\left( r\right) $ acting on a test
particle around the BI-NED AdS black hole surrounded by PFDM. Here we take: $%
m_{0}=1$, $q=0.5$, $L=10$ and $\Lambda=-1$}
\label{fig:force}
\end{figure}

The effective force governing the motion of a test particle plays a crucial
role in determining the nature of its trajectory, as it characterizes
whether the particle is subject to gravitational attraction or experiences
repulsive effects in the vicinity of the black hole. The effective force is
derived from Eq. (\ref{ef18}) by taking the radial derivative of the
effective potential, and is given by%
\begin{eqnarray}
F\left( r\right) &=&-\frac{1}{2}\frac{d}{dr}V_{eff}\left( r\right) =\left( 
\frac{1}{r}-\frac{2m_{0}}{r^{2}}+\frac{2\Lambda r}{3}+\frac{3b}{2r^{2}}\ln
\left( \frac{r}{\left\vert b\right\vert }\right) -\frac{b}{2r^{2}}+\frac{%
2q^{2}\mathfrak{F}_{2}}{r^{3}}\right) \frac{L^{2}}{r^{2}}  \notag \\
&&-\frac{m_{0}}{r^{2}}+\frac{\Lambda r}{3}-\frac{b}{2r^{2}}+\frac{b}{2r^{2}}%
\ln \left( \frac{r}{\left\vert b\right\vert }\right) -\frac{2\beta
^{2}r\left( 1-\mathfrak{F}_{1}\right) }{3}+\frac{2q^{2}\mathfrak{F}_{2}}{%
r^{3}}.
\end{eqnarray}%
The effective force $F\left( r\right) $ as a function of the radial
coordinate $r$ is displayed in Fig.\ref{fig:force}. In the left panel,
increasing the PFDM parameter $b$ leads to a noticeable enhancement of the
peak amplitude of the force, indicating a stronger gravitational influence
due to the surrounding dark matter. In contrast, as shown in the right
panel, variations of the BI parameter $\beta $ produce only marginal
modifications of the force profile, suggesting that the nonlinear
electrodynamic effects remain subdominant in the considered parameter range.

\section{Null geodesics}

In general relativity, photon trajectories are described by the null
geodesics of the background spacetime. However, in NED, photons do not
propagate along the null geodesics of the original metric. Instead, due to
self-interaction of the electromagnetic field, they follow null geodesics of
an effective geometry determined by the nonlinear structure of the theory.
This effective metric depends explicitly on the particular NED model under
consideration. Following the formalism of Novello et al. \cite{Novel}, the
photon trajectory is obtained by analyzing the propagation of
discontinuities in the first derivatives of the electromagnetic field, which
propagate along the wave vector $k^{\mu }$ satisfying the null condition%
\begin{equation}
\hat{g}^{\mu \nu }k_{\mu }k_{\nu }=0,
\end{equation}

where the effective metric is given by

\begin{equation}
\hat{g}^{\mu \nu }=\left( L_{\mathcal{F}}g^{\mu \nu }-4L_{\mathcal{FF}%
}F^{\mu \alpha }F_{\alpha }^{\nu }\right) ,
\end{equation}%
with $L_{\mathcal{FF}}=\frac{d^{2}}{d\mathcal{F}^{2}}L\left( \mathcal{F}%
\right) $. Using the non-vanishing components of the background metric and
the electromagnetic field, the effective metric for the BI-NED model takes
the form%
\begin{equation}
ds^{2}=\hat{g}_{00}dt^{2}+\hat{g}_{rr}dr^{2}+\hat{g}_{\theta \theta }d\theta
^{2}+\hat{g}_{\varphi \varphi }d\varphi ^{2},
\end{equation}%
where the inverse components are

\begin{equation}
\hat{g}^{00}=-\frac{L_{\mathcal{F}}+4L_{\mathcal{FF}}H\left( r\right) ^{2}}{%
\psi \left( r\right) }=\frac{1}{\psi \left( r\right) }\left( 1-\frac{q^{2}}{%
\beta ^{2}r^{4}}\right) \sqrt{1+\frac{q^{2}}{\beta ^{2}r^{4}}},
\end{equation}%
\begin{equation}
g^{rr}=\psi \left( r\right) \left( L_{\mathcal{F}}g^{rr}+4L_{\mathcal{FF}%
}H\left( r\right) ^{2}\right) =-\psi \left( r\right) \left( 1-\frac{q^{2}}{%
\beta ^{2}r^{4}}\right) \sqrt{1+\frac{q^{2}}{\beta ^{2}r^{4}}},
\end{equation}%
\begin{equation}
\hat{g}^{\theta \theta }=\frac{L_{\mathcal{F}}}{r^{2}}=-\frac{1}{r^{2}}\sqrt{%
1+\frac{q^{2}}{\beta ^{2}r^{4}}},
\end{equation}%
\begin{equation}
\hat{g}^{\varphi \varphi }=-\frac{1}{r^{2}\sin ^{2}\theta }\sqrt{1+\frac{%
q^{2}}{\beta ^{2}r^{4}}}.
\end{equation}%
We now investigate photon motion in the background of a BI-NED AdS black
hole surrounded by PFDM. The dynamics are governed by the null geodesics of
the effective metric, which can be derived from the Lagrangian

\begin{equation}
\mathcal{L}=-\frac{1}{2}\frac{1}{\sqrt{1+\frac{q^{2}}{\beta ^{2}r^{4}}}}%
\left[ -\frac{\psi \left( r\right) }{\left( 1-\frac{q^{2}}{\beta ^{2}r^{4}}%
\right) }\dot{t}^{2}+\frac{1}{\left( 1-\frac{q^{2}}{\beta ^{2}r^{4}}\right)
\psi \left( r\right) }\dot{r}^{2}+r^{2}\dot{\theta}^{2}+r^{2}\sin ^{2}\theta 
\dot{\varphi}^{2}\right] .
\end{equation}%
which, upon substitution of the effective metric components, yields the
modified radial and angular equations of motion. Since the metric components
are independent of $t$ and $\varphi $, two conserved quantities arise: the
energy $E$ and angular momentum $L$, given by

\begin{equation}
E=\frac{\psi \left( r\right) }{\left( 1-\frac{q^{2}}{\beta ^{2}r^{4}}\right) 
\sqrt{1+\frac{q^{2}}{\beta ^{2}r^{4}}}}\dot{t},
\end{equation}%
\begin{equation}
L=\frac{r^{2}\sin ^{2}\theta }{\sqrt{1+\frac{q^{2}}{\beta ^{2}r^{4}}}}\dot{%
\varphi}.
\end{equation}%
Imposing the null condition $\hat{g}_{\mu \nu }\dot{x}^{\mu }\dot{x}^{\nu
}=0 $, and restricting to the equatorial plane, the radial equation of
motion can be expressed in the standard form 
\begin{equation}
\dot{r}^{2}+\mathcal{V}_{eff}\left( r\right) =0,
\end{equation}%
where the effective potential is%
\begin{equation}
\mathcal{V}_{eff}\left( r\right) =\left( 1-\frac{q^{4}}{\beta ^{4}r^{8}}%
\right) \left[ \psi \left( r\right) \frac{L^{2}}{r^{2}}-\left( 1-\frac{q^{2}%
}{\beta ^{2}r^{4}}\right) E^{2}\right] .  \label{33}
\end{equation}%
This expression explicitly shows how nonlinear electrodynamics modifies
photon motion through multiplicative corrections depending on $\beta $,
thereby altering the structure of null trajectories relative to the
Maxwellian case.

\begin{figure}[ht]
\centering
\includegraphics[width=80mm]{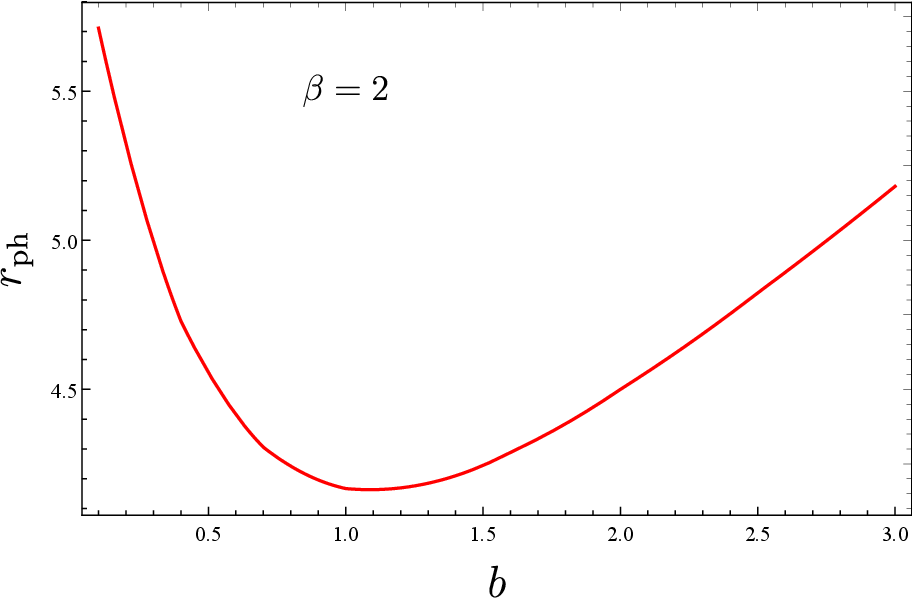} %
\includegraphics[width=80mm]{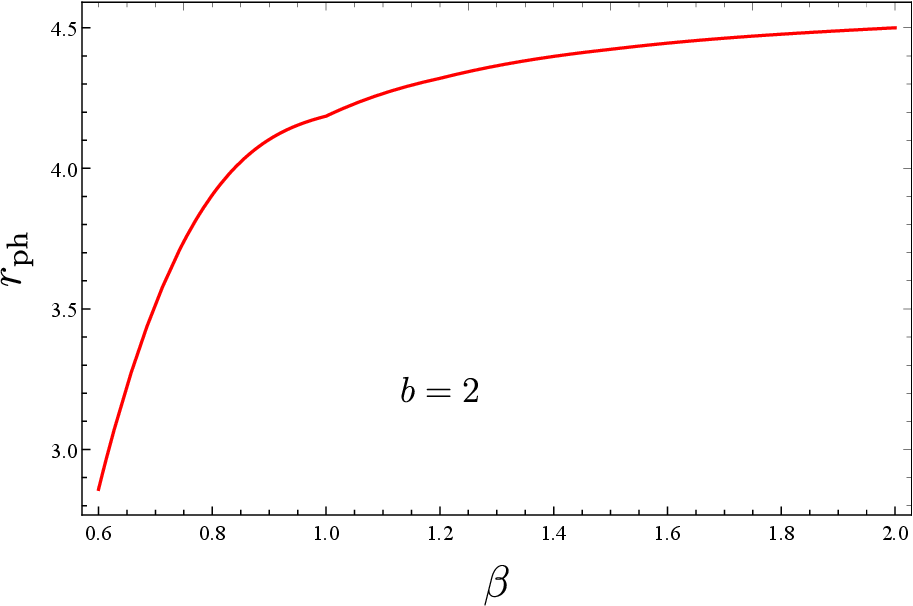}
\caption{Radius of the photon sphere $r_{\mathrm{ph}}$ as a function of the
PFDM parameter $b$ (left) and the BI parameter $\protect\beta $ (right).
Here we tahe: $m_{0}=2.5$, $q=1.8$ and $\Lambda=-1$.}
\label{fig:photon}
\end{figure}

In order to find the unstable circular orbits we impose the conditions%
\begin{equation}
\left. \mathcal{V}_{eff}\left( r\right) \right\vert _{r=r_{\mathrm{ph}}}=0;%
\text{ }\left. \frac{\partial }{\partial r}\mathcal{V}_{eff}\left( r\right)
\right\vert _{r=r_{\mathrm{ph}}}=0,
\end{equation}%
and check whether $\mathcal{V}_{eff}\left( r\right) $ is a maxima at $r=r_{%
\mathrm{ph}}$, that is%
\begin{equation}
\left. \frac{\partial ^{2}}{\partial r^{2}}\mathcal{V}_{eff}\left( r\right)
\right\vert _{r=r_{\mathrm{ph}}}<0,
\end{equation}%
where $r_{\mathrm{p}}$ is the radius of the photon sphere. Using Eq. (\ref%
{33}), the first condition leads to the critical impact parameter

\begin{equation}
\xi _{\mathrm{ph}}=\frac{L}{E}=r_{\mathrm{ph}}\sqrt{\frac{1-\frac{q^{2}}{%
\beta ^{2}r_{\mathrm{ph}}^{4}}}{\psi \left( r_{\mathrm{ph}}\right) }}.
\end{equation}%
An ingoing photon from infinity with $\xi <\xi _{\mathrm{ph}}$ falls into
the black hole, while photons with $\xi >\xi _{\mathrm{ph}}$ aare scattered
and can reach an observer at infinity. The second condition leads to

\begin{equation}
\left. \left( 1-\frac{q^{2}}{\beta ^{2}r^{4}}\right) \frac{d}{dr}\ln \left( 
\frac{\psi \left( r\right) }{r^{2}}\right) -\frac{4q^{2}}{\beta ^{2}r^{5}}%
\right\vert _{r=r_{\mathrm{ph}}}=0.  \label{ph}
\end{equation}%
In general Eq.(\ref{ph}) cannot be solved analytically, so we proceed
numerically. Figure \ref{fig:photon} shows the equatorial photon sphere
radius for a BI AdS black hole surrounded by PFDM. For fixed $\beta $ and
varying the PFDM parameter $b$, the photon sphere radius exhibits a
non-monotonic dependence on the PFDM parameter $b$ : for small $b$,
increasing $b$ reduces $r_{\mathrm{ph}}$ indicating enhanced gravitational
attraction near the black hole. Beyond a critical value $b_{c}$, however, $%
r_{\mathrm{ph}}$ increases, reflecting the modification of spacetime
geometry by dark matter that shifts the unstable null orbit outward. For
fixed $b=2$ varying BI-NED parameter $\beta $, results in a photon sphere
radius that increases monotonically with $\beta $ and approaches a constant
for large $\beta $ indicating that NED effects weaken as $\beta $ grows. In
the large-$\beta $ limit, the solution reduces to the Reissner-Nordstr\"{o}m
geometry, leading to saturation of the photon sphere radius. The initial
growth of $r_{\mathrm{ph}}$ demonstrates that strong NED effects compress
the photon orbit inward, while the Maxwell limit restores standard charged
black hole behavior.

\begin{figure}[ht]
\centering
\includegraphics[width=80mm]{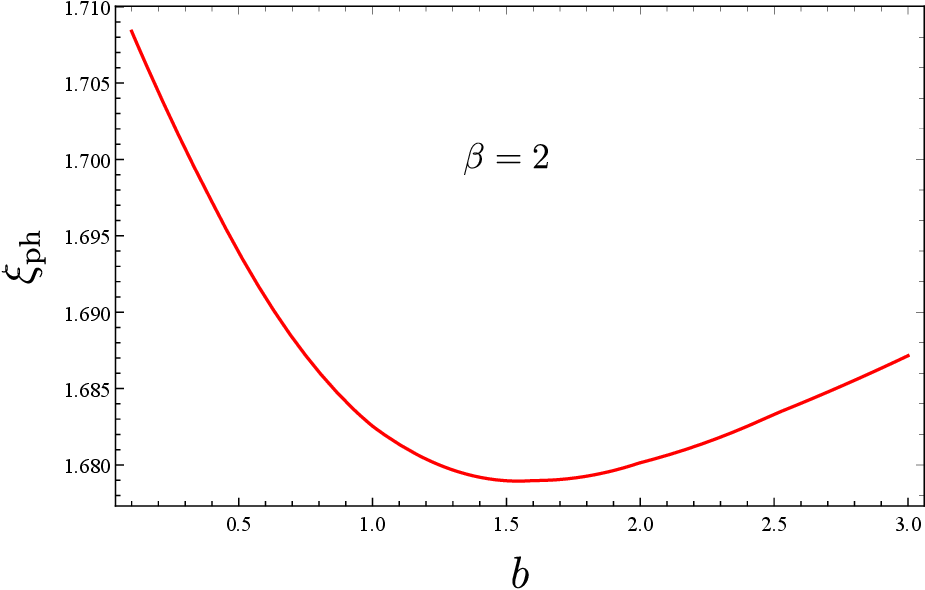} %
\includegraphics[width=80mm]{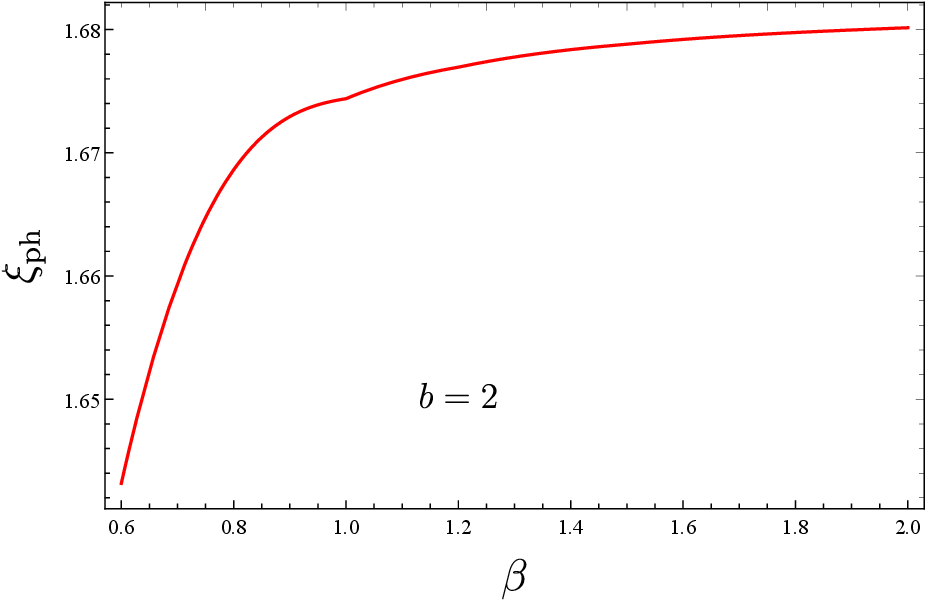}
\caption{Critical impact parameter $\protect\xi _{\mathrm{ph}}$ as a
function of the PFDM parameter $b$ (left) and the BI parameter $\protect%
\beta $ (right). Here we tahe: $m_{0}=2.5$, $q=1.8$ and $\Lambda=-1$.}
\label{fig:impact}
\end{figure}

Figure \ref{fig:impact} shows the critical impact parameter as a function of
the radial coordinate. For fixed $\beta $, the critical impact parameter
exhibits a non-monotonic dependence on $b$, initially decreasing and then
increasing beyond a critical value $b_{c}$,reflecting the competing
gravitational effects of PFDM. For fixed $b$ the impact parameter increases
monotonically with $\beta $ and saturates at large $\beta $, signaling the
recovery of the Maxwell limit as nonlinear electrodynamics effects vanish.

For black hole spacetimes possessing a cosmological horizon, such as the
Kottler solution, the angular size of the black hole shadow depends
explicitly on the radial position of the observer and on the observer's
state of motion. For a static observer located at a radial coordinate $r_{%
\mathrm{o}}$, the angular radius of the shadow, $\alpha _{\mathrm{sh}}$, is
given by

\begin{equation}
\sin \alpha _{\mathrm{sh}}=\sqrt{\frac{g_{\varphi \varphi }\left( r_{\mathrm{%
ph}}\right) g_{rr}\left( r_{\mathrm{o}}\right) }{g_{rr}\left( r_{\mathrm{ph}%
}\right) g_{\varphi \varphi }\left( r_{\mathrm{o}}\right) }}=\frac{r_{%
\mathrm{ph}}}{r_{\mathrm{o}}}\sqrt{\frac{1-\frac{q^{2}}{\beta ^{2}r_{\mathrm{%
ph}}^{4}}}{1-\frac{q^{2}}{\beta ^{2}r_{\mathrm{o}}^{4}}}\frac{\psi \left( r_{%
\mathrm{o}}\right) }{\psi \left( r_{\mathrm{ph}}\right) }},
\end{equation}%
which can equivalently be written as%
\begin{equation}
\sin \alpha _{\mathrm{sh}}=\frac{\xi _{_{\mathrm{ph}}}}{r_{\mathrm{o}}}\sqrt{%
\frac{\psi \left( r_{\mathrm{o}}\right) }{1-\frac{q^{2}}{\beta ^{2}r_{%
\mathrm{o}}^{4}}}}.
\end{equation}

It is important to emphasize that this relation is originally derived in
asymptotically flat spacetimes, where null geodesics either escape to
infinity or are captured by the black hole. In asymptotically AdS
spacetimes, however, the conformal boundary reflects null signals, thereby
modifying the global causal structure. Nevertheless, for an observer located
at a finite radial distance $r_{\mathrm{o}}$, the shadow is still
predominantly determined by the photon sphere, since the critical impact
parameter $\xi _{_{\mathrm{ph}}}$ governs photon capture. In this regime,
the influence of the AdS boundary on the angular size of the shadow is
negligible, and the above expression provides a reliable approximation.
Noticeable deviations would arise only for observers situated at very large $%
r_{\mathrm{o}}$, where multiple reflections from the AdS boundary may
contribute to the observed image.

\begin{figure}[ht]
\centering
\includegraphics[width=80mm]{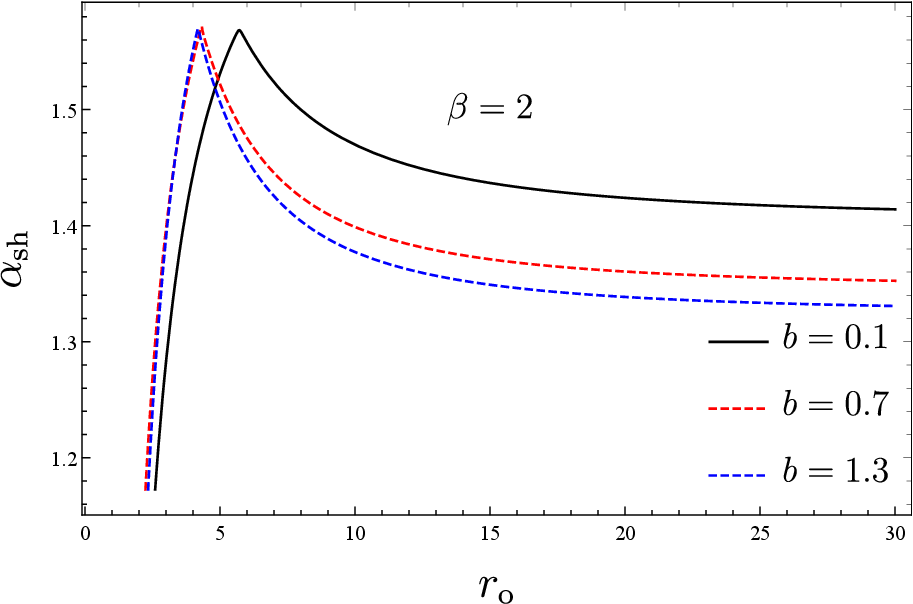} \includegraphics[width=80mm]{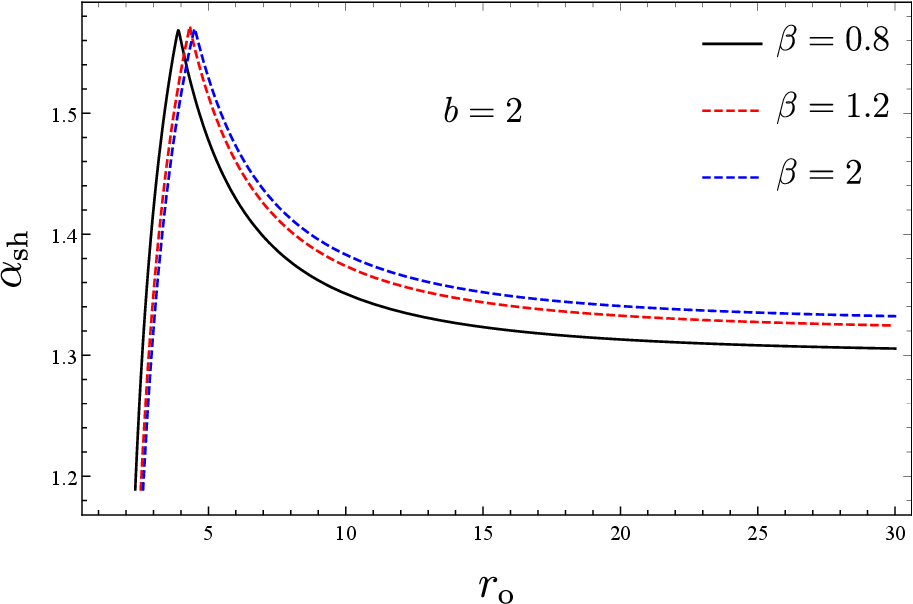}
\newline
\caption{Angular radius of the black hole shadow $\protect\alpha _{\mathrm{sh%
}}$ versus the PFDM parameter $b$ (left) and the BI parameter $\protect\beta 
$ (right).Here we tahe: $m_{0}=2.5$, $q=1.8$ and $\Lambda=-1$.}
\label{fig:alpha}
\end{figure}

In the limiting case $r_{\mathrm{o}}=r_{_{\mathrm{ph}}}$, i.e. when the
observer is located at the photon sphere, the angular radius satisfies $%
\alpha _{\mathrm{sh}}=\pi /2$, meaning that the shadow occupies exactly one
half of the observer's sky. The behaviour of the static angular radius $%
\alpha _{\mathrm{sh}}$ as a function of $r_{\mathrm{o}}$ is displayed in
Fig.~\ref{fig:alpha}. In all cases, the shadow angle increases rapidly with
increasing $r_{\mathrm{o}}$, reaches a maximum at an intermediate distance,
and subsequently approaches a constant value for large $r_{\mathrm{o}}$.
This behaviour indicates that for observers close to the black hole the
apparent shadow size is strongly affected by local spacetime curvature,
whereas for distant observers the angular radius converges to an asymptotic
value determined primarily by the photon sphere structure.

\bigskip

On the other hand, for a static observer located at a radial
coordinate $r_{\mathrm{o}}$, the radius of the black hole shadow is
expressed as%
\begin{equation}
R_{\mathrm{sh}}=\xi _{_{\mathrm{ph}}}\sqrt{\frac{1-\frac{2m_{0}}{r_{\mathrm{o%
}}}-\frac{\Lambda r_{\mathrm{o}}^{2}}{3}+\frac{2\beta ^{2}r_{\mathrm{o}%
}^{2}\left( 1-\mathfrak{F}_{1}\right) }{3}+\frac{b}{r_{\mathrm{o}}}\ln
\left( \frac{r_{\mathrm{o}}}{\left\vert b\right\vert }\right) }{1-\frac{q^{2}%
}{\beta ^{2}r_{\mathrm{o}}^{4}}}}.
\end{equation}%
This relation shows that the shadow radius depends explicitly on the
observer's position as well as the parameters of the spacetime, including
the cosmological constant, the BI-NED parameter, the electric charge, and
the PFDM parameter. In the asymptotic limit $r_{\mathrm{o}}\rightarrow
\infty $, the shadow radius $R_{\mathrm{sh}}$ does not reduce to the
critical impact parameter $\xi _{_{\mathrm{ph}}}$, indicating that the
presence of the cosmological constant and NED modifies the asymptotic
behavior of the shadow compared to the standard asymptotically flat case.

\section{Discussion and Conclusion}

In this study, we examined Einstein's theory of gravity in the context of a
negative cosmological constant and Born-Infeld Nonlinear Electrodynamics
(BI-NED), all within a framework that includes a Perfect Fluid Dark Matter
(PFDM) component. By applying this gravitational framework to a static,
spherically symmetric scenario in four-dimensional spacetime, we derived
exact solutions. These solutions exhibited a singularity at $r=0$, which was
encompassed by at least one event horizon. As a result, we interpreted these
solutions as BI-NED AdS black holes surrounded by PFDM.

We analyzed the influence of the PFDM parameter ($b$) and the BI-NED
parameter ($\beta$) on the roots of AdS black hole configurations. The
results revealed that the horizon structure of these black holes was highly
sensitive to variations in $b$ and $\beta$. For small $b$, two horizons
appeared, while small $\beta$ yielded only a single event horizon (similar
to Schwarzschild-like black hole). At intermediate values of both
parameters, three distinct horizons were observed, representing
multi-horizon black holes. When $b$ became large, the geometry approached
the Schwarzschild-like limit with one event horizon, whereas large $\beta$
generated two horizons, resembling the Reissner\^{a}\euro ``Nordstr\"{o}m
case. Furthermore, increasing $b$ led to an expansion of the event horizon,
while increasing $\beta$ resulted in its reduction. These findings
underscored the coupled role of dark matter effects and nonlinear
electrodynamics in shaping the geometric properties of AdS black holes.

By considering BI-NED AdS black holes surrounded by PFDM, we calculated
their conserved and thermodynamical quantities$-$such as Hawking
temperature, entropy, charge, electric potential, and total mass. We then
examined how the parameters of PFDM and BI-NED affected these quantities and
showed that they satisfied the first law of thermodynamics. The effects of $%
b $ and $\beta$ on the total mass was very interesting. Indeed, we found
that the asymptotic behavior of total mass $M$ remained similar for various
values of $b$, $q$, and $\beta$, as it depended only on the cosmological
constant. For the negative values of $\Lambda$, the total mass increased
with the size of the black hole. In the high-energy limit the behavior of $M$
varied with $b$: small $b$ yielded a finite positive mass, medium $b$ made $%
M $ approach zero, and large $b$ produced black holes with $M=0$ at finite
radii shifting outward with increasing $b$. This demonstrated that the mass
of small black holes was highly sensitive to PFDM, differing from
Schwarzschild and Reissner\^{a}\euro ``Nordstr\"{o}m cases. Conversely, the
BI-NED parameter affected $M$ oppositely$-$smaller $\beta$ removed mass
divergence and led to finite values, while very large $\beta$ caused
divergent behavior like Reissner\^{a}\euro ``Nordstr\"{o}m black holes.
Overall, the divergence of total mass at $r_{+}=0$ disappeared when $\beta$
decreased or $b$ increased, indicating that PFDM and BI-NED exerted opposite
influences on the mass evolution of AdS black holes.

The local thermodynamic stability of BI-NED AdS black holes surrounded by
PFDM was analyzed through the simultaneous study of heat capacity and
temperature. The results showed that for $b<b_{critical}$, two roots and one
divergence point appeared in the heat capacity, where the divergence point
was located between the two roots. In this regime, large black holes ($%
r_{+}>r_{0_{2}}$) satisfied the local stability condition since both heat
capacity and temperature were positive. When $b>b_{critical}$, only one
divergence point was found, shifting toward larger horizon radii as $b$
increased, and the stability region reduced accordingly. Similarly, for the
BI-NED parameter, a critical value existed where two roots and a divergence
point occurred when $\beta>\beta_{critical}$, indicating stability for $%
r_{+}>r_{0_{2}}$. For $\beta<\beta_{critical}$, one divergence point was
observed, with $C$ and $T$ positive beyond that point. Overall, the study
demonstrated that large BI-NED AdS black holes maintained local
thermodynamic stability, while the stability region varied with both PFDM
and BI-NED parameters.

We studied the global thermodynamic stability of these black holes through
the Helmholtz free energy analysis. For $b<b_{critical}$, two roots of $F$
appeared, where the free energy was negative before the first and after the
second root, implying that both small and large black holes satisfied the
global stability condition. When $b>b_{critical}$, the free energy remained
negative at all radii, meaning the black holes were globally stable. For
very large $b<<b_{critical}$, only large black holes met the global
stability requirement. Regarding the BI-NED parameter, increasing $\beta$
reduced the global stability region, and between two roots, $F$ became
positive, indicating instability for medium-sized black holes. As $\beta \to
\infty$, there was a single root where $F$ turned negative beyond it,
showing that large BI-NED AdS black holes exhibited global thermodynamic
stability for sufficiently large $\beta$.

We compared the heat capacity and the Helmholtz free energy together, and we
found that the large BI-NED AdS black holes surrounded by PFDM could satisfy
the local and global stability conditions simultaneously.

We extended our study to including the cosmological constant as a
thermodynamics pressure. In this space (extended phase space), the conserved
and thermodynamic quantities were calculated and were found to satisfy the
first law of thermodynamics. The specific heat at constant pressure ($C_{P}$%
), the volume expansion coefficient ($\alpha$), and the isothermal
compressibility coefficient ($\kappa _{T}$) were then examined. The results
showed that all three quantities shared a common factor in their
denominators and diverged at the critical point. The Ehrenfest equations
were analytically verified, and both relations were confirmed to be
satisfied. Furthermore, the Prigogine\^{a}\euro ``Defay ratio was computed
and was found to be exactly equal to one, indicating that the phase
transition at the critical point was a second-order transition.

In the extended phase space, the BI-NED AdS black hole surrounded by PFDM
was treated as a thermodynamic heat engine. The effects of the PFDM and
BI-NED parameters on the heat engine efficiency were examined, and it was
found that increasing $b$ led to a decrease in efficiency when $S_{2}$
remained constant. The efficiency was shown to depend sensitively on the
BI-NED parameter $\beta$ because of the hypergeometric function in the
denominator of $\eta$. The analysis revealed that $\eta$ increased with
increasing $\beta$ for a fixed $S_{2}$. Since $S_{2}>S_{1}$, an increase in $%
S_{2}$ resulted in higher heat engine efficiency for all values of $b$ and $%
\beta$.

The effects of PFDM and BI-NED parameters on the Carnot efficiency were also
studied. The results indicated that as $b$ increased, the heat engine
efficiency decreased at constant $S_{2}$, while the Carnot efficiency
increased and approached unity as $S_{2}$ became very large. Furthermore, $%
\eta_{c}$ increased with increasing $\beta$ for fixed $S_{2}$ and also
approached unity for large $S_{2}$.

To compare the efficiencies, the ratio $\frac{\eta}{\eta_c}$ was plotted for
different values of $b$ and $\beta$. This ratio was found to remain below
one, implying a constraint on PFDM parameters: very large values of $b$ were
not physically acceptable since$\frac{\eta}{\eta_c}<1$. However, the BI-NED
parameter $\beta$ imposed no such restriction, as the ratio stayed less than
one for all its values.

We have also performed the topological analysis and classification of our new solution, resulting in an equilibrium between stable and unstable states, with four distinct phases, two of them locally stable and two of them locally unstable.

We further explored the motion of massive and massless particles. For
timelike geodesics, the effective potential approach shows that the PFDM
parameter $b$ significantly modifies the global orbital structure, leading
to non-trivial behavior of the stable and unstable circular orbits,
including a non-monotonic dependence of the ISCO radius on $b$. In contrast,
the BI-NED parameter $\beta $ induces a smooth deformation of the
strong-field region and governs the transition toward the Maxwell limit. For
null geodesics, photon propagation in the effective geometry demonstrates
that both PFDM and nonlinear electrodynamics corrections alter the photon
sphere radius and consequently the shadow characteristics. The PFDM
background mainly affects the large-scale structure of the effective
potential, whereas the BI-NED parameter controls near-horizon deviations
from linear electrodynamics. These results indicate that the combined
effects of dark matter and nonlinear electrodynamics leave distinct imprints
on particle dynamics and optical observables, which may provide potential
signatures in strong gravitational lensing and black hole shadow
measurements.

\begin{acknowledgements}
B. Eslam Panah thanks University of Mazandaran. MER thanks Conselho Nacional de Desenvolvimento Cient\'{\i}fico e Tecnol\'{o}gico $-$ CNPq, Brazil, for partial financial support.
\end{acknowledgements}

\end{document}